\documentclass[aps,prd,a4paper,twocolumn,showpacs,nofootinbib,eqsecnum]{revtex4}
 
\usepackage{amssymb,amsmath,latexsym,mathrsfs}
\usepackage{graphicx}
\usepackage{epsfig}
\usepackage{varioref,xr-hyper}

\usepackage[hypertex,draft=false,verbose=false,debug=false
            hyperindex=true,hypertexnames=true]{hyperref}

\newcommand{\hide}[1]{{}}

\newcommand{\si}[1]{{\scriptscriptstyle{#1}}}

\newcommand{\ads}{\ensuremath{\mathrm{AdS}_{\si{5}}\,}}

\newcommand{\kappafive}{\kappa_{{_5}}}
\newcommand{\kappareduct}{\kappa_{{_4}}}
\newcommand{\tension}{\mathcal{T}}

\newcommand{\bb}{{\si{\bullet}}}
\newcommand{\ve}[1]{\boldsymbol{#1}}
\newcommand{\p}{\partial}
\newcommand{\ud}{\mathrm{d}}
\newcommand{\ub}{\mathrm{b}}

\newcommand{\yreg}{y_{{\mathrm{s}}}}
\newcommand{\ybr}{y_\ub}

\newcommand{\Hb}{h_{\bb}}

\newcommand{\PiTb}{\Pi^{\si{(T)}}_{\bb}}
\newcommand{\TKKR}{\widehat{\mathrm{m}}}  

\newcommand{\HH}{\mathcal{H}}

\newcommand{\vv}{{v}}

\newcommand{\ga}{\gamma}

\newcommand{\dd}{\partial}
\newcommand{\ra}{\rightarrow}
\newcommand{\al}{\alpha}
\newcommand{\om}{\omega}

\def\lsim{\;\raise 0.4ex\hbox{$<$}\kern -0.8em\lower 0.62 ex\hbox{$\sim$}\;}
\def\gsim{\;\raise 0.4ex\hbox{$>$}\kern -0.7em\lower 0.62 ex\hbox{$\sim$}\;}

\begin{document}

\title{Dynamical Casimir effect for gravitons in bouncing braneworlds}
\author{Marcus Ruser}
\email{marcus.ruser@physics.unige.ch}
\author{Ruth Durrer}
\email{ruth.durrer@physics.unige.ch}
\affiliation{D\'epartement de Physique Th\'eorique, Universit\'e de
Gen\`eve, 24 quai Ernest Ansermet, 1211 Gen\`eve 4, Switzerland.}


\begin{abstract}
We consider a two-brane system in five-dimensional anti-de Sitter
space-time. We study particle creation due to the  motion of
the physical brane which first approaches the second static brane
(contraction) and then recedes from it (expansion). The
spectrum and the energy density of the generated gravitons are
calculated. 
We show that the massless gravitons have a blue spectrum and that their energy
density satisfies the nucleosynthesis bound with very mild constraints
on the parameters. We also show that the Kaluza-Klein modes cannot
provide the dark matter in an anti-de-Sitter braneworld.
However, for natural choices of parameters, backreaction from the 
Kaluza-Klein gravitons may well become important.
The main findings of this work have been published in form 
of a Letter [R.Durrer and M.Ruser, Phys. Rev. Lett. {\bf 99}, 071601
  (2007), arXiv:0704.0756]. 
\end{abstract}


\maketitle

\section{Introduction}
\begin{figure}
\centering
\includegraphics[height=5cm]{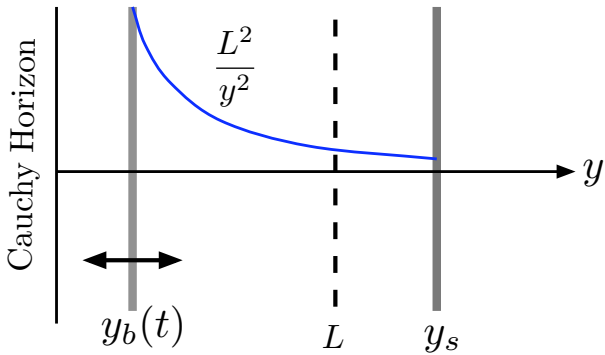}
\caption{\label{f:moving brane} Two branes in an AdS$_5$ spacetime, with 
  $y$ denoting the fifth dimension and $L$ the AdS curvature scale. The physical
  brane is on the left at time dependent position $y_b(t)$. 
  While it is approaching the static brane its
  scale factor is decreasing and when it moves away from the static
  brane it is expanding [cf.~Eq.~(\ref{e:a of y})]. 
  The value of the scale factor of the brane
  metric as function of the extra dimension $y$ is also indicated.}
\end{figure}

%
In recent times, the possibility that our observed Universe might
represent a hypersurface in a higher-dimensional space-time has
received considerable attention. The main motivation for this idea is
the fact, that string
theory~\cite{Polchinski:1998rq,Polchinski:1998rr}, which is consistent
only in ten spac-etime dimensions (or 11 for M--theory) allows for solutions where
the standard model particles (like fermions and gauge bosons) are
confined to some hypersurface, called the {\it brane}, 
and only the graviton can propagate in the whole space-time, the
{\it bulk}~\cite{Polchinski:1998rr,Polchinski:1995mt}. Since gravity is
not well constrained at small distances, the dimensions normal 
to the brane, the extra dimensions, can be as large as 0.1mm. 
\\
Based on this feature, Arkani-Hamed, Dimopoulos and Dvali (ADD)
proposed a braneworld model where the presence 
of two or more flat extra-dimensions 
can provide a solution to the  hierarchy problem, the
problem of the huge difference between the Planck scale and the
electroweak scale~\cite{Arkani-Hamed:1998rs,Arkani:1999}.
\\
In 1999 Randall and Sundrum (RS) introduced a model with one 
extra dimension, where the bulk is a slice of five-dimensional 
anti de-Sitter (AdS) space.
Such curved extra dimensions are also referred to as 
{\it warped extra dimensions}.
While in the RS I model \cite{Randall:1999ee} with two flat branes 
of opposite tension at the edges of the bulk the warping
leads to an interesting solution of the hierarchy problem, 
it localizes four-dimensional gravity on a single positive tension
brane in the RS II model \cite{Randall:1999vf}. 
\\
Within the context of warped braneworlds, cosmological evolution, 
i.e., the expansion of the Universe, can be understood as 
the motion of the brane representing our Universe through
the AdS bulk.
Thereby the Lanczos-Sen-Darmois-Israel-junction
conditions~\cite{Lanczos:1924,Sen:1924,Darmois:1927,Israel:1966},
relate the energy-momentum tensor on the brane to the extrinsic
curvature and hence to the brane motion which is described
by a modified Friedmann equation.
At low energy, however, the usual Friedmann equations for the expansion 
of the Universe are recovered \cite{Kraus:1999it,Binetruy:1999hy}.
\\
\\
Since gravity probes the extra dimension, gravitational 
perturbations on the brane, i.e. in our Universe, 
carry five-dimensional effects in form of massive 
four-dimensional gravitons, the so-called Kaluza-Klein 
(KK) tower.  
Depending on the particular brane trajectory, 
these perturbations may be significantly amplified
leading to observable consequences, for example,  
a stochastic gravitational wave background.
(For a review of stochastic gravitational 
waves see \cite{mm}.)
This amplification mechanism is identical to the dynamical 
Casimir effect for the electromagnetic field in 
cavities with dynamical walls (moving mirrors); see 
\cite{Ruser:2004,Ruser:2005xg,Ruser:2006xg}
and references therein.
In the quantum field theoretical language, such 
an amplification corresponds to the creation 
of particles out of vacuum fluctuations.
Hence, in the same way a moving mirror leads to production 
of photons, the brane moving through the bulk causes 
creation of gravitons.
Thereby, not only the usual four-dimensional graviton might be 
produced, but also gravitons of the KK 
tower can be excited.
Those massive gravitons are of particular interest, since 
their energy density could dominate the energy density
of the Universe and spoil the phenomenology if their production 
is sufficiently copious. 
\\
\\
The evolution of cosmological perturbations under 
the influence of a moving brane has been the subject 
of many studies during recent years.  
Since one has to deal with partial 
differential equations and time-dependent boundary
conditions, the investigation of the 
evolution of perturbations in the background 
of a moving brane is quite complicated.
Analytical progress has been made based on approximations like
the ``near brane limit'' and a slowly moving brane
\cite{Battye:2004a,Battye:2004b,Easther:2003,Kobayashi:2004ana}.
\\
The case of de Sitter or quasi-de Sitter inflation
on the brane has been investigated analytically in
\cite{Gorbunov:2001ge,Kobayashi:2003,Maartens:2000,
Langlois:2000,Frolov:2002qm}.
In \cite{Langlois:2000} it is demonstrated that during 
slow-roll inflation (modeled as a period of quasi-de Sitter 
expansion) the standard four-dimensional result for the
amplitude of perturbations is recovered at low energies 
while it is enhanced at high energies. 
\\
However, most of the effort has gone into 
numerical simulations
\cite{Hiramatsu:2004,Hiramatsu:2005,Hiramatsu:2006,Koyama:2004cf,
Ichiki:2004a,Ichiki:2004b,Kobayashi:2005,Kobayashi:2006a,
Kobayashi:2006b,Seahra:2006}, 
in particular in order to investigate the high-energy regime.
Thereby different coordinate systems have been used 
for which the brane is at rest, and different 
numerical evolution schemes have been employed 
in order to solve the partial differential equation. 
\\
\\
In this work we chose a different way of looking at 
the problem. 
We shall apply a formalism used to describe the dynamical Casimir effect 
to study the production of gravitons in braneworld cosmology.
This approach and its numerical implementation 
offers many advantages.
The most important one is the fact that this approach
deals directly with the appearing mode couplings by means of 
coupling matrices. (In \cite{Battye:2004b} a similar approach
involving coupling matrices has been used. 
However, perturbatively only, and not 
in the complexity presented here.)
Hence, the interaction between the four-dimensional
graviton and the KK modes is not hidden within 
a numerical simulation but can directly be investigated 
making it possible to reveal the underlying physics 
in a very transparent way.
\\
\\
We consider a five-dimensional anti-de Sitter spacetime
with two branes in it; a moving positive tension brane
representing our Universe and a second brane 
which, for definiteness, is kept at rest.   
This setup is depicted in Fig.~\ref{f:moving brane}.
For this model we have previously shown that in a radiation dominated
Universe, where the second, fixed brane is arbitrarily far away, no
gravitons are produced~\cite{Cartier:2005}. 
\\
\\
The particular model which we shall consider 
is strongly motivated by the {\it ekpyrotic} 
or {\it cyclic Universe} and similar ideas 
\cite{Khoury:2001,Kallosh:2001,Neronov:2001,Steinhardt:2002,
Khoury:2002a,Khoury:2002b,Khoury:2003,
Khoury:2004,Tolley:2004}.
In this model, roughly speaking, the hot big bang corresponds
to the collision of two branes; a moving bulk brane which 
hits ``our'' brane, i.e. the observable Universe.
Within such a model, it seems to be possible to address 
all major cosmological problems (homogeneity, origin of density
perturbations, monopole problem) without invoking the 
paradigm of inflation.
For more details see \cite{Khoury:2001} but also 
\cite{Kallosh:2001} for critical comments.
\\
One important difference between the ekpyrotic model 
and standard inflation is that in the latter one 
tensor perturbations have a nearly scale invariant spectrum.
The ekpyrotic model, on the other hand, predicts a strongly blue
gravitational wave spectrum with spectral tilt
$n_T \simeq 2$ \cite{Khoury:2001}.
This blue spectrum is a key test for the ekpyrotic 
scenario since inflation always predicts a slightly red spectrum
for gravitational waves.
One method to detect a background of primordial 
gravitational waves of wavelengths comparable to 
the Hubble horizon today is the 
polarization of the cosmic microwave background.
Since a strongly blue spectrum of gravitational 
waves is unobservably small for large length scales, 
the detection of gravitational waves in the cosmic 
microwave background polarization would falsify 
the ekpyrotic model \cite{Khoury:2001}.
\\
\\
Here we consider a simple specific model which 
is generic enough to cover important main 
features of the generation and evolution of gravitational waves 
in the background of a moving brane whose 
trajectory involves a bounce.
First, the physical brane moves towards the static brane, 
initially the motion is very slow. 
During this phase our Universe is contracting, i.e. the scale 
factor on the brane decreases, the energy density on the brane
increases and the motion becomes faster. 
We suppose that the evolution of the brane is driven by a radiation 
component on the brane, and that at some more or less close encounter of the 
two branes which we call the {\it bounce}, some high-energy mechanism which we do not 
want to specify in any detail, turns around the motion of
the brane leading to an expanding Universe.
Modeling the transition from contraction to 
subsequent expansion in any detail would require 
assumptions about unknown physics.
We shall therefore ignore results which depend on the details of the
transition. 
Finally the physical brane moves away from the static brane
back towards the horizon with expansion first fast 
and then becoming slower as the energy density drops.
This model is more similar to the {\it pyrotechnic Universe} 
of Kallosh, Kofman and Linde \cite{Kallosh:2001} where 
the observable Universe is also represented by a positive 
tension brane rather than to the ekpyrotic model where our 
brane has negative tension.
\\
We address the following questions: What is the
spectrum and energy density of the produced gravitons, the massless
zero mode and the KK modes? Can the
graviton production in such a brane Universe lead to limits, e.g. on
the  AdS curvature scale via the nucleosynthesis bound? Can the
KK modes provide the dark matter or lead to stringent limits
on these models? 
Similar results could be obtained for the free gravi-photon
and gravi-scalar, i.e. when we neglect the perturbations of the brane
energy momentum tensor which also couple to these gravity wave modes
which have spin-1 respectively spin-0 on the brane.
\\
\\
The reminder of the paper is organized as follows. 
After reviewing the basic equations of braneworld 
cosmology and tensor perturbations 
in Sec. II, we discuss the dynamical Casimir effect 
approach in Sec. III. 
In Sec. IV we derive expressions for the energy density and the power
spectrum of gravitons.
Thereby we show that, very generically, KK gravitons cannot play the role
of dark matter in warped braneworlds. 
This is explained by the localization of gravity on the 
moving brane which we discuss in detail.
Section~V is devoted to the presentation and discussion
of our numerical results.
In Sec.~VI we reproduce some of the numerical results with 
analytical approximations and we derive fits for the number of 
produced gravitons. 
We discuss our main results and their implications for
bouncing braneworlds in Sec.~VII and 
conclude in Sec.~VIII.
Some technical aspects are collected in appendices. 
\\
The main and most important results of this rather long and technical 
paper are published in the Letter~\cite{letter}. 
%
\section{Gravitons in moving braneworlds}

\subsection{A moving brane in AdS$_5$}
%
We consider a AdS-5 spacetime. In Poincar\'e coordinates, the
bulk metric is given by
\begin{equation}\label{e:bulk-metric}
 ds^2
 = g_{\si{AB}} d x^{\si{A}} d  x^{\si{B}}
 = \frac{L^2}{y^2} \left[-d t^2 + \delta_{ij} d x^i d x^j + \ud
 y^2\right]~.
\end{equation}
The physical brane (our Universe) is located at some time dependent
position $y=y_b(t)$,
while the 2nd brane is at fixed position $y=y_s$ (see Fig.~\ref{f:moving brane}).
The induced metric on the  physical brane is given by
\begin{eqnarray}
 ds^2
 &=& \frac{L^2}{y_b^2(t)}
   \left[-\left(1- \left(\frac{dy_b}{dt}\right)^2 \right)d t^2
   +\delta_{ij}d x^i d x^j\right] \nonumber \\
 &=& a^2(\eta)\left[ -\ud\eta^2 +\delta_{ij}d x^id x^j\right]~,
 \label{eq:branemetric}
\end{eqnarray}
where 
\begin{equation}
a(\eta)=\frac{L}{y_b(t)}
\label{e:a of y}
\end{equation}
is the scale factor and $\eta$ denotes the conformal time 
of an observer on the brane,
\begin{equation}
 d\eta = \sqrt{1- \left(\frac{dy_b}{dt}\right)^2}dt
         \equiv \ga^{-1}dt~.
\end{equation}
We have introduced the brane velocity
\begin{eqnarray}
 v &\equiv& \frac{dy_b}{dt} = -\frac{LH}{\sqrt{1+L^2H^2}}\qquad \mbox{
 and } \label{e:latetime}\\
\ga  &=& \frac{1}{\sqrt{1-v^2}} = {\sqrt{1+L^2H^2}}~.
\end{eqnarray}
Here $H$ is the usual Hubble parameter,
\begin{equation}
H \equiv \dot a/a^2 \equiv a^{-1}{\cal H}= -L^{-1}\ga v~,
\end{equation}
and an overdot denotes the derivative with respect to conformal time $\eta$.
The bulk cosmological constant $\Lambda$ is related to the curvature
scale $L$ by $\Lambda = -6/L^2$. 
The junction conditions on the brane
lead to~\cite{CR,Cartier:2005}
\begin{align}
\kappafive({\rho}+\tension)
&= 6\frac{\sqrt{1+L^2H^2}}{L}~, 
\label{e:hub1} \\
\kappafive({\rho}+ {P})
&= -\frac{2L\dot{H}}{a\sqrt{1+L^2H^2}}~.
\label{e:ij}
 \end{align}
Here ${\cal T}$ is the brane tension and $\rho$ and $P$ denote the 
energy density and pressure of the matter confined on the brane.
Combining (\ref{e:hub1}) and (\ref{e:ij}) results in
\begin{equation}
\dot{{\rho}} =-3Ha({\rho}+P)~,
\label{e:T:continuity}
\end{equation}
while taking the square of (\ref{e:hub1}) leads to 
\begin{equation}
H^2 = \frac{\kappafive^2}{18}\tension{\rho}
\left(1+\frac{{\rho}}{2\tension}\right)
+\frac{\kappafive^2\tension^2}{36}-\frac{1}{L^2}~.
\label{e:T:Hubble}
\end{equation}
These equations form the basis of brane cosmology and have been
discussed at length in the literature (for reviews
see~\cite{Maartens:2003tw,Durrer:2005dj}). 
The last equation is called the {\it modified Friedmann equation} for brane
cosmology~\cite{Binetruy:1999hy}. 
For usual matter with $\rho+P>0$,
$\rho$ decreases during expansion and at sufficiently late time
$\rho\ll \tension$. 
The ordinary four-dimensional Friedmann equation is
then recovered if
\begin{equation}
 \frac{\kappafive^2 \tension^2}{12} = \frac{3}{L^2}
 \quad\text{and we set}\quad
 \kappareduct =8\pi G_4 = \frac{\kappafive^2 \tension}{6}~.
 \label{e:fine}
\end{equation}
Here we have neglected a possible four-dimensional cosmological
constant. 
The first of these equations is the RS fine tuning implying
\begin{equation}
\kappa_5=\kappa_4\,L~. 
\label{e:RS fine tuning 0}
\end{equation}
Defining the string and Planck scales by 
\begin{equation}
\kappa_5  =\frac{1}{M_5^3} = L_s^3~, \qquad
\kappa_4 = \frac{1}{M_{\rm Pl}^2}= L_{\rm Pl}^2~,
\label{e:string and Pl scale}
\end{equation}
respectively, the RS fine tuning condition
leads to 
\begin{equation}
\frac{L}{L_s} = \left( \frac{L_s}{L_{\rm Pl}} \right)^2.
\label{e:RS fine tuning}
\end{equation}
\\
As outlined in the introduction, we shall be interested mainly in a 
radiation dominated low-energy phase, 
hence in the period where
\begin{equation}
P = \frac{1}{3}\rho ~ \mbox{ and }\quad |v|\ll 1~ \mbox{ so that }\quad
 \ga \simeq 1~, d\eta \simeq dt~.
\end{equation}
In such a period, the solutions to the above equations are of the form
\begin{eqnarray}
a(t) &=& \frac{|t| + t_b}{L}\,, \label{e:solexp}\\
y_b(t) &=& \frac{L^2}{|t| + t_b}\,,\label{e:yb}\\
v(t) &=& -\frac{\mathrm{sgn}(t)L^2}{(|t|+t_b)^2} \simeq -HL~.
\label{e:vb}
\end{eqnarray}
Negative times ($t < 0$) describe a contracting phase, 
while positive times ($t > 0$)
describe radiation dominated expansion. 
At $t=0$, the scale factor
exhibits a kink and the evolution equations are singular. 
This is the bounce which we shall not model in detail, but we will have to
introduce a cutoff in order to avoid ultraviolet
divergencies in the total particle number and energy density
which are due to this unphysical kink.
We shall show, that when the kink is smoothed out at some
length scale, the production of particles (KK gravitons)
of masses larger than this scale is exponentially suppressed,
as it is expected.
The (free) parameter $t_b>0$ determines the value of the scale 
factor at the bounce $a_b$, i.e. the minimal interbrane distance,
as well as the velocity at the bounce $v_b$
\begin{equation}
a_b = a(0) = \frac{1}{\sqrt{v_b}}\;,\;\;
|v(0)| \equiv v_b = \frac{L^2}{t_b^2}~.
\label{e:bounce values}
\end{equation}
Apparently we have to demand $t_b > L$ which implies $y_b(t)<L$.
%
\subsection{Tensor perturbations in AdS$_5$}
%
We now consider tensor perturbations on this background.
Allowing for tensor perturbations $h_{ij}(t,{\bf x},y)$
of the spatial three-dimensional geometry at fixed $y$, 
the bulk metric reads 
\begin{align}
 \ud s^2 = \frac{L^2}{y^2}
 \left[-\ud t^2+(\delta_{ij}+2h_{ij})\ud x^i \ud x^j+\ud y^2 \right]~.
\end{align}
Tensor modes satisfy the traceless and transverse conditions, $h_i^i =
\p_ih^i_j = 0$. 
These conditions imply that $h_{ij}$ has only two independent degrees 
of freedom, the two polarization states $\bullet=\times,+$.
We decompose $h_{ij}$ into spatial Fourier modes,
\begin{equation}
 h_{ij}(t,\ve{x},y)
 = \int \frac{d^3k}{(2\pi)^{3/2}} \sum_{\bb=+,\times}
  e^{i\ve{k}\cdot\ve{x}}e_{ij}^{\bb}({\bf k})\Hb(t,y;{\bf k})~,
\label{e:h fourier decomposition}
\end{equation}
where $e_{ij}^{\bb}({\bf k})$ are unitary constant transverse-traceless
polarization tensors which form a basis of the two polarization
states $\bullet = \times,+$.
For $h_{ij}$ to be real we require
\begin{equation}
h_\bullet^*(t, y; {\bf k}) = h_\bullet(t, y; -{\bf k}).
\label{reality condition for h}
\end{equation}
The perturbed Einstein equations yield the equation of motion for the mode functions
$\Hb$, which obey the Klein-Gordon equation for minimally coupled
massless scalar fields in
\ads~\cite{Hawking:2000kj,Hawking:2000bb,Langlois:2000}
\begin{equation}
 \left[\p_t^2 +k^2 -\p_y^2 + \frac{3}{y}\p_y \right] \Hb(t,y;{\bf k}) = 0~.
\label{e:T-bulk-eq}
\end{equation}
In addition to the bulk equation of motion the modes also satisfy
a boundary condition at the brane coming from the second junction
condition,
\begin{eqnarray}
\hspace{-7mm} \left.
\left[LH\p_t \Hb -\sqrt{1+L^2H^2}\p_y\Hb\right]\right|_{\ybr}
 &=& \nonumber\\
-\left.\ga\left(\vv\dd_t +\dd_y\right)\Hb \right|_{\ybr}
 &=& \frac{\kappafive}{2}aP\PiTb   \,.
 \label{e:T-JC-general}
\end{eqnarray}
Here $\PiTb$ denotes possible anisotropic stress perturbations in the 
brane energy momentum tensor.
We are interested in the quantum production of free gravitons, not in
the coupling of gravitational waves to matter. 
Therefore we shall set $\PiTb=0$ in the sequel, i.e. we make 
the assumption that the Universe is filled with a 
perfect fluid.
Then, (\ref{e:T-JC-general}) reduces to 
\footnote{In Equations (4) and (8) of our Letter \cite{letter} 
two sign mistakes have creeped in.} 
\begin{equation}
\left . \left(\vv\dd_t +\dd_y\right)\Hb \right|_{y_b(t)}=0~.
\label{e:T-JC-simple}
\end{equation}
This is not entirely correct for the
evolution of gravity modes since at late times, when matter on the
brane is no longer a perfect fluid (e.g., free-streaming neutrinos)
and anisotropic stresses 
develop which slightly modify the evolution of gravitational waves. We
neglect this subdominant effect in our treatment.
(Some of the difficulties which appear when $\PiTb\neq0$ are 
discussed in \cite{CR}.)
\\
The wave equation (\ref{e:T-bulk-eq}) together with the boundary condition
(\ref{e:T-JC-simple}) can also be obtained by variation of the action 
\begin{eqnarray}
{\cal S}_h = 2\,\frac{L^3}{2\kappa_5} \sum_\bullet \int dt \int d^3k 
\int_{y_b(t)}^{y_s} \frac{dy}{y^3} \times \nonumber \\
\times \Big[|\partial_t h_\bullet|^2 
- |\partial_y h_\bullet|^2
-k^2|h_\bullet|^2 \Big]~,
\label{e:action h}
\end{eqnarray}
which follows from the second order perturbation of the
gravitational Lagrangian.
The factor 2 in the action is due to ${\mathbb Z}_2$ symmetry.
Indeed, Equation (\ref{e:T-JC-simple}) is the only  
boundary condition for the perturbation amplitude $h_\bullet$
which is compatible with the variational principle
$\delta {\cal S}_h = 0$, except if $h_\bullet$ is constant 
on the brane.
Since this issue is important in the following, 
it is discussed more detailed in Appendix 
\ref{a:variation}. 
%
\subsection{Equations of motion in the late time/low energy limit}
%
In this work we restrict ourselves to relatively late 
times, when 
\begin{equation}
\rho\tension\gg \rho^2\;\; 
{\rm and}\;\;{\rm therefore} \;\;|v|\ll 1. 
\label{e:low energy approach}
\end{equation}
In this limit the conformal time on
the brane agrees roughly with the 5D time coordinate, 
$d\eta \simeq dt$ and we shall therefore
not distinguish these times; we set $t=\eta$. 
\\
We want to study the quantum mechanical evolution of 
tensor perturbations within a canonical formulation
similar to the dynamical Casimir effect for the 
electromagnetic field in dynamical cavities 
\cite{Ruser:2004,Ruser:2006xg,Ruser:2005xg}. 
In order to pave the way for canonical quantization, 
we have to introduce a suitable set of functions 
allowing the expansion of the perturbation amplitude $h_\bullet$ 
in canonical variables. 
More precisely, we need a complete and orthonormal 
set of eigenfunctions $\phi_\alpha$ of the spatial
part $-\partial_y^2 + \frac{3}{y} \partial_y
=-y^3\partial_y\left[y^{-3} \partial_y\right]$ 
of the differential operator \eqref{e:T-bulk-eq}.
The existence of such a set depends on the 
boundary conditions and is ensured if the 
problem is of Sturm-Liouville type 
(see, e.g.,\cite{partial}).
For the junction condition (\ref{e:T-JC-simple}),
such a set does unfortunately not exist due to the time
derivative.  
One way to proceed would be to introduce 
other coordinates along the lines of \cite{Crocce} 
for which the junction condition reduces to a 
simple Neumann boundary condition leading to a problem
of Sturm-Liouville type. 
This transformation is, however, relatively complicated to 
implement without approximations and is the subject 
of future work. 
\\
Here we shall proceed otherwise, harnessing the fact that 
we are interested in low energy effects only, i.e.
in small brane velocities. 
Assuming that one can neglect the time derivative 
in the junction condition since $|v| \ll 1$,  
Eq.~\eqref{e:T-JC-general} reduces to a simple Neumann 
boundary condition.
We shall therefore work with the boundary conditions 
\begin{equation}
\left. \p_y\Hb\right|_{\ybr} = \left. \p_y\Hb\right|_{y_s} =  0~.
\label{boundary conditions}
\end{equation}
Then, at any time $t$ the eigenvalue problem for the spatial
part of the differential operator \eqref{e:T-bulk-eq}
\begin{eqnarray}
\left[ -\partial_y^2 + \frac{3}{y} \partial_y\right] \phi_\alpha(t,y) &=&
-y^3\partial_y\left[y^{-3} \partial_y \phi_\alpha(t,y)\right]  \nonumber \\
&=& m_\alpha^2(t)\phi_\alpha(t,y)~
\label{laplacian equation}
\end{eqnarray}
is of Sturm-Liouville type if we demand that the $\phi_\alpha$'s
are subject to the boundary conditions (\ref{boundary conditions}).
Consequently, the set of eigenfunctions 
$\{\phi_\alpha(t,y)\}_{\alpha = 0}^\infty$ is 
complete, 
\begin{equation}
2\,\sum_\alpha \phi_\alpha(t,y) \phi_\alpha(t,\tilde{y}) =
\delta(y-\tilde{y}) y^3~,
\end{equation}
and orthonormal with respect to the inner-product
\begin{equation}
(\phi_\alpha, \phi_\beta) = 2\,\int_{y_b(t)}^{y_s} \frac{dy}{y^3} \phi_\alpha(t,y)\phi_\beta(t,y)
= \delta_{\alpha\beta}.
\label{e:T:inner-product}
\end{equation}
Note the factor 2 in front of both expressions which is
necessary in order to take the ${\mathbb Z}_2$ 
symmetry properly into account.
\\
The eigenvalues $m_\alpha(t)$ are time-dependent and discrete 
due to the time-dependent but finite distance between 
the branes and 
the eigenfunctions $\phi_\alpha(t,y)$ are time-dependent 
in particular because of the time dependence of the boundary 
conditions (\ref{boundary conditions}).
The case $\alpha = 0$ with $m_{0} = 0$  is the
zero mode, i.e.  the massless four-dimensional graviton.
Its general solution in accordance with the boundary conditions is
just a constant with respect to the extra dimension,
$\phi_0(t,y) = \phi_0(t)$, and is fully determined by 
the normalization condition
$(\phi_0, \phi_0) =1$:
\begin{equation}
\phi_0(t) = \frac{y_s y_b(t)}{\sqrt{y_s^2 - y_b^2(t)}}.
\label{zero mode phi}
\end{equation}
For $\alpha = i \in \{ 1,2,3,\cdots,\}$ with eigenvalues
$m_i > 0$, the general solution of \eqref{laplacian equation} is a combination
of the Bessel functions $J_2\left(m_i(t)\,y\right)$ and
$Y_2\left(m_i(t)\,y\right)$.
Their particular combination is determined by the boundary condition at the moving brane.
The remaining boundary condition at the static brane selects the possible values for the
eigenvalues $m_i(t)$, the KK masses.
For any three-momentum ${\bf k}$ these masses build up an entire tower
of momenta in the $y$-direction; the fifth dimension.
Explicitely, the solutions $\phi_i(t,y)$ for the KK modes read
\footnote{Note that we have changed the parameterization of the
  solutions with respect to \cite{Cartier:2005} for technical reasons.
There, we also did not take into account the factor 2 related to 
${\mathbb Z}_2$ symmetry.}
\begin{equation}\label{e:phij}
\phi_i(t,y) = N_i (t) y^2 {\cal C}_2\left(m_i(t) \,y\right)
\end{equation}
with
\begin{equation}
{\cal C}_\nu (m_i y) = Y_1(m_i y_b) J_\nu(m_i y) - J_1(m_i y_b) Y_\nu(m_i y).
\end{equation}
The normalization reads
\begin{equation}
N_i(t,y_b,y_s) = \left[ \frac{1}{y_s^2{\cal C}_2^2(m_i\,y_s) -
\left(2/(m_i\pi)\right)^2}\right]^\frac{1}{2}
\label{e:normalization 1}
\end{equation}
where we have used that
\begin{equation}
{\cal C}_2(m_i\,y_b) = \frac{2}{\pi\,m_i\,y_b}~.
\label{C2 on brane}
\end{equation}
It can be simplified further by using
\begin{equation}
{\cal C}_2(m_i\,y_s) = \frac{Y_1(m_i\,y_b)}{Y_1(m_i\,y_s)}\frac{2}{\pi\,m_i\,y_s}
\end{equation}
leading to
\begin{equation}
N_i = \frac{m_i \pi}{2}\left[ \frac{Y_1^2(m_i y_s)}
{Y_1^2(m_i y_b) - Y_1^2(m_i y_s)}\right]^{\frac{1}{2}}.
\label{e:normalization 2}
\end{equation}
Note that it is possible to have $Y^2_1(m_i\,y_s) -
Y^2_1(m_i\,y_b)=0$. But then both $Y_1^2(m_i y_s)=Y_1^2(m_i y_b)=0$ and
Eq.~(\ref{e:normalization 2}) has to be understood as a limit.
For that reason, the expression (\ref{e:normalization 1}) for the
normalization is used in the numerical simulations later on. 
Its denominator remains always finite.
\\
The time-dependent KK masses $\{ m_i(t) \}_{i=1}^\infty$ 
are determined by the condition
\begin{equation}
{\cal C}_1 \left(m_i(t) y_s\right) = 0~.
\label{e:zero equation}
\end{equation}
Because the zeros of the cross product of the Bessel functions $J_1$
and $Y_1$ are not known analytically in closed form,
the KK-spectrum has to be determined by solving
Eq. \eqref{e:zero equation} numerically \footnote{Approximate
expressions for the zeros can be  found in~\cite{Abramowitz}.}.
An important quantity which we need below is the rate of change 
$\dot{m}_i/m_i$ of a KK mass given by
\begin{equation}
\hat{m}_i \equiv \frac{\dot{m}_i}{m_i} = \hat{y}_b\,\frac{4}{m_i^2\,\pi^2}\,N_i^2
\label{e:m hat}
\end{equation}
where the rate of change of the brane motion 
$\hat{y}_b$ is just the Hubble parameter on the brane
\begin{equation}
\hat{y}_b(t) \equiv \frac{\dot{y}_b(t)}{y_b(t)} \simeq -Ha =
  -\frac{\dot a}{a} = -\HH ~.
\end{equation}
\\
\\
On account of the completeness of the eigenfunctions
$\phi_\alpha(t,y)$ the gravitational wave amplitude
$h_\bullet(t,y;{\bf k})$ subject to the boundary conditions 
(\ref{boundary conditions}) can now be expanded as
\begin{equation}
h_\bullet (t,y;{\bf k}) = \sqrt{\frac{\kappa_5}{L^3}}\sum_{\alpha = 0}^\infty
q_{\alpha,{\bf k}, \bullet}(t)\phi_\alpha(t,y)~.
\label{e: mode expansion}
\end{equation}
The coefficients  $q_{\alpha,{\bf k}, \bullet}(t)$
are canonical variables describing the time evolution of 
the perturbations and the factor $\sqrt{\kappa_5/L^3}$ 
has been introduced in order to render the 
$q_{\alpha,{\bf k}, \bullet}$'s canonically normalized.
In order to satisfy (\ref{reality condition for h}) we have to impose
the same condition for the canonical variables, 
i.e. 
\begin{equation}
q_{\alpha,{\bf k}, \bullet}^* = q_{\alpha,{\bf -k}, \bullet}.
\label{e:reality for q}
\end{equation}
\\
One could now insert the expansion (\ref{e: mode expansion}) 
into the wave equation (\ref{e:T-bulk-eq}), 
multiplying it by $\phi_\beta(t,y)$ and integrating out the 
$y-$dependence by using the orthonormality 
to derive the equations of motion for the variables
$q_{\alpha, {\bf k},\bullet}$.
However, as we explain in Appendix 
\ref{a:variation}, a Neumann boundary condition 
at a moving brane is not compatible with a free 
wave equation. 
The only consistent way to implement the boundary 
conditions (\ref{boundary conditions}) is therefore
to consider the action (\ref{e:action h}) of the perturbations
as the starting point to derive the equations of
motion for $q_{\alpha,{\bf k},\bullet}$.  
Inserting (\ref{e: mode expansion}) into (\ref{e:action h}) 
leads to the canonical action 
\begin{align}
{\cal S} = \frac{1}{2}\sum_\bullet \int dt \int d^3k &\Big\{
\sum_\alpha \left[ |\dot{q}_{\alpha,{\bf k},\bullet}|^2 -
\omega_{\alpha,k}^2|q_{\alpha,{\bf k},\bullet}|^2\right] \nonumber \\
+\sum_{\alpha\beta}&\big[M_{\alpha\beta} \left(q_{\alpha,{\bf k},\bullet}
\dot{q}_{\beta,{\bf -k},\bullet} + q_{\alpha,{\bf -k},\bullet}
\dot{q}_{\beta,{\bf k},\bullet}\right) \nonumber \\
&+ N_{\alpha\beta}
q_{\alpha,{\bf k},\bullet}q_{\beta,{\bf -k},\bullet}\big]\Big\}~.
\end{align}
We have introduced the time-dependent frequency of a graviton mode
\begin{equation}
\om_{\al, k}^2 = \sqrt{ k^2 +m_\al^2}\;\;,\;\;k = |{\bf k}|\;,
\end{equation}
and the time-dependent coupling matrices 
\begin{eqnarray}
M_{\alpha\beta} &=& (\partial_t \phi_\alpha, \phi_\beta)~,\\
N_{\alpha\beta} &=& (\partial_t \phi_\alpha, \partial_t
\phi_\beta)=\sum_\gamma M_{\alpha\gamma}M_{\beta\gamma}~
\end{eqnarray}
which are given explicitely in Appendix \ref{a:MN}
(see also \cite{Cartier:2005}).
Consequently, the equations of motion for the canonical variables  
are
\begin{eqnarray}
\ddot{q}_{\alpha,{\bf k}, \bullet} &+& \omega_{\alpha, k}^2 q_{\alpha,{\bf k}, \bullet} + \sum_\beta
\left[M_{\beta\alpha} - M_{\alpha\beta}\right]\dot{q}_{\beta,{\bf k}, \bullet}
\nonumber \\
&+&\sum_\beta\left[\dot{M}_{\alpha\beta} -
  N_{\alpha\beta}\right]q_{\beta,{\bf k}, \bullet} = 0 ~.
\label{deq for q}
\end{eqnarray}
\\
\\
The motion of the brane through the bulk, i.e. the expansion 
of the Universe, is encoded in the time-dependent coupling matrices 
$M_{\alpha\beta}, N_{\alpha\beta}$.
The mode couplings are caused by the time-dependent boundary condition
$\partial_y h_\bullet(t,y)|_{y_b} = 0$ which forces the eigenfunctions
$\phi_\alpha(t,y)$ to be explicitly time-dependent.
In addition, the frequency of a KK mode $\omega_{\alpha,k}$
is also time-dependent since the distance between the two branes 
changes when the brane is in motion.
Both time-dependencies can lead to the amplification 
of tensor perturbations and, within a quantum theory 
which is developed in the next section, 
to graviton production from vacuum.
\\
Because of translation invariance with respect to the 
directions parallel to the brane, modes with different 
${\bf k}$ do not couple in (\ref{deq for q}).
The three-momentum ${\bf k}$ enters the equation of motion 
for the perturbation only via the frequency $\omega_{\alpha,k}$, 
i.e. as a global quantity. 
Equation (\ref{deq for q}) is similar to the equation
describing the time-evolution of electromagnetic field modes 
in a three-dimensional dynamical cavity \cite{Ruser:2005xg} 
and may effectively be described by a massive scalar field 
on a time-dependent interval \cite{Ruser:2006xg}.
For the electromagnetic field, the dynamics of the 
cavity, or more precisely the motion of one of its walls,
leads to photon creation from vacuum fluctuations.
This phenomenon is usually referred to as dynamical 
Casimir effect.
Inspired by this, we shall call the production of gravitons
by the moving brane as {\it dynamical Casimir effect for 
gravitons}.
%
\subsection{Remarks and comments}
%
In \cite{Cartier:2005} we have already shown that 
in the limit where the fixed brane is sent off to infinity, 
$y_s \rightarrow \infty$, only the $M_{00}$ matrix element 
survives with $ M_{00} = -{\cal H}[1+{\cal O}(\epsilon)]$
and $\epsilon = y_b/y_s$.
$M_{00}$ expresses the coupling of the zero mode 
to the brane motion. 
Since all other couplings disappear for $\epsilon \rightarrow 0$
all modes decouple from each other and, in addition, the 
canonical variables for the KK modes decouple from 
the brane motion itself.
This has led to the result that at late times and 
in the limit $y_s \gg y_b$, the KK modes 
with non-vanishing mass evolve trivially, and only
the massless zero mode is coupled to the brane motion
with
\begin{equation}  
  \ddot{q}_{0,{\bf k},\bullet} + \left[k^2 -\dot{\cal H} -{\cal H}^2\right] 
  q_{0,{\bf k},\bullet}=0~.
\label{e:ddotq0}
\end{equation} 
Since $\phi_0 \propto 1/a$ 
[cf.~Eqs.~(\ref{e:phi on brane}),(\ref{e:cal Y})]
we have found in \cite{Cartier:2005} that
the gravitational zero mode on the brane 
$h_{0,\bullet}(t;{\bf k})\equiv\sqrt{\kappa_5/L^3} 
q_{0,{\bf k},\bullet}\phi_0(t,y_b)$ evolves according 
to 
\begin{equation}
 \ddot{h}_{0,\bullet}(t;{\bf k}) + 
 2{\cal H}\dot{h}_{0,\bullet}(t;{\bf k}) + 
 k^2h_{0,\bullet}(t;{\bf k}) = 0~,
\label{e:4d gravity wave from brane}
\end{equation}
which explicitely demonstrates that at low energies (late times) the 
homogeneous tensor perturbation equation in brane cosmology 
reduces to the four-dimensional tensor 
perturbation equation. 
\\
\\
An important comment is in order here concerning 
the RS II model. 
In the limit $y_s\rightarrow \infty$ the fixed brane is
sent off to infinity and one ends up with a single
positive tension brane in AdS, i.e. the 
RS II model.  
Even though we have shown that all couplings 
except $M_{00}$ vanish in this limit, that does not
imply that this is necessarily the case for the RS II setup.
Strictly speaking, the above arguments are only valid
in a two brane model with $y_s \gg 1$.
Starting with the RS II model from the beginning, 
the coupling matrices do in general not vanish when 
calculated with the corresponding eigenfunctions 
which can be found in, e.g., \cite{Gorbunov:2001ge}. 
One just has to be careful when taking those limits. 
But what the above consideration demonstrates is that, 
if the couplings of the zero mode to the KK modes 
vanish, like in the $y_s \gg 1$ limit or in the
low energy RS II model as observed in numerical simulations
(see below)
the standard evolution equation for the zero mode
emerges automatically from five-dimensional perturbation 
theory. 
\\
\\ 
Starting from five-dimensional perturbation theory, 
our formalism does imply the 
usual evolution equation for the four-dimensional 
graviton in a FLRW-Universe in the limit of
vanishing couplings. 
This serves as a very strong indication (but certainly 
not proof!) for the fact that the approach based on the 
approximation (\ref{boundary conditions}) and the expansion 
of the action in canonical variables rather than the wave equation
is consistent and leads to results which should reflect 
the physics at low energies.   
As already outlined, if one would expand the wave equation 
(\ref{e:T-bulk-eq}) in the set of functions $\phi_\alpha$, 
the resulting equation of motion for the corresponding 
canonical variables is different from 
Eq.~(\ref{deq for q}) and cannot be derived from a 
Lagrangian or Hamiltonian (see Appendix \ref{a:variation}).
Moreover, in \cite{Koyama:2004cf} the low energy 
RS II scenario has been studied numerically 
including the full junction condition 
(\ref{e:T-JC-simple}) 
without approximations (see also \cite{Hiramatsu:2004}).
Those numerical results show that the evolution of tensor
perturbations on the brane is four-dimensional, i.e.
described by Eq.~(\ref{e:4d gravity wave from brane})
derived here analytically.  
Combining these observations gives us confidence that 
the used approach based on the Neumann boundary condition 
approximation and the action as starting point for the 
canonical formulation is adequate for the study 
of tensor perturbations in the low energy limit.
The many benefits this approach offers will become 
visible in the following. 
%
\section{Quantum generation of tensor perturbations}
\label{sec:III}
\subsection{Preliminary remarks}
%
We now introduce a treatment of quantum generation of tensor 
perturbations. 
This formalism is an advancement of the 
method which is presented in \cite{Ruser:2004,Ruser:2006xg,Ruser:2005xg}
for the dynamical Casimir effect for a scalar field 
and the electromagnetic field to gravitational perturbations 
in the braneworld scenario.   
\\
The following method is very general and not restricted 
to a particular brane motion as long as it complies with the 
low energy approach [cf.~Eq.~(\ref{e:low energy approach})].
We assume that asymptotically, i.e. for $t\rightarrow \pm \infty$,
the physical brane approaches the Cauchy horizon
($y_b\rightarrow 0$), moving very slowly.
Then, the coupling matrices vanish and the KK masses
are constant (for $y_b$ close to zero, Eq.~\eqref{e:zero equation}
reduces to $J_1(m_iy_s)=0$):
\begin{equation}
\lim_{t\ra \pm \infty}M_{\alpha\beta}(t)=0\;\;,\;\;
\lim_{t\ra \pm \infty} m_\alpha(t) = {\rm const.}
\;\;\forall \alpha,\beta
\;\;.
\end{equation}
In this limit, the system~(\ref{deq for q}) reduces to an infinite set of uncoupled
harmonic oscillators.
This allows to introduce an unambiguous and meaningful particle
concept, i.e. notion of (massive) gravitons.
\\
As a matter of fact, in the numerical simulations, the brane 
motion has to be switched on and off at finite times.
These times are denoted by $t_{\rm in}$ and $t_{\rm out}$, 
respectively.
We introduce vacuum states with respect to 
times $t < t_{\rm in} < 0$  and $t > t_{\rm out} > 0$.
In order to avoid spurious effects influencing the particle
creation, we have to chose $t_{\rm in}$ small, 
respectively $t_{\rm out}$ large enough such that the couplings
are effectively zero at these times.
Checking the independence of the numerical results on the choice of
$t_{\rm in}$ and ${\rm t_{\rm out}}$ guarantees that these 
times correspond virtually to the real asymptotic 
states of the brane configuration.
%
\subsection{Quantization, initial and final state }
%
Canonical quantization of the gravity wave amplitude is performed by replacing the
canonical variables $q_{\alpha, {\bf k}, \bullet}$ by the corresponding operators
$\hat{q}_{\alpha, {\bf k}, \bullet}$
\begin{equation}
\hat{h}_\bullet(t,y;{\bf k}) = \sqrt{\frac{\kappa_5}{L^3}}\sum_\alpha
\hat{q}_{\alpha,{\bf k}, \bullet}(t) \phi_\alpha(t,y)~.
\label{e:expansion of h bullet in q}
\end{equation}
Adopting the Heisenberg picture to describe the quantum
time-evolution, it follows that  $\hat{q}_{\alpha, {\bf k}, \bullet}$
satisfies the same equation (\ref{deq for q}) as the canonical variable
$q_{\alpha, {\bf k}, \bullet}$.
\\
Under the assumptions outlined above, the operator $\hat{q}_{\alpha, {\bf k}, \bullet}$
can be written for times $t < t_{\rm in}$ as
\begin{align}\label{e:initial q}
&\hat{q}_{\alpha, {\bf k}, \bullet}(t < t_{\rm in}) = \\
&\frac{1}{\sqrt{2\omega_{\alpha, k}^{\rm in}}}
\left[ \hat{a}^{\rm in}_{\alpha, {\bf k}, \bullet} e^{-i\,\omega_{\alpha,k}^{\rm in}\,t}
+\hat{a}^{{\rm in} \dagger}_{\alpha, -{\bf k}, \bullet} e^{i\,\omega_{\alpha,k}^{\rm in}\,t}
\right]
\nonumber
\end{align}
where we have introduced the initial-state frequency
\begin{equation}
\omega_{\alpha,k}^{\rm in}  \equiv  \omega_{\alpha,k}(t < t_{\rm in})~.
\end{equation}
This expansion ensures that Eq.~(\ref{e:reality for q}) is satisfied.
The set of annihilation and creation operators
$\{\hat{a}^{\rm in}_{\alpha, {\bf k}, \bullet}$,
$\hat{a}^{{\rm in} \dagger}_{\alpha, {\bf k}, \bullet} \}$
corresponding to the notion of gravitons for
$t < t_{\rm in}$ is subject to the usual commutation relations
\begin{eqnarray}
\left[\hat{a}^{\rm in}_{\alpha, {\bf k}, \bullet},
\hat{a}^{{\rm in}\dagger}_{\alpha', {\bf k}', \bullet'}\right]
&=&\delta_{\alpha\alpha'} \delta_{\bullet\bullet'}\delta^{(3)}({\bf k} - {\bf k'})\;,\\
\Big[\hat{a}^{\rm in}_{\alpha, {\bf k}, \bullet},
\hat{a}^{\rm in}_{\alpha', {\bf k'}, \bullet'}\Big]
&=&
\left[\hat{a}^{{\rm in} \dagger}_{\alpha, {\bf k}, \bullet},
\hat{a}^{{\rm in} \dagger}_{\alpha', {\bf k'}, \bullet'}\right]
=0.
\end{eqnarray}
For times $t > t_{\rm out}$, i.e. after the motion of the brane has
ceased, the operator
$\hat{q}_{\alpha, {\bf k}, \bullet}$  can be expanded in a similar manner,
\begin{align}\label{e:final q}
&\hat{q}_{\alpha, {\bf k}, \bullet}(t > t_{\rm out}) = \\
&\frac{1}{\sqrt{2\omega_{\alpha, k}^{\rm out}}}
\left[ \hat{a}^{\rm out}_{\alpha, {\bf k}, \bullet} e^{-i\,\omega_{\alpha,k}^{\rm out}\,t}
+\hat{a}^{{\rm out}\,\dagger}_{\alpha, -{\bf k}, \bullet} e^{i\,\omega_{\alpha,k}^{\rm out}\,t}
\right] \nonumber
\end{align}
with final state frequency
\begin{equation}
\omega_{\alpha,k}^{\rm out}  \equiv  \omega_{\alpha,k}(t > t_{\rm out})~.
\end{equation}
The annihilation and creation operators
$\{\hat{a}^{\rm out}_{\alpha, {\bf k}, \bullet} ,
\hat{a}^{{\rm out} \, \dagger}_{\alpha, {\bf k}, \bullet} \}$
correspond to a meaningful definition of final state gravitons
(they are associated with positive and negative
frequency solutions for $t\ge t_{\rm out}$)
and satisfy the same commutation relations as the initial state operators.
\\
Initial $|0,{\rm in}\rangle \equiv |0,t <  t_{\rm in} \rangle$ and final
$|0,{\rm out}\rangle \equiv |0,t >  t_{\rm out} \rangle$ vacuum
states are uniquely defined via 
\footnote{Note that the notations $|0,t <  t_{\rm in} \rangle$ and $|0,t >  t_{\rm out} \rangle$
do not mean that the states are time-dependent; 
states do not evolve in the Heisenberg picture.}
\begin{equation}
\hat{a}^{\rm in}_{\alpha,{\bf k}, \bullet} |0,{\rm in}\rangle = 0 \;,\;
\hat{a}^{\rm out}_{\alpha,{\bf k}, \bullet} |0,{\rm out}\rangle = 0 \;,\;\;
\forall \;\alpha,\;{\bf k},\;\bullet ~.
\label{vacuum definitions}
\end{equation}
The operators counting the number of particles defined with respect to the initial
and final vacuum state, respectively, are
\begin{equation}
\hat{N}^{\rm in}_{\alpha, {\bf k},\bb} =  \hat{a}^{{\rm in}\,\dagger}_{\alpha,{\bf k},\bb}
\hat{a}^{\rm in}_{\alpha,{\bf k},\bb}\;,\;\;
\hat{N}^{\rm out}_{\alpha, {\bf k},\bb} =  \hat{a}^{{\rm out} \,\dagger}_{\alpha,{\bf k},\bb}
\hat{a}^{\rm out}_{\alpha,{\bf k},\bb}~.
\end{equation}
The number of gravitons created during the motion of the brane
for each momentum ${\bf k}$, quantum number $\alpha$ and polarization
state $\bullet$ is given by the expectation value of the number operator
$\hat{N}^{\rm out}_{\alpha,{\bf k},\bb}$
of final-state gravitons with respect to the initial
vacuum state $| 0,{\rm in} \rangle$:
\begin{equation}
{\cal N}^{\rm out}_{\alpha,{\bf k},\bb} = \langle 0,{\rm in}|
\hat{N}^{\rm out}_{\alpha,{\bf k},\bb}|0,{\rm in}\rangle.
\label{e:graviton number definition}
\end{equation}
If the brane undergoes a non-trivial dynamics between
$t_{\rm in} < t < t_{\rm out}$ it is
$\hat{a}^{\rm out}_{\alpha,{\bf k},\bb}|0,{\rm in}\rangle \neq 0$ 
in general, i.e. graviton production from vacuum fluctuations 
takes place.
\\
\\
From (\ref{e:h fourier decomposition}), the expansion
(\ref{e:expansion of h bullet in q}) and
Eqs.(\ref{e:initial q}), (\ref{e:final q}) it follows that the
quantized tensor perturbation with respect to 
the initial and final state can be written as
\begin{align} 
\hat{h}_{ij}(t < t_{\rm in},{\bf x},&y) = 
\sqrt{\frac{\kappa_5}{L^3}} \sum_{\bb\alpha}
 \int \frac{d^3k}{(2\pi)^{3/2}}  \frac{\hat{a}^{\rm in}_{\alpha, {\bf k},
 \bullet} \,e^{-i\,\omega_{\alpha,k}^{\rm in}\,t}}
 {\sqrt{2\omega_{\alpha, k}^{\rm in}}} \times \nonumber \\
&\times u^\bb_{ij, \alpha}(t<t_{\rm in},{\bf x},y,{\bf k})
 + {\rm h.c.}
\label{e:full initial perturbation}
\end{align}
and
\begin{align}  
\hat{h}_{ij}(t > t_{\rm out},{\bf x},&y) = \sqrt{\frac{\kappa_5}{L^3}} \sum_{\bb\alpha}
 \int \frac{d^3k}{(2\pi)^{3/2}}  \frac{\hat{a}^{\rm out}_{\alpha, {\bf k},
 \bullet} \,e^{-i\,\omega_{\alpha,k}^{\rm out}\,t}}
 {\sqrt{2\omega_{\alpha, k}^{\rm out}}} \times \nonumber \\
&\times u^\bb_{ij, \alpha}(t>t_{\rm out},{\bf x},y,{\bf k})
 + {\rm h.c.}~.
\label{e:full final perturbation}
\end{align}
We have introduced the basis functions
\begin{equation}
u^\bb_{ij, \alpha}(t,{\bf x},y,{\bf k}) = e^{i\ve{k}\,\cdot\ve{x}}\,e_{ij}^{\bb}({\bf k})\,\phi_\alpha(t,y).
\end{equation}
which, on account of $(e_{ij}^{\bb}({\bf k}))^*=e_{ij}^{\bb}({\bf -k})$,
satisfy $(u^\bb_{ij,\alpha}(t,{\bf x},y,{\bf k}))^*=
u^\bb_{ij,\alpha}(t,{\bf x},y,{\bf -k})$.
%
\subsection{Time evolution}
%
During the motion of the brane the time evolution of the field modes is described by
the system of coupled differential equations (\ref{deq for q}).
To account for the inter-mode couplings mediated by the coupling matrix $M_{\alpha\beta}$
the operator $\hat{q}_{\alpha,{\bf k},\bullet}$ is decomposed as
\begin{equation}
\hat{q}_{\alpha,{\bf k},\bullet}(t) = \sum_\beta \frac{1}{\sqrt{2\omega_{\beta,k}^{\rm in}}}
\left[  \hat{a}^{\rm in}_{\beta, {\bf k}, \bullet} \epsilon_{\alpha,k}^{(\beta)}(t)
+\hat{a}^{{\rm in} \dagger}_{\beta, -{\bf k}, \bullet}
\epsilon_{\alpha,k}^{(\beta)^*}(t) \right].
\label{q expansion}
\end{equation}
The complex functions $\epsilon_{\alpha,k}^{(\beta)}(t)$ also satisfy the
system of coupled differential equations (\ref{deq for q}).
With the ansatz (\ref{q expansion}) the quantized tensor perturbation
at any time during the brane motion reads
\begin{align}
&\hat{h}_{ij}(t,{\bf x},y)=
\label{e:any t field expansion}\\
&\sqrt{\frac{\kappa_5}{L^3}}\sum_{\bb\alpha\beta}
  \int \frac{d^3k}{(2\pi)^\frac{3}{2}}
  \frac{\hat{a}^{\rm in}_{\beta, {\bf k}, \bullet}}{\sqrt{2\omega_{\beta,k}^{\rm in}}}
  \epsilon_{\alpha,k}^{(\beta)}(t)
   u^\bb_{ij, \alpha}(t,{\bf x},y,{\bf k})+ {\rm h.c.}\;.
  \nonumber
\end{align}
Due to the time-dependence of the eigenfunctions $\phi_\alpha$, the
time-derivative of  the gravity wave amplitude contains additional
mode coupling contributions.
Using the completeness and orthnormality of the $\phi_\alpha$'s 
it is readily shown that
\begin{equation}
\dot{\hat{h}}_\bullet(t,y;{\bf k})=
\sqrt{\frac{\kappa_5}{L^3}}\sum_\alpha
\hat{p}_{\alpha,{\bf -k}, \bullet}(t) \phi_\alpha(t,y)
\label{e:h dot as function of p}
\end{equation}
where
\begin{equation}
\hat{p}_{\alpha,{\bf -k}, \bullet}(t)  = \dot{\hat{q}}_{\alpha,{\bf k},
  \bullet}(t) + \sum_\beta M_{\beta\alpha}\hat{q}_{\beta,{\bf k}, \bullet}(t).
\label{e:definition of momentum p}
\end{equation}
The coupling term arises from the time dependence of the mode functions
$\phi_\alpha$.
Accordingly, the time derivative $\dot{\hat{h}}_{ij}$ reads
\begin{align}
\dot{\hat{h}}_{ij} (t,{\bf x},y)  =
&\sqrt{\frac{\kappa_5}{L^3}}\sum_{\bb\alpha\beta}
\int \frac{d^3k}{(2\pi)^\frac{3}{2}} \frac{\hat{a}^{\rm in}_{\beta, {\bf
k},\bullet}}{\sqrt{2\omega_{\beta,k}^{\rm in}}}\times
\label{e:any t field expansion derivative}\\
& \times f_{\alpha,k}^{(\beta)}(t)\,
u^\bb_{ij, \alpha}(t,{\bf x},y,{\bf k})
+ {\rm h.c.}\; \nonumber
\end{align}
where we have introduced the function 
\begin{equation}
f_{\alpha,k}^{(\beta)}(t) = \dot{\epsilon}_{\alpha,k}^{(\beta)}(t)
 + \sum_\gamma M_{\gamma\alpha}(t)
\epsilon_{\gamma,k}^{(\beta)}(t)~.
\end{equation}
By comparing Eq.~(\ref{e:full initial perturbation}) and its
time-derivative with Eqs. (\ref{e:any t field expansion}) and
(\ref{e:any t field expansion derivative}) at $t=t_{\rm in}$
one can read off the initial conditions for the functions
$\epsilon_{\alpha,k}^{(\beta)}$:
\begin{align}
&\epsilon_{\alpha,k}^{(\beta)}(t_{\rm in}) =
\delta_{\alpha\beta}\;\Theta^{\rm in}_{\alpha,k}\;,
\label{e:initial conditions for epsilon}\\
&\dot{\epsilon}_{\alpha,k}^{(\beta)}(t_{\rm in}) =
\left[ -i \omega_{\alpha,k}^{\rm in} \delta_{\alpha\beta} -
M_{\beta\alpha}(t_{\rm in})\right] \;\Theta^{\rm in}_{\beta,k}
\label{e:initial conditions for epsilon dot}
\end{align}
with phase
\begin{equation}\label{e:Theta}
\Theta^{\rm in}_{\alpha,k} = e^{-i \omega^{\rm in}_{\alpha,k}\,t_{\rm in}}.
\end{equation}
The choice of this  phase for the initial condition is in principle
arbitrary, we could as well set $\Theta^{\rm in}_{\alpha,k} =1$. But
with this choice, $\epsilon_{\alpha,k}^{(\beta)}(t)$ is independent of
$t_{\rm in}$ for $t<t_{\rm in}$ and therefore it is also at later times
independent of $t_{\rm in}$ if only we choose $t_{\rm in}$
sufficiently early. This is especially useful for the numerical work.
%
\subsection{Bogoliubov transformations}
%
The two sets of annihilation and creation operators
$\{\hat{a}^{\rm in}_{\alpha,{\bf k},\bullet}$,
$\hat{a}^{{\rm in}\,\dagger}_{\alpha,{\bf k},\bullet} \}$ and
$\{\hat{a}^{\rm out}_{\alpha,{\bf k},\bullet}$,
$\hat{a}^{{\rm out}\,\dagger}_{\alpha,{\bf k},\bullet} \}$
corresponding to the notion of initial-state and final-state 
gravitons are related via a Bogoliubov transformation.
Matching the expression for the tensor perturbation
Eq.~(\ref{e:any t field expansion})
and its time-derivative
Eq.~(\ref{e:any t field expansion derivative}) 
with the final state expression
Eq.~(\ref{e:full final perturbation}) and its corresponding
time-derivative at $t=t_{\rm out}$ one finds
\begin{equation}
\hat{a}^{\rm out}_{\beta, {\bf k},\bullet}
= \sum_\alpha\left[{\cal A}_{\alpha\beta,k}(t_{\rm out})
\hat{a}^{\rm in}_{\alpha,{\bf k}, \bullet} + {\cal B}_{\alpha\beta,k}^* (t_{\rm out})
\hat{a}^{{\rm in}\,\dagger}_{\alpha, {\bf -k}, \bullet}\right]
\label{bogoliubov trafo}
\end{equation}
with
\begin{equation} \label{e:BogA}
{\cal A}_{\beta\alpha,k}(t_{\rm out}) =
       \frac{\Theta^{{\rm out}^*}_{\alpha,k}}{2}
       \sqrt{
     \frac{\omega_{\rm \alpha,k}^{\rm out}}
              {\omega_{\rm \beta,k}^{\rm in}}
       }
       \left[
       \epsilon_{\alpha,k}^{(\beta)}(t_{\rm out})
       + \frac{i}{\omega_{\alpha,k}^{\rm out}}
       f_{\alpha,k}^{(\beta)}(t_{\rm out})
       \right]
\end{equation}
and
\begin{equation}\label{e:BogB}
{\cal B}_{\beta\alpha,k}(t_{\rm out}) =
       \frac{\Theta^{\rm out}_{\alpha,k}}{2}
       \sqrt{
     \frac{\omega_{\rm \alpha,k}^{\rm out}}
              {\omega_{\rm \beta,k}^{\rm in}}
       }
       \left[
       \epsilon_{\alpha,k}^{(\beta)}(t_{\rm out})
       - \frac{i}{\omega_{\alpha,k}^{\rm out}}
       f_{\alpha,k}^{(\beta)}(t_{\rm out})
       \right]
\end{equation}
where we shall stick to the phase $\Theta^{\rm out}_{\alpha,k}$ defined like
$\Theta^{\rm in}_{\alpha,k}$ in
(\ref{e:Theta}) for completeness.
Performing the matching at $t_{\rm out} = t_{\rm in}$ the
Bogoliubov transformation should become trivial, i.e.
the Bogoliubov coefficients are subject to vacuum initial conditions
\begin{equation}
{\cal A}_{\alpha\beta,k} ( t_{\rm in}) = \delta_{\alpha\beta} \;\;,\;\;
{\cal B}_{\alpha\beta,k} ( t_{\rm in}) = 0.
\label{e:vacuum initial conditions for A,B}
\end{equation}
Evaluating the Bogoliubov coefficients (\ref{e:BogA}) and (\ref{e:BogB}) for
$t_{\rm out} = t_{\rm in}$ by making use of the initial conditions
(\ref{e:initial conditions for epsilon}) and (\ref{e:initial
  conditions for epsilon dot}) shows the consistency.
Note that the Bogoliubov transformation (\ref{bogoliubov trafo}) is not diagonal
due to the inter-mode coupling.
If during the motion of the brane the graviton field departs form its vacuum state one has
${\cal B}_{\alpha\beta,k} (t_{\rm out}) \neq 0$, i.e. gravitons have been generated.
\\
\\
By means of Eq.~(\ref{bogoliubov trafo}) the number of generated
final state gravitons (\ref{e:graviton number definition}),
which is the same for every polarization state, is given by
\begin{eqnarray}
{\cal N}^{\rm out}_{\alpha, k} (t \ge t_{\rm out}) &=&
\sum_{\bb=+,\times} \langle 0, {\rm in} |
\hat{N}^{\rm out}_{\alpha, {\bf k},\bb} | 0, {\rm in}\rangle \nonumber \\
&=& 2\sum_\beta |{\cal B}_{\beta\alpha,k}(t_{\rm out})|^2.
\label{particle number}
\end{eqnarray}
Later we will sometimes interpret $t_{\rm out}$
as a continuous variable $t_{\rm out} \rightarrow t$ such that
${\cal N}_{\alpha,k}^{\rm out} \rightarrow {\cal N}_{\alpha,k}(t)$,
i.e. it becomes a continuous function of time. 
We shall call ${\cal N}_{\alpha,k}(t)$ the {\it instantaneous particle number}
[see Appendix \ref{a:power}], however, a physical interpretation 
should be made with caution.
%
\subsection{The first order system}
%
From the solutions of the system of differential equations
(\ref{deq for q}) for the complex functions $\epsilon_{\alpha,k}^{(\beta)}$,
the Bogoliubov coefficient ${\cal B}_{\alpha\beta,k}$, and
hence the number of created final state gravitons 
(\ref{particle number}), can now be calculated.
It is however useful to introduce auxiliary functions 
$\xi_{\alpha,k}^{(\beta)}(t),\eta_{\alpha,k}^{(\beta)}(t)$ 
through 
\begin{eqnarray}
\xi_{\alpha,k}^{(\beta)}(t) &=& \epsilon_{\alpha,k}^{(\beta)}(t) +
\frac{i}{\omega_{\alpha,k}^{\rm in}} f_{\alpha,k}^{(\beta)}(t)\\
\eta_{\alpha,k}^{(\beta)}(t) &=& \epsilon_{\alpha,k}^{(\beta)}(t) -
\frac{i}{\omega_{\alpha,k}^{\rm in}} f_{\alpha,k}^{(\beta)}(t)~.
\end{eqnarray}
These are related to the Bogoliubov coefficients via
\begin{align}
&{\cal A}_{\beta\alpha,k}(t_{\rm out}) = \\
       &\frac{\Theta^{{\rm out}^*}_{\alpha,k}}{2}
       \sqrt{
     \frac{\omega_{\rm \alpha,k}^{\rm out}}
              {\omega_{\rm \beta,k}^{\rm in}}
       }
       \left[
       \Delta^+_{\alpha,k}(t_{\rm out})\xi_{\alpha,k}^{(\beta)}(t_{\rm
       out}) + \Delta^-_{\alpha,k}(t_{\rm out})\eta_{\alpha,k}^{(\beta)}(t_{\rm
       out})
       \right]\nonumber
\end{align}
\begin{align}\label{e:Bout}
&{\cal B}_{\beta\alpha,k}(t_{\rm out}) = \\
       &\frac{\Theta^{\rm out}_{\alpha,k} }{2}
       \sqrt{
     \frac{\omega_{\rm \alpha,k}^{\rm out}}
              {\omega_{\rm \beta,k}^{\rm in}}
       }
       \left[
       \Delta^-_{\alpha,k}(t_{\rm out})\xi_{\alpha,k}^{(\beta)}(t_{\rm
       out}) + \Delta^+_{\alpha,k}(t_{\rm out})\eta_{\alpha,k}^{(\beta)}(t_{\rm
       out})
       \right]\nonumber
\end{align}
where we have defined
\begin{equation}
\Delta^\pm_{\alpha,k}(t) =
\frac{1}{2}\left[1\pm \frac{\omega^{\rm in}_{\alpha,k}}{\omega_{\alpha,k}(t)}
\right]~,
\end{equation}
Using the second order differential equation for
$\epsilon_{\alpha,k}^{(\beta)}$, it is readily shown that the functions
$\xi_{\alpha,k}^{(\beta)}(t)$, $\eta_{\alpha,k}^{(\beta)}(t)$
satisfy the following system of first order differential equations:
\begin{align}
\dot{\xi}_{\alpha,k}^{({\beta})}(t)=-i\left[a^+_{\alpha\alpha,k}(t)\xi_{\alpha,k}^{({\beta})}(t)-
a^-_{\alpha\alpha,k}(t)\eta_{\alpha,k}^{({\beta})}(t)\right]\nonumber\\
-\sum_{\gamma}\left[c^-_{\alpha\gamma,k}(t)\xi_{\gamma,k}^{({\beta})}(t)+
c^+_{\alpha\gamma,k}(t)\eta_{\gamma,k}^{({\beta})}(t)\right]
\label{deq for xi}
\end{align}
\begin{align}
\dot{\eta}_{\alpha,k}^{({\beta})}(t)=-i\left[a^-_{\alpha\alpha,k}(t)\xi_{\alpha,k}^{({\beta})}(t)-
a^+_{\alpha\alpha,k}(t)\eta_{\alpha,k}^{({\beta})}(t)\right]\nonumber\\
-\sum_{\gamma}\left[c^+_{\alpha\gamma,k}(t)\xi_{\gamma,k}^{({\beta})}(t)+
c^-_{\alpha\gamma,k}(t)\eta_{\gamma,k}^{({\beta})}(t)\right]
\label{deq for eta}
\end{align}
with
\begin{eqnarray}
a_{\alpha\alpha,k}^\pm(t)&=&\frac{\omega_{\alpha,k}^{\rm in}}{2}
\left\{1 \pm \left[\frac{\omega_{\alpha,k}(t)}{\omega_{\alpha,k}^{\rm in}}\right]^2\right\},
\label{def a matrix} \\
c_{\gamma\alpha,k}^\pm(t)&=&\frac{1}{2}\left[M_{{\alpha\gamma}}(t) \pm
\frac{\omega_{\alpha,k}^{\rm in}}{\omega_{\gamma,k}^{\rm in}}M_{\gamma\alpha}(t)\right].
\label{def c matrix}
\end{eqnarray}
The vacuum initial conditions
(\ref{e:vacuum initial conditions for A,B})
entail the initial conditions
\begin{equation}
\xi_{\alpha,k}^{(\beta)} (t_{\rm in}) =
2\,\delta_{\alpha\beta}\,\Theta_{\alpha,k}^{\rm in}
\;,\;\;
\eta_{\alpha,k}^{(\beta)} (t_{\rm in}) = 0.
\label{e:vacuum initial conditions for xi,eta}
\end{equation}
With the aid of Eq. \eqref{e:Bout}, the coefficient 
${\cal B}_{\alpha\beta,k}(t_{\rm out})$,
and therefore the number of produced gravitons,
can be directly deduced from the solutions to this system of coupled
first order differential equations which can be solved
using standard numerics. 
\\
\\
In the next section we will show how
interesting observables like the power spectrum and the
energy density of the amplified gravitational waves are
expressed in terms of the number of created gravitons.
The system (\ref{deq for xi},~\ref{deq for eta}) of coupled
differential equations forms the basis of our numerical simulations.
Details of the applied numerics are collected in Appendix~\ref{a:numerics}.
%
\section{Power spectrum, energy density and localization 
of gravity}
%
\subsection{Perturbations on the brane}
%
By solving the system of coupled differential equations
formed by Eqs.~(\ref{deq for xi}) and (\ref{deq for eta})
the time evolution of the quantized tensor 
perturbation $\hat{h}_{ij}(t,{\bf x},y)$ 
can be completely reconstructed at any position $y$ in the 
bulk.
Accessible to observations is the imprint which the 
perturbations leave on the brane, i.e. in our Universe.
Of particular interest is therefore the part of 
the tensor perturbation which resides on the brane.
It is given by evaluating Eq.~(\ref{e:h fourier decomposition}) at the 
brane position $y=y_b$ (see also \cite{Seahra:2006})
\begin{equation}
\hat{h}_{ij}(t,{\bf x},y_b) = \int \frac{d^3k}{(2\pi)^{3/2}} 
\sum_{\bullet=+,\times}
e^{i{\bf k\cdot x}} e^\bullet_{ij}({\bf k}) \hat{h}_\bullet(t,y_b,{\bf k})~.
\label{e:perturbation on brane}
\end{equation}
The motion of the brane (expansion of the Universe)
enters this expression via the eigenfunctions 
$\phi_\alpha(t,y_b(t))$.  
We shall take (\ref{e:perturbation on brane}) as the starting 
point to define observables on the brane.
\\
The zero-mode function $\phi_0(t)$ [cf.~Eq.~(\ref{zero mode phi})] 
does not depend on the extra dimension $y$. 
Using Eq.~(\ref{C2 on brane}), one reads 
off from Eq.~(\ref{e:phij}) that the
eigenfunctions on the brane $\phi_\alpha(t,y_b)$ are
\begin{equation}
\phi_\alpha(t,y_b) = y_b\,{\cal Y}_\alpha(y_b) 
= \frac{L}{a}\,{\cal Y}_\alpha(a)
\label{e:phi on brane}
\end{equation}
where we have defined 
\begin{eqnarray}
{\cal Y}_0(a) &=& \sqrt{\frac{y_s^2}{y_s^2 - y_b^2}}
\;\;\;\;{\rm and} \\
{\cal Y}_n(a) &=& \sqrt{\frac{Y_1^2(m_ny_s)}{Y_1^2(m_ny_b) - Y_1^2(m_ny_s)}},
\end{eqnarray}
for the zero- and KK modes, respectively.
One immediately is confronted with an interesting observation:
the function ${\cal Y}_\alpha(a)$ behaves differently with the
expansion of the Universe for the zero mode $\alpha = 0$ and
the KK modes $\alpha = n$.
This is evident in particular in the asymptotic regime $y_s \gg y_b$, i.e.
$y_b \rightarrow 0$ ($|t|,a \rightarrow \infty$) where,
exploiting the asymptotics of $Y_1$ (see \cite{Abramowitz}), 
one finds
\begin{equation}
{\cal Y}_0(a) \simeq 1\;,\;\;
{\cal Y}_n(a) \simeq \frac{L}{a} \frac{\pi m_n}{2} |Y_1(m_n y_s)|
\simeq \frac{L}{a}\sqrt{\frac{m_n\,\pi}{2\,y_s}}
\label{e:cal Y}
\end{equation}
Ergo, ${\cal Y}_0$ is constant while ${\cal Y}_n$ decays
with the expansion of the Universe as $1/a$.
For large $n$ one can approximate $m_n\simeq n\pi/y_s$ and 
$Y_1(m_ny_s) \simeq Y_1(n\pi) \simeq (1/\pi)\sqrt{2/n}$
\cite{Abramowitz}, so that
\begin{equation}
{\cal Y}_n(a) \simeq \frac{Lm_n}{\sqrt{2\,n}a},~~
{\cal Y}_n^2(a) \simeq \frac{\pi L^2m_n}{2\,y_sa^2}~.
\label{e:Yfunclargen}
\end{equation}
In summary, the amplitude of the KK modes on the brane 
decreases faster with the expansion of the Universe than the amplitude of the
zero mode. 
This leads to interesting consequences for the observable power
spectrum and energy density and has a clear physical
interpretation: It manifest the localization of usual
gravity on the brane. 
As we shall show below, KK gravitons which are traces of the
five-dimensional nature of gravity escape rapidly from 
the brane.
%
\subsection{Power spectrum}
%
We define the power spectrum ${\cal P}(k)$ of gravitational waves 
on the brane as in four-dimensional cosmology 
by using the restriction of the tensor amplitude
to the brane position (\ref{e:perturbation on brane}):
\begin{align}\label{def power spectrum}
&\frac{(2\pi)^3}{k^3}{\cal P}(k) \delta^{(3)}({\bf k} - {\bf k'})\\
&= \sum_{\bullet = \times, +}
\left \langle 0,{\rm in} \Big |\hat{h}_\bullet(t,y_b;{\bf k})
\hat{h}^\dagger_\bullet(t,y_b;{\bf k'})
\Big|0,{\rm in} \right \rangle,
\nonumber
\end{align}
i.e. we consider the expectation value of the field operator $\hat{h}_\bullet$
with respect to the initial vacuum state at the position of the brane
$y = y_b(t)$. 
In order to get a physically meaningful power spectrum, averaging
over several oscillations of the gravitational wave amplitude has to
be performed.
Equation (\ref{def power spectrum}) describes the observable
power spectrum imprinted in our Universe by the four-dimensional 
spin-2 graviton component of the five-dimensional tensor
perturbation.
\\
The explicit calculation of the expectation value involving a
``renormalization'' of a divergent contribution is carried out in
detail in Appendix \ref{a:power}.
The final result reads
\begin{equation}
{\cal P}(k) = \frac{1}{a^2} \frac{k^3}{(2\pi)^3} \frac{\kappa_5}{L}
\sum_\alpha {\cal R}_{\alpha,k}(t) \, {\cal Y}_\alpha^2(a).
\label{e:power spectrum with R}
\end{equation}
The function ${\cal R}_{\alpha,k}(t)$ can be expressed in terms of
the Bogoliubov coefficients (\ref{e:BogA}) and (\ref{e:BogB}) 
if one considers $t_{\rm out}$ as a continuous variable $t$:
\begin{equation}
{\cal R}_{\alpha,k}(t) = \frac{{\cal N}_{\alpha,k}(t) +
{\cal O}^{{\cal N}}_{\alpha,k}(t)}{\omega_{\alpha,k}(t)}.
\label{e:R function with inst particle number}
\end{equation}
${\cal N}_{\alpha,k}(t)$ is the instantaneous particle number
[cf.~Appendix \ref{a:inst}] 
and the function ${\cal O}^{{\cal N}}_{\alpha,k}(t)$
is defined in Eq.~(\ref{e:O in terms of A and B}).
\\
It is important to recall that ${\cal N}_{\alpha,k}(t)$ can in
general not be interpreted as a physical particle number. 
For example zero modes with wave numbers
such that $kt<1$ cannot be considered as particles. They have not
performed several oscillations and their energy density cannot be
defined in a meaningful way.
\\
Equivalently, expressed in terms of the complex functions
$\epsilon_{\alpha,k}^{(\beta)}$, one finds
\begin{equation}
{\cal R}_{\alpha,k}(t) = \sum_\beta
\frac{|\epsilon_{\alpha,k}^{(\beta)}(t)|^2}
{\omega_{\beta,k}^{\rm in}}
-\frac{1}{\omega_{\alpha,k}(t)} + {\cal O}_{\alpha,k}^{\epsilon}(t),
\label{e:R function with epsilon}
\end{equation}
with ${\cal O}_{\alpha,k}^{\epsilon}$ given in
Eq.~(\ref{e:O in terms of epsilon}).
Equation (\ref{e:power spectrum with R}) together with
(\ref{e:R function with inst particle number}) or
(\ref{e:R function with epsilon})
holds at all times.
\\
If one is interested in the power spectrum at early times
$kt \ll 1$, it is not sufficient to take only the
instantaneous particle number ${\cal N}_{\alpha,k}(t)$
in Eq.~(\ref{e:R function with inst particle number}) into account.
This is due to the fact that even if the mode functions
$\epsilon_{\alpha,k}^{(\beta)}$ are already oscillating,
the coupling matrix entering the Bogoliubov
coefficients might still undergo a non-trivial 
time dependence [cf.~Eq.~(\ref{e:Bogoliubov B00})].
In the next section we shall show explicitly, that in a radiation
dominated bounce particle creation, especially of
the zero mode, only stops on sub-Hubble times, $kt > 1$, even
if the mode functions are plane waves right after the bounce
[cf, e.g., Figs.~\ref{genfig1}, \ref{genfig2}, \ref{f:zeromode numbers}].
Therefore, in order to determine the perturbation spectrum of
the zero mode, one has to make use of the full expression
expression \eqref{e:R function with epsilon} and may not use
\eqref{e:Ps}, given below.
\\
At late times, $kt \gg 1$ ($t\ge t_{\rm out}$) when 
the brane moves slowly, the couplings
$M_{\alpha\beta}$ go to zero and particle creation has come to an end,
both functions ${\cal O}^{{\cal N}}_{\alpha,k}$
and ${\cal O}_{\alpha,k}^{\epsilon}$ do not contribute to the
observable power spectrum after averaging over
several oscillations. Furthermore, the instantaneous particle
number then equals the (physically meaningful) number of
created final state gravitons ${\cal N}_{\alpha,k}^{\rm out}$ and
the KK masses are constant.
Consequently, the observable power spectrum at late times
takes the form
\begin{equation}\label{e:Ps}
{\cal P}(k,t \ge t_{\rm out}) =  \frac{\kappa_4}{a^2}
\frac{k^3}{(2\pi)^3}  \sum_\alpha
\frac{{\cal N}_{\alpha,k}^{\rm out}}{\omega_{\rm \alpha,k}^{\rm out}}
     {\cal Y}^2_\alpha(a)~,
\end{equation}
where we have used that $\kappa_5/L = \kappa_4$. 
Its dependence on the wave number $k$ is completely determined
by the spectral behavior of the number of created gravitons
${\cal N}_{\alpha,k}^{\rm out}$.
\\
It is useful to decompose the power spectrum in its zero-mode
and KK-contributions:
\begin{equation}
{\cal P} = {\cal P}_0 + {\cal P}_{KK}.
\end{equation}
In the late time regime, using Eqs.~(\ref{e:Ps}) and (\ref{e:cal Y}), the
zero-mode power spectrum reads
\begin{equation}
{\cal P}_0(k,t \ge t_{\rm out}) =  \frac{\kappa_4}{a^2}
\frac{k^2}{(2\pi)^3} {\cal N}_{0,k}^{\rm out}.
\end{equation}
As expected for a usual four-dimensional tensor perturbation
(massless graviton), on sub-Hubble scales
the power spectrum decreases with the expansion 
of the Universe as $1/a^2$.
\\
In contrast, the KK mode power spectrum for late times, given by
\begin{equation}
{\cal P}_{\rm KK}(k,t \ge t_{\rm out}) = \frac{k^3}{a^4}
\frac{\kappa_4 L^2}{32 \pi}\sum_n {\cal N}_{n,k}^{\rm out}\,\frac{m_n^2}
{\omega_{\rm n,k}^{\rm out}}  Y^2_1(m_n y_s),
\label{late time KK power spectrum}
\end{equation}
decreases as $1/a^4$, i.e. with a factor $1/a^2$ faster than
${\cal P}_0$. The gravity wave power spectrum at late times
is therefore dominated by the zero-mode power spectrum and
looks four dimensional. Contributions to it
arising from five-dimensional effects are scaled away
rapidly as the Universe expands due to the $1/a^4$
behavior of ${\cal P}_{\rm KK}$.
In the limit of large masses $m_n y_s \gg 1$, $n \gg 1$
and for wave lengths $k \ll m_n$ such that
$\omega_{n,k} \simeq m_n$, the late-time KK-mode power spectrum 
can be approximated by
\begin{equation}
{\cal P}_{\rm KK}(k,t \ge t_{\rm out}) =  \frac{k^3}{a^4}
\frac{\kappa_4 L^2}{16 \pi^2 y_s}\sum_n {\cal N}_{n,k}^{\rm out}\,
\label{late time large mass KK power spectrum}
\end{equation}
where we have inserted Eq.~(\ref{e:Yfunclargen}) for ${\cal Y}_n^2(a)$.
\\
Note that the formal summations over the particle number 
might be ill defined if the brane 
trajectory contains unphysical features like discontinuities 
in the velocity.
An appropriate regularization is then necessary, for example,
by introducing a physically motivated cutoff.
%
\subsection{Energy density}
%
For a usual four-dimensional tensor perturbation $h_{\mu\nu}$
on a background metric $g_{\mu\nu}$ an associated effective
energy momentum tensor can be defined unambiguously by
(see, e.g., \cite{Straumann:CMB,mm})
\begin{equation}
T_{\mu\nu} = \frac{1}{\kappa_4} \langle
h_{\alpha\beta\|\mu}h^{\alpha\beta}_{\;\;\;\;\|\nu}\rangle ~,
\end{equation}
where the bracket stands for averaging over 
several periods of the wave and ``$\|$'' denotes
the covariant derivative with respect to the unperturbed
background metric.
The energy density of gravity waves is the
$00$-component of the effective energy momentum tensor.
We shall use the same effective energy momentum tensor to calculate
the energy density corresponding to the 
four-dimensional spin-2 graviton component of the five-dimensional tensor
perturbation on the brane, i.e. for
the perturbation $h_{ij}(t,{\bf x}, y_b)$ 
given by Eq.~(\ref{e:perturbation on brane}).
For this it is important to remember that in our low energy
approach, and in particular at very late times for which we 
want to calculate the energy density, the conformal 
time $\eta$ on the brane is identical 
to the conformal bulk time $t$. 
The energy density of four-dimensional spin-2 gravitons 
on the brane produced during the brane 
motion is then given by [see also \cite{Seahra:2006}]
\begin{equation}
\rho = \frac{1}{\kappa_4\,a^2} \left \langle \left \langle 0, {\rm in} |
\dot{\hat{h}}_{ij} (t, {\bf x}, y_b)\dot{\hat{h}}^{ij} (t, {\bf x}, y_b)
|0,{\rm in}\right \rangle \right \rangle.
\end{equation}
Here the outer bracket denotes averaging over several 
oscillations, which (in contrast to the
power spectrum) we embrace from the very beginning. 
The factor $1/a^2$ comes from the
fact that an over-dot indicates the derivative with respect $t$.
A detailed calculation is carried out in Appendix \ref{a:energy}
leading to
\begin{equation}
\rho = \frac{1}{a^4}\sum_{\alpha}
\int \frac{d^3k}{(2\pi)^{3}}
\omega_{\alpha,k}{\cal N}_{\alpha,k}(t){\cal Y}^2_{\alpha}(a)~
\label{energy density}
\end{equation}
where again ${\cal N}_{\alpha,k}(t)$ is the instantaneous 
particle number.
At late times $t > t_{\rm out}$ after particle creation has ceased, 
the energy density is therefore given by
\begin{equation}
\rho = \frac{1}{a^4}\sum_{\alpha}
\int \frac{d^3k}{(2\pi)^{3}}
\omega_{\rm \alpha,k}^{\rm out} \;{\cal N}_{\alpha,k}^{\rm out}\;
{\cal Y}^2_{\alpha}(a).
\label{e:energy density late time out number}
\end{equation}
This expression looks at first sight very similar to a ``naive'' 
definition of energy
density as integration over momentum space and summation over all
quantum numbers $\alpha$ of the energy
$\omega_{\rm \alpha,k}^{\rm out} \;{\cal N}_{\alpha,k}^{\rm out}$
of created gravitons. 
(Note that the graviton number
${\cal N}_{\alpha,k}^{\rm out}$ already contains the contributions of
both polarizations [see Eq.~(\ref{particle number})].)
However, the important difference is the appearance of the function
${\cal Y}^2_{\alpha}(a)$ which exhibits a different dependence on
the scale factor for the zero mode compared to the KK modes.
\\
Let us decompose the energy density into zero-mode and
KK contributions
\begin{equation}
\rho = \rho_0 + \rho_{KK}.
\end{equation}
For the energy density of the massless zero mode 
one then obtains 
\begin{equation}\label{4.15}
\rho_0 = \frac{1}{a^4}\int \frac{d^3k}{(2\pi)^{3}}
\,k \,{\cal N}_{0,k}^{\rm out}~.
\end{equation}
This is the expected behavior;
the energy density of standard four-dimensional gravitons 
scales like radiation.
\\
On contrast, the energy density of the KK modes 
at late times is found to be
\begin{equation}
\rho_{\rm KK} = \frac{L^2}{a^6}\frac{\pi^2}{4} \sum_n
\int \frac{d^3k}{(2\pi)^{3}}
\omega_{n,k}^{\rm out} \;{\cal N}_{n,k}^{\rm out}\,m_n^2 Y_1^2(m_ny_s),
\end{equation}
which decays like $1/a^6$. 
As the Universe expands, the energy density
of massive gravitons on the brane is therefore rapidly diluted.
The total energy density of gravitational waves in our Universe
at late times is dominated by the standard four-dimensional 
graviton (massless zero mode).
In the large mass limit $m_n y_s \gg 1$,$n \gg 1$
the KK-energy density can be approximated by
\begin{equation}
\rho_{{\rm KK}} \simeq \frac{\pi L^2}{2a^6y_s}
\sum_n
\int \frac{d^3k}{(2\pi)^{3}}
\;{\cal N}_{n,k}^{\rm out}\,\omega_{n,k}^{\rm out}m_n ~.
\label{late time large mass KK energy density}
\end{equation}
Due to the factor $m_n$ coming from the function 
${\cal Y}_n^2$, i.e. from the normalization of the functions $\phi_n(t,y)$, 
for the summation over the KK-tower to converge,
the number of produced gravitons ${\cal N}^{\rm out}_{n,k}$
has to decrease faster than $1/m_n^3$ for large masses and
not just faster than $1/m_n^2$ as one might naively expect.
%
\subsection{Escaping of massive gravitons and localization of gravity}
\label{ss:escape}
\begin{figure}
\begin{center}
\includegraphics[height=6.5cm]{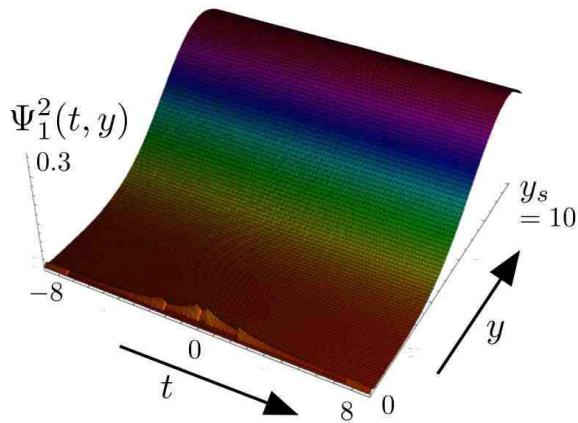}
\caption{Evolution of $\Psi^2_1(t,y) = \phi_1^2(t,y)/y^{3}$ 
corresponding to the probability to find the first KK graviton
at time $t$ at the position $y$ in the AdS-bulk. The static
brane is at $y_s=10L$ and the maximal brane velocity is given by
$v_b=0.1$.
\label{f:prob 1}}
\end{center}
\end{figure}

\begin{figure}
\begin{center}
\includegraphics[height=6.5cm]{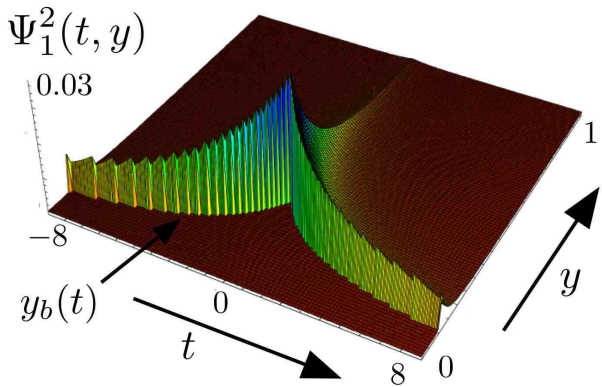}
\caption{Evolution of $\Psi^2_1(t,y)$ as in Fig. \ref{f:prob 1}
but zoomed into the bulk-region close to the moving brane.
\label{f:prob 2}}
\end{center}
\end{figure}

\begin{figure}
\begin{center}
\includegraphics[height=6.8cm]{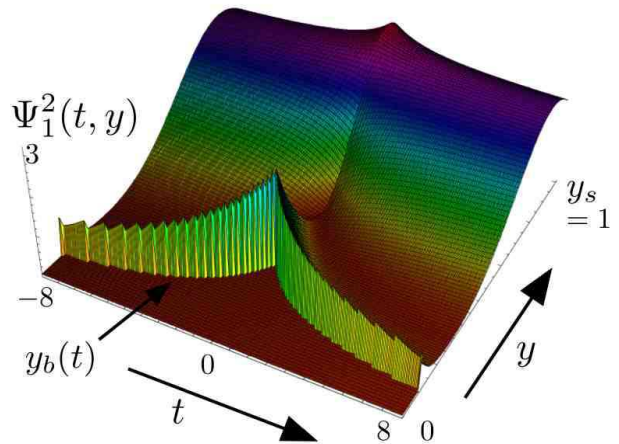}
\caption{Evolution of $\Psi^2_1(t,y)$ for $y_s=L$ and $v_b=0.1$.
\label{f:prob 3}}
\end{center}
\end{figure}

\begin{figure}
\begin{center}
\includegraphics[height=6.8cm]{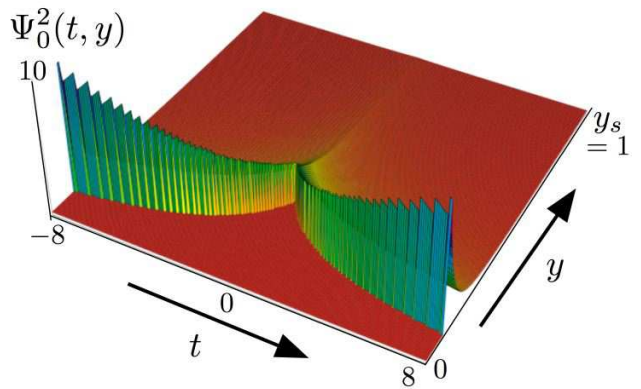}
\caption{Localization of four-dimensional gravity on a 
moving brane: Evolution of $\Psi^2_0(t,y)$ for $y_s=L=1$ 
and $v_b=0.1$ which should be compared with $\Psi^2_1(t,y)$
shown in Fig.~\ref{f:prob 3}.
\label{f:prob 4}}
\end{center}
\end{figure}

%
As we have shown, the power spectrum and energy density 
of the KK modes scale, at late times when particle 
production has ceased, with the expansion of the 
Universe like
\begin{equation}
{\cal P}_{\rm KK} \propto1/a^4\;,\;\; {\rho}_{\rm KK} \propto1/a^6.
\label{e:KK scalings}
\end{equation}
Both quantities decay by a factor $1/a^2$ faster than the 
corresponding expressions for the zero-mode graviton.
In particular, the energy density of the KK particles on 
the brane behaves effectively like stiff matter. 
Mathematically, this difference arises from the 
distinct behavior of the functions ${\cal Y}_0(a)$
and ${\cal Y}_n(a)$ [cf.~Eq.~(\ref{e:cal Y})] and
is a direct consequence of the warping of the 
fifth dimension.
But what is the underlying physics?
As we shall discuss now, this scaling behavior 
for the KK particles has indeed a 
very appealing physical interpretation which is 
in the spirit of the RS model.
\\
First, the mass $m_n$ is a comoving mass. 
The (instantaneous) 'comoving' frequency or energy 
of a KK graviton is $\omega_{n,k} =\sqrt{k^2+m_n^2}$, 
with comoving wave number $k$. 
The physical mass of a KK mode measured
by an observer on the brane with cosmic time $d\tau =adt$ is therefore
$m_n/a$, i.e. the KK masses are redshifted with the expansion of the
Universe. This comes from the fact that $m_n$ is the wave number
corresponding to the $y$-direction with respect to the bulk time $t$
which corresponds to {\it conformal time} $\eta$ on the brane and
not to physical time. It implies that the energy of KK particles on
a moving AdS brane is redshifted like that of massless
particles. 
From this alone one would expect that the energy density of 
KK modes on the brane decays like $1/a^4$ 
(see also Appendix D of \cite{Gorbunov:2001ge}).
\\
\\
Now, let us define the ``wave function'' for a graviton
\begin{equation}
\Psi_\alpha(t,y) = \frac{\phi_\alpha(t,y)}{y^{3/2}}
\end{equation}
which, by virtue of $(\phi_\alpha,\phi_\alpha)=1$,  
satisfies 
\begin{equation}
2\,\int_{y_b}^{y_s} dy \Psi_\alpha^2(t,y)=1
\end{equation}
From the expansion of the gravity wave amplitude Eq.~(\ref{e: mode expansion})
and the normalization condition  
it is clear that $\Psi_\alpha^2(t,y)$ gives the probability to find 
a graviton of mass $m_\alpha$ for a given (fixed) time $t$ at position 
$y$ in the ${\mathbb Z}_2$-symmetric AdS-bulk.
Since $\phi_\alpha$ satisfies Equation (\ref{laplacian equation}), 
the wave function $\Psi_\alpha$ satisfies the 
Schr{\"o}dinger like equation
\begin{equation}
-\partial_y^2 \Psi_\alpha + \frac{15}{4\,y^2} \Psi_\alpha = m_\alpha^2\Psi_\alpha
\label{e:graviton wave equation} 
\end{equation}
and the junction conditions (\ref{boundary conditions}) translate into
\begin{equation}
\left(\partial_y + \frac{3}{2\,y}\right)\Psi_\alpha|_{y=\{y_b,y_s\}} = 0.
\label{e:graviton wave junction}
\end{equation}
\\
In Fig.~\ref{f:prob 1} we plot the evolution of $\Psi_1^2(t,y)$
under the influence of the brane motion Eq.~(\ref{e:yb})
with $v_b= 0.1$. 
For this motion, the physical brane starting at $y_b \rightarrow 0$
for $t \rightarrow -\infty$ moves towards the static 
brane, corresponding to a contracting Universe. 
After a bounce, it moves back to the Cauchy horizon, 
i.e. the Universe expands.
The second brane is placed at $y_s = 10L$ 
and $y$ ranges from $y_b(t)$ to $y_s$. 
We set $\Psi^2_1 \equiv 0$ for $y < y_b(t)$ . 
The time-dependent KK mass $m_1$ is determined numerically 
from Eq.~(\ref{e:zero equation}).
As it is evident from this Figure, $\Psi_1^2$
is effectively localized close to the static brane, i.e. 
the weight of the KK-mode wave function lies
in the region of less warping, far from the physical brane.
Thus the probability to find a KK mode is larger 
in the region with less warping.
Since the effect of the brane motion on $\Psi_1^2$ is hardly visible
in Fig.~\ref{f:prob 1}, we show the behavior of $\Psi_1^2$ close to
the physical brane in Fig.~\ref{f:prob 2}. 
This shows that $\Psi_1^2$ peaks also at the physical 
brane but with an amplitude roughly ten times smaller 
than the amplitude at the static brane. 
While the brane, coming from $t\rightarrow -\infty$, approaches 
the point of closest encounter $\Psi_1^2$ slightly increases and 
peaks at the bounce $t=0$ where, as we shall show in the next 
Section, the production of KK particles
takes place. 
Afterwards, for $t\rightarrow \infty$, when the brane is
moving back towards the Cauchy horizon, the amplitude
$\Psi_1^2$ decreases again and so does the  probability to find a
KK particle at the position of the physical brane, 
i.e. in our Universe.
The parameter settings used in Figures \ref{f:prob 1} and \ref{f:prob 2}
are typical parameters which we use in the numerical simulations
described later on.
However, the effect is illustrated much 
better if the second brane is closer to the moving brane. 
In Figure \ref{f:prob 3} we show $\Psi_1^2$ for the same parameters 
as in Figures \ref{f:prob 1} and \ref{f:prob 2} but now with $y_s=L$. 
In this case, the probability to find a KK particle on the physical brane
is of the same order as in the region close to the second brane
during times close to the bounce.
However, as the Universe expands, $\Psi_1^2$ rapidly decreases
at the position of the physical brane.
\\
\\
From Eqs.~(\ref{e:phi on brane}) and (\ref{e:cal Y})
it follows that $\Psi_n^2(t,y_b) \propto 1/a$. 
The behavior of the KK-mode wave function suggests 
the following interpretation:
If KK gravitons are created on the brane, or equivalently 
in our Universe, they escape from the brane into the bulk 
as the brane moves back to the Cauchy horizon, i.e.
when the Universe undergoes expansion. 
This is the reason why the power spectrum and the energy density 
imprinted by the KK modes on the brane decrease faster with the 
expansion of the Universe than for the massless zero mode. 
\\
\\
The zero mode, on the other hand, is localized at the 
position of the moving brane. 
The profile of $\phi_0$ does not depend on the extra 
dimension, but the zero-mode wave function $\Psi_0$ does.
Its square is  
\begin{equation}
\Psi_0^2(t,y) = \frac{y_s^2 y_b^2}{y_s^2 - y_b^2}\frac{1}{y^3}
\rightarrow \frac{y_b^2}{y^3} = \left(\frac{L}{a}\right)^2\frac{1}{y^3}
\;\;{\rm if}\;\;y_s \gg y_b~,
\label{e:zero mode in bulk}
\end{equation} 
such that on the brane ($y=y_b)$ it behaves as 
\begin{equation}
\Psi_0^2(t,y_b) \simeq \frac{a}{L}.
\label{e:zero mode on brane}
\end{equation}
Equation (\ref{e:zero mode in bulk}) shows that, 
at any time, the zero mode is localized 
at the position of the moving brane. 
For a better illustration we show 
Eq.~(\ref{e:zero mode in bulk}) in Fig.~\ref{f:prob 4} 
for the same parameters as in Fig.~\ref{f:prob 3}.
This is the ``dynamical analog'' of the localization mechanism 
for four-dimensional gravity discussed in \cite{Randall:1999vf}.
\\
\\
To establish contact with \cite{Randall:1999vf} and to 
obtain a intuitive physical description, we rewrite the boundary value 
problem (\ref{e:graviton wave equation}), 
(\ref{e:graviton wave junction}) as a Schr{\"o}dinger-like equation
\begin{equation}
-\partial_y^2 \Psi_\alpha(t,y) + V(y,t) \Psi_\alpha(y,t) = m_\alpha(t)\Psi_\alpha(y,t)
\end{equation}
with 
\begin{eqnarray}
V(y,t) &=& \frac{15}{4\,y^2} - \frac{3}{y_b(t)} \,\delta(|y| - y_b(t))
\nonumber \\ 
&=& \frac{15}{4\,y^2} - 3\frac{a(t)}{L} \,\delta(|y| - y_b(t))~,
\label{e:volcano time}
\end{eqnarray}
where we have absorbed the boundary condition at the moving brane into 
the (instantaneous) {\it volcano potential} $V(y,t)$ and made 
use of ${\mathbb Z}_2$ symmetry. 
Similar to the static case \cite{Randall:1999vf}, 
at any time the potential (\ref{e:volcano time}) supports 
a single bound state, the four-dimensional graviton 
(\ref{e:zero mode in bulk}), and acts as a barrier 
for the massive KK modes.
The potential, ensuring localization of 
four-dimensional gravity on the brane and the repulsion 
of KK modes, moves together with the brane 
through the fifth dimension. 
Note that with the expansion of the Universe, the 
``depth of the delta-function'' becomes larger, 
expressing the fact that the localization of four-dimensional
gravity becomes stronger at late times 
[cf.~Eq.~(\ref{e:zero mode on brane}), Fig.~\ref{f:prob 4}].
\\
\\
In summary, the different scaling behavior for the zero- and KK modes 
on the brane is entirely a consequence
of the geometry of the bulk space-time, i.e. of the warping
$L^2/y^2$ of the metric (\ref{e:bulk-metric}) 
\footnote{Note that it does not depend on a particular type 
of brane motion and is expected to be true also in the high
energy case which we do not consider here.}.
It is simply a manifestation of the localization of gravity on
the brane: as time evolves, the KK gravitons, which are traces of the
five-dimensional nature of gravity, escape into the bulk 
and only the zero mode which corresponds to the usual 
four-dimensional graviton remains on the brane.
\\
\\
This, and in particular the scaling behavior
(\ref{e:KK scalings}), remains  also true if the second 
brane is removed, i.e. in the limit $y_s\rightarrow \infty$,
leading to the original RS II model.
By looking at (\ref{late time large mass KK power spectrum}) and 
(\ref{late time large mass KK energy density}) one could at first 
think that then the KK-power spectrum and energy density
vanish and no traces of the KK gravitons could be 
observed on the brane since both expressions behave 
as $1/y_s$.
But this is not the case since the spectrum of KK masses 
becomes continuous. In the continuum limit 
$y_s \rightarrow \infty$ the summation 
over the discrete spectrum $m_n$ has to be replaced 
by an integration over continuous masses $m$ 
in the following way:  
\begin{equation}
\frac{1}{y_s} \sum_n f(m_n) \longrightarrow \frac{1}{\pi} 
\int dm \,f(m)~.
\label{e:dis to cont}
\end{equation}
$f$ is some function depending on the spectrum, 
for example $f(m_n) = {\cal N}_{n,k}^{\rm out}$.
The pre-factor $1/y_s$ in 
(\ref{late time large mass KK power spectrum}) and 
(\ref{late time large mass KK energy density})
therefore ensures the existence of the proper 
continuum limit of both expressions.
\\
Another way of seeing this is to repeat the same 
calculations but using the eigenfunctions 
for the case with only one brane from the beginning.
Those are $\delta$-function normalized and can be 
found in, e.g., \cite{Gorbunov:2001ge}. 
They are basically the same as (\ref{e:phij}) 
except that the normalization is different since 
it depends on whether the fifth dimension is 
compact or not.
In particular, on the brane, they have the same 
scale factor dependence as (\ref{e:phi on brane}). 
\\
\\
At the end, the behavior found for the KK modes 
should not come as a surprise, since the 
RS II model has attracted lots of attention 
because of exactly this; it localizes usual 
four-dimensional gravity on the brane.
As we have shown here, localization of standard four-dimensional
gravity on a moving brane via a warped geometry automatically 
ensures that the KK modes escape into the bulk as the 
Universe expands because their wave function has its weight 
in the region of less warping, resulting in an 
KK-mode energy density on the brane which scales like stiff matter.
\\
An immediate consequence of this particular scaling behavior is
that KK gravitons in an AdS braneworld
cannot play the role of dark matter.
Their energy density in our Universe decays much faster
with the expansion than that of ordinary matter which is 
restricted to reside on the brane. 

%
\section{Numerical simulations}\label{sec:num}
%
\subsection{Preliminary remarks}
%
In this section we present results of numerical
simulations for the bouncing model described by
the equations \eqref{e:solexp}-\eqref{e:vb}.
\\
In the numerical simulations we set $L=1$, i.e. all
dimensionful quantities are measured in units of
the AdS$_5$ curvature scale.
Starting at initial time $t_{\rm in} \ll 0$ where the initial vacuum
state $|0,{\rm in}\rangle$ is defined, the 
system~(\ref{deq for xi},\ref{deq for eta}) is evolved numerically
up to final time $t_{\rm out}$.
Thereby we set $t_{\rm in} = - 2 \pi N_{\rm in} / k$ with
$1 \le N_{\rm in}  \in {\mathbb N}$, such that 
$\Theta_{0,k}^{\rm in} = 1$ [cf.~Eq.~(\ref{e:Theta})].
This implies $\xi_{0}^{(0)}(t_{\rm in})=2$, i.e.
independent of the three-dimensional momentum $k$ a (plane wave) zero-mode
solution always performs a fixed number of oscillations between
$t_{\rm in}$ and the bounce at $t=0$ 
[cf.~Eq.~(\ref{e:vacuum initial conditions for xi,eta})].
The final graviton spectrum at 
${\cal N}_{\alpha,k}^{\rm out}$ is calculated at late times 
$t_{\rm out} \gg 1$ when the brane approaches the Cauchy horizon
and graviton creation has ceased. 
This quantity is physically well defined and leads to the
late-time power spectrum ~(\ref{e:Ps}) and energy 
density (\ref{e:energy density late time out number}) 
on the brane.
For illustrative purposes, we also plot
the instantaneous particle number ${\cal N}_{\alpha,k,\bullet}(t)$ 
which also determines the power spectrum at all times
[cf~Eq.(\ref{e:R function with inst particle number})].
In this section we shall use the term 
particle number respectively graviton number
for both, the instantaneous particle number
${\cal N}_{\alpha,k,\bullet}(t)$ as well as the final state graviton
number ${\cal N}_{\alpha,k,\bullet}^{\rm out}$, keeping in mind that
only the latter one is physically meaningful.
\\
\\
There are two physical input parameters for the numerical simulation;
the maximal brane velocity $v_b$ (i.e. $t_b$)
and the position of the static brane $y_s$.
The latter determines the number of KK modes which fall
within a particular mass range.
On the numerical side one has to specify $N_{\rm in}$ and
$t_{\rm out}$, as well as the maximum number of KK modes
$n_{\rm max}$ which one takes into account, i.e. after which KK mode
the system of differential equations is truncated.
The independence of the numerical results on the choice of the time
parameters is checked and the convergence of the particle spectrum
with increasing $n_{\rm max}$ is investigated.
More detailed information on numerical issues including 
accuracy considerations are collected in Appendix 
\ref{a:numerics}.
\\
One strong feature of the brane motion (\ref{e:yb}) is its kink at the
bounce $t=0$.
In order to study how particle production depends 
on the kink, we shall compare the motion (\ref{e:yb}) with the
following motion which has a smooth transition from 
contraction to expansion $(L=1)$:
\begin{equation}
y_b(t) = \left \{
\begin{array}{ll}
(|t|+t_b-t_s)^{-1} & {\rm if}\;\; |t| > t_s   \label{e:yb smooth} \\
a +(b/2)t^2 + (c/4)t^4 & {\rm if}\;\; |t| \le t_s
\end{array}
\right .
\end{equation}
with the new parameter $t_s$ in the range  $0<t_s<t_b$.
This motion is constructed such that its velocity
at $|t|=t_s$ is the same as the velocity of the kink motion at
the bounce. 
This will be the important quantity determining the number of
produced gravitons.
For $t_s\rightarrow 0$ the motion with smooth transition
approaches (\ref{e:yb}).
The parameters $a,b$ and $c$ are obtained by matching the motions
and the first and second derivatives.
Matching also the second derivative guarantees that possible spurious
effects contributing to particle production are avoided.
The parameter $t_s$ has to be chosen small enough, $t_s \ll 1$,
such that the maximal velocity of the smooth motion
is not much larger than $v_b$ in order to have comparable situations.
\\
For reasons which will become obvious in the next two 
sections we shall discuss the cases of long $k\ll 1$ and 
short wavelengths $k\gg 1$, separately.
%
\subsection{Generic results and observations for long wavelengths $k \ll 1$}
\begin{figure}
\begin{center}
\includegraphics[height=6cm]{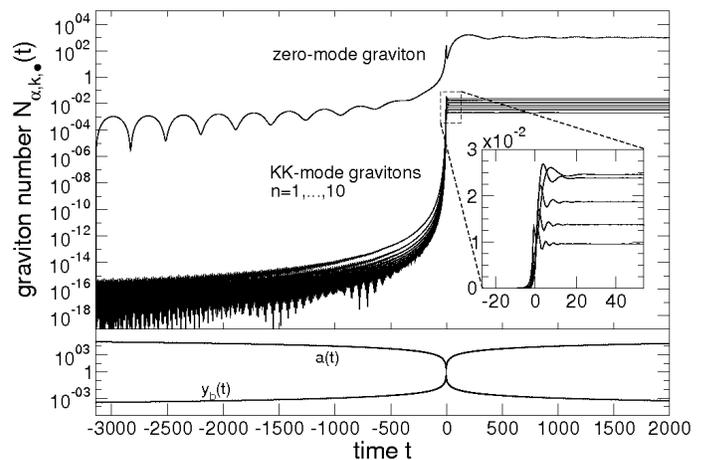}
\caption{Evolution of the graviton number 
${\cal N}_{\alpha,k,\bullet}(t)$ for the zero mode and the first
ten KK modes for three-momentum $k=0.01$ and
$v_b=0.1$, $y_s=10$.
\label{genfig1}}
\end{center}
\end{figure}

%
Figure \ref{genfig1} displays the results of a numerical simulation
for three-momentum $k=0.01$, static brane position
$y_s = 10$ and maximal brane velocity $v_b = 0.1$.
Depicted is the graviton number for one polarization
${\cal N}_{\alpha,k,\bullet}(t)$ for the zero mode and the first
ten KK modes as well as the evolution of the scale factor
$a(t)$ and the position of the physical brane $y_b(t)$.
Initial and final times are $N_{\rm in} = 5$ 
and $t_{\rm out} = 2000$, respectively.
The KK-particle spectrum will be discussed in detail below.
\begin{figure}
\begin{center}
\includegraphics[height=6cm]{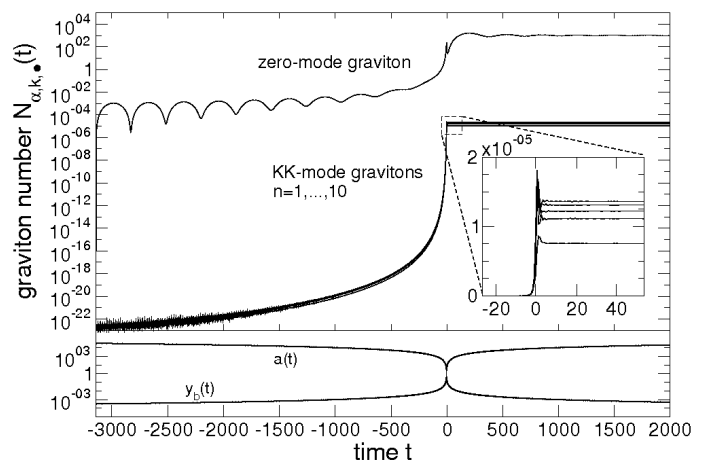}
\caption{${\cal N}_{n,k,\bullet}(t)$ for the zero mode and the
first ten KK modes for the parameters of Fig.~\ref{genfig1},
but without coupling of the zero mode to the KK modes, i.e.
$M_{i0}\equiv 0$.
\label{genfig2}}
\end{center}
\end{figure}
One observes that the zero-mode particle number increases slightly
with the expansion of the Universe towards the bounce at $t=0$.
Close to the bounce ${\cal N}_{0,k,\bullet}(t)$
increases drastically, shows a local peak at the bounce and,
after a short decrease, grows again until the mode is sub-horizon
($kt \gg 1$). Inside the horizon ${\cal N}_{0,k,\bullet}(t)$
is oscillating around a mean value with diminishing amplitude.
This mean value which is reached asymptotically 
for $t\rightarrow \infty$ corresponds
to the number of generated final state zero-mode gravitons
${\cal N}_{0,k,\bullet}^{\rm out}$.
Production of KK-mode gravitons takes effectively place only at
the bounce in a step-like manner and the graviton number remains
constant right after the bounce.
\\
In Fig.~\ref{genfig2} we show the numerical
results obtained for the same parameters as in Fig.~\ref{genfig1}
but without coupling of the zero mode to the KK modes, i.e.
$M_{i0}=0$ (and thus also $N_{i0}=N_{0i}=0$).
One observes that the production of zero-mode gravitons
is virtually not affected by the artificial decoupling
\footnote{Quantitatively it is
${\cal N}_{0,k,\bullet}(t=2000) = 965.01$ with and
${\cal N}_{0,k,\bullet}(t=2000) = 965.06$ without $M_{i0}$.
Note that this difference lies indeed within the accuracy 
of our numerical simulations (see Appendix \ref{a:numerics}.)}.
Note that even if $M_{0j} \equiv 0$ (see Eqs.~\ref{e:T:Mij-exact 0j}),
which is in general true for Neumann boundary conditions,
the zero mode $q_{0,{\bf k},\bullet}$ couples in Eq.~(\ref{deq for q})
to the KK modes via $N_{0j}=M_{00}M_{j0}$ and through the
anti-symmetric combination $M_{\alpha\beta} - M_{\beta\alpha}$.
\\
In contrast, the production of the first ten KK modes is heavily 
suppressed if $M_{i0} = 0$.
The corresponding final-state graviton numbers 
${\cal N}_{n,k,\bullet}^{\rm out}$ are reduced by four orders of
magnitude. 
This shows that the coupling to the zero mode is essential 
for the production of massive gravitons.
Later we will see that this is true for light
KK gravitons only. 
If the KK masses exceed $m_i\sim 1$, they evolve independently of
the four-dimensional graviton and their evolution is entirely driven by 
the intermode couplings $M_{ij}$.
It will also turn out that the time-dependence of the KK mass 
$m_i$ plays only an inferior role for the generation of massive KK modes.
On the other hand, the effective decoupling of the evolution of the
zero mode from the KK modes occurs in general as long as 
$k \ll 1$ is satisfied, i.e. for long-wavelengths.  
We will see that it is no longer true for short wavelengths
$k \gg 1$.
\\
The effective decoupling of the zero-mode evolution 
from the KK modes makes it possible to derive analytical 
expressions for the number of zero-mode gravitons, their power spectrum
and energy density.
The calculations are carried out in section 
\ref{ss:zm long ana}
\\
In summary we emphasize the important observation that
for long wavelengths the amplification of the four dimensional 
gravity wave amplitude during the bounce is 
not affected by the evolution of the KK gravitons.
We can therefore study the zero mode separately from the KK modes
in this case.  
%
\subsection{Zero mode: long wavelengths $k \ll 1$}
\begin{figure}
\begin{center}
\includegraphics[height=6cm]{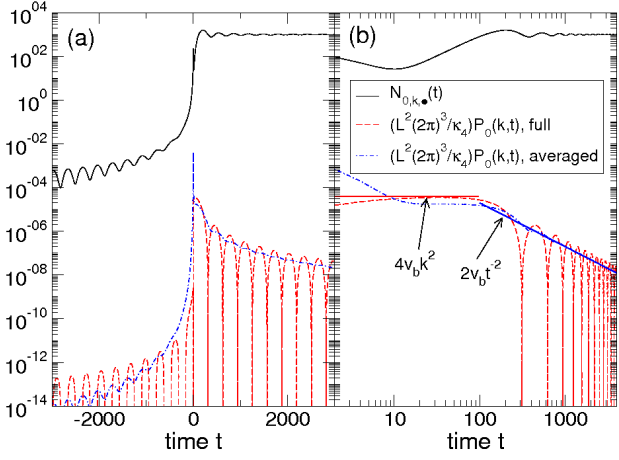}
\caption{Time evolution of the number of created zero-mode
gravitons ${\cal N}_{0,k,\bullet}(t)$
and of the zero-mode power spectrum (\ref{e:power spectrum with R}):
(a) for the entire integration time; (b) for $t > 0$ only.
Parameters are $k=0.01$, $y_s=10$ and $v_b=0.1$.
Initial and final time of integration are given by $N_{\rm in} = 10$ and
$t_{\rm out}=4000$, respectively. The power spectrum is shown
with and without the term ${\cal O}^{{\cal N}}_{0,k,\bullet}$,
i.e. before and after averaging, respectively, and compared with the
analytical results. 
\label{figure pow}}
\end{center}
\end{figure}

\begin{figure}
\begin{center}
\includegraphics[height=6cm]{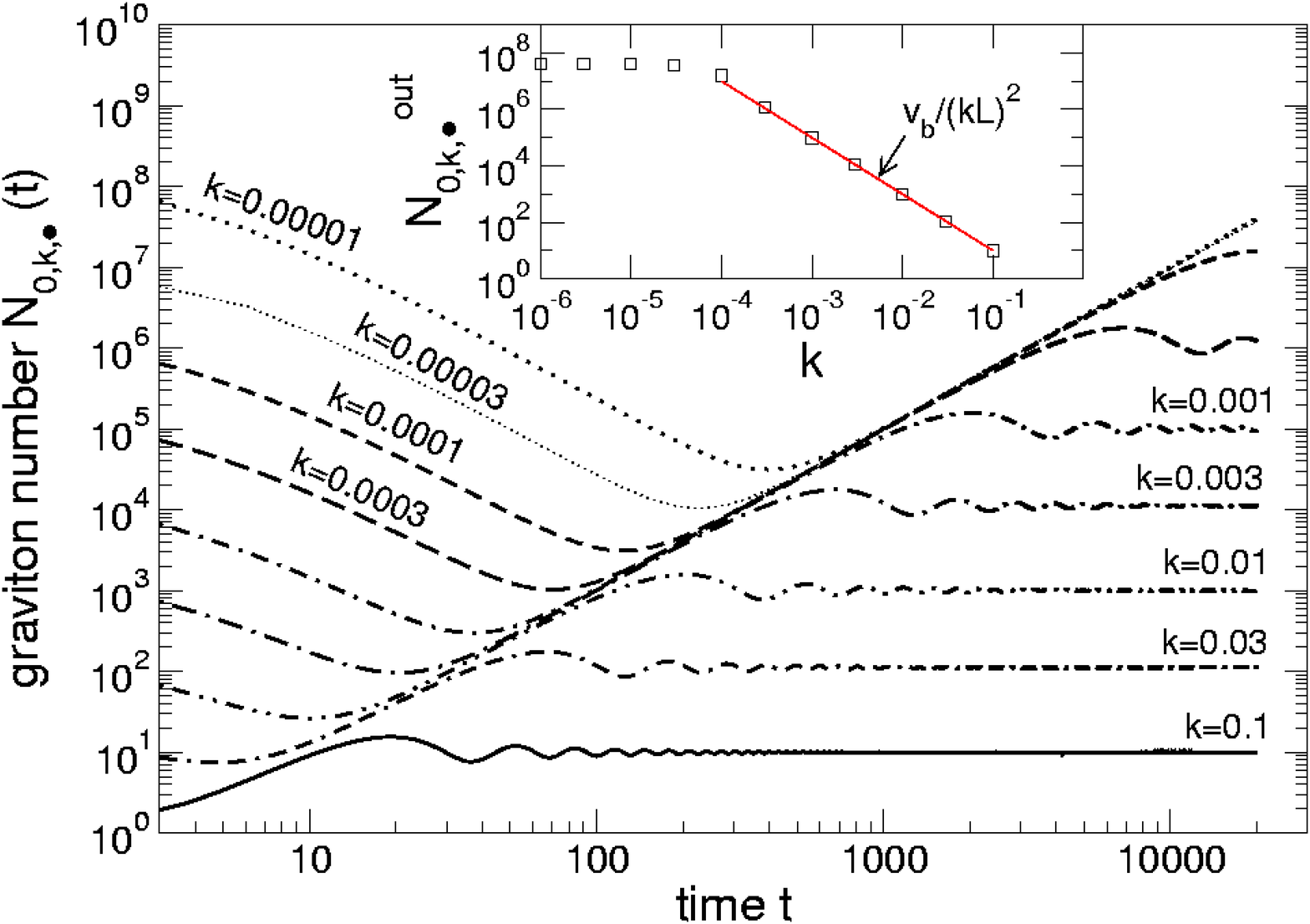}
\caption{Numerical results for the time evolution of the number
of created zero-mode gravitons ${\cal N}_{0,k,\bullet}(t)$
after the bounce $t>0$ for different
three-momenta $k$. The maximal brane velocity at the
bounce is $v_b = 0.1$ and the second brane is positioned at
$y_s = 10$. In the final particle spectrum the numerical values are
compared with the analytical prediction Eq.~\eqref{N0}. Initial and final
time of integration are given by $N_{\rm in} = 5$ and
$t_{\rm out}=20000$, respectively.
\label{f:zeromode numbers}}
\end{center}
\end{figure}

\begin{figure}
\begin{center}
\includegraphics[height=6cm]{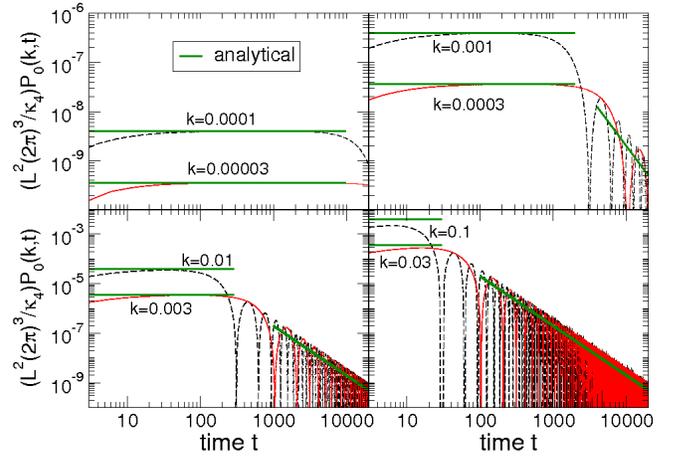}
\caption{Evolution of the zero-mode power spectrum after the bounce
$t>0$ corresponding to the values and parameters of
  Fig.~\ref{f:zeromode numbers}. 
The numerical results are compared to the analytical predictions
Eqs.~\eqref{e:Ps late sub} and \eqref{P0super}.
\label{f:zeromode powerspectra}}
\end{center}
\end{figure}

\begin{figure}
\begin{center}
\includegraphics[height=6cm]{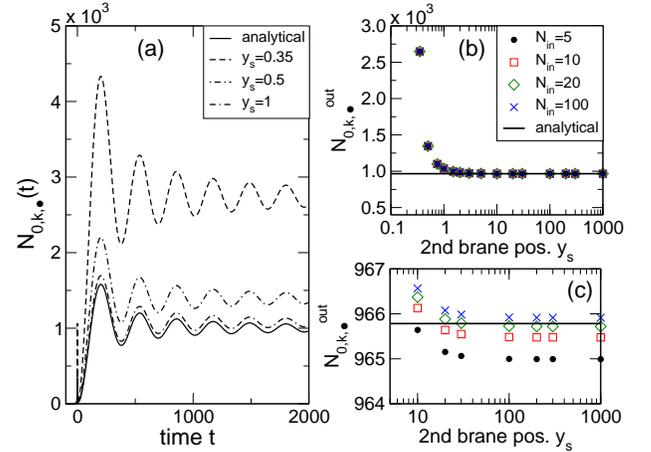}
\caption{Dependence of the zero-mode particle number on
inter-brane distance and initial integration time
for momentum $k=0.01$, maximal brane velocity
$v_b=0.1$ in comparison with the analytical expression
Eq.~\eqref{e:Bogoliubov B00}.
(a) Evolution of the instantaneous particle number
${\cal N}_{0,k,\bullet}(t)$ with initial integration time given by
$N_{\rm in} = 5$ for $y_s = 0.35,0.5$ and $1$.
(b) Final zero-mode graviton spectrum
${\cal N}_{0,k,\bullet}(t_{\rm out} = 2000)$ for various values of $y_s$ and
$N_{\rm in}$.
(c) Close-up view of (b) for large $y_s$.
\label{f:zeromode dep}}
\end{center}
\end{figure}

%
In Figure \ref{figure pow} we show the numerical results
for the number of generated zero-mode gravitons 
${\cal N}_{0,k,\bullet}(t)$ and the evolution
of the corresponding power spectrum ${\cal P}_0(k)$ 
on the brane for momentum $k=0.01$, position of the static brane $y_s=10$
and maximal brane velocity $v_b=0.1$.
The results have been obtained by solving the equations for the zero
mode alone, i.e. without the couplings to the KK modes, since, as we have
just shown, the evolution of the four-dimensional graviton for long 
wavelengths is not influenced by the KK modes.
Thereby the power spectrum is shown before and after averaging over
several oscillations, i.e. employing
Eq.~(\ref{e:R function with inst particle number}) with and without the term
${\cal O}_{0,k}^{{\cal N}}$, respectively.
Right after the bounce where the generation of gravitons is
initiated and which is responsible for the peak in ${\cal N}_{0,k,\bullet}$ at
$t=0$, the number of gravitons first decreases again.
Afterwards ${\cal N}_{0,k,\bullet}$ grows further until the mode enters
the horizon at $kt=1$. Once on sub-horizon scales $kt \gg 1$,
the number of produced gravitons oscillates with a diminishing
amplitude and asymptotically approaches the final state
graviton number ${\cal N}_{0,k,\bullet}^{\rm out}$.
During the growth of ${\cal N}_{0,k,\bullet}$ after the bounce, the power
spectrum remains practically constant. 
Within the range of validity it is in good agreement
with the analytical prediction \eqref{P0super} yielding
$(L^2(2\pi)^3/\kappa_4){\cal P}_0(k,t)=4v_b(kL)^2 $. 
When particle creation has ceased, the
full power spectrum Eq.(\ref{e:power spectrum with R}) starts to
oscillate with an decreasing amplitude. The time-averaged power
spectrum obtained by using Eq. (\ref{e:R function with inst particle number})
without the ${\cal O}_{0,k}^{\cal N}$-term is perfectly in agreement
with the analytical expression Eq.~(\ref{e:Ps late sub}) which gives
$(L^2(2\pi)^3/\kappa_4){\cal P}_0(k,t)=2v_b/t^2 $.
Note that at early times, the time-averaged power spectrum behaves
not in the same way as the full one, demonstrating the importance of the
term ${\cal O}_{0,k}^{{\cal N}}$. 
\\
Figure \ref{f:zeromode numbers} shows a summary 
of numerical results for the number
of created zero-mode gravitons ${\cal N}_{0,k,\bullet}(t)$ for different
values of the three-momentum $k$.
The maximum velocity at the bounce is $v_b = 0.1$ and the
second brane is at $y_s=10$. These values are representative.
Other values in accordance with the considered low-energy
regime do not lead to a qualitatively different behavior.
Note that the evolution of the zero mode 
does virtually not depend on the value of $y_s$ as long as 
$y_s \gg y_b(0)$ (see below).
Initial and final integration times are given by $N_{\rm in} = 5$
and $t_{\rm out} = 20000$, respectively.
\\
For sub-horizon modes we compare the final graviton 
spectra with the analytical prediction \eqref{N0}. 
Both are in perfect agreement.
On super-horizon scales where particle creation has not ceased yet
${\cal N}_{0,k,\bullet}$ is independent of $k$.
The corresponding time-evolution of the power spectra
${\cal P}_{0}(k,t)$ is depicted in 
Fig.~\ref{f:zeromode powerspectra}.
For the sake of clarity, only the results for $t>0$, i.e. after the
bounce, are shown in both figures.
\\
The numerical simulations and the calculations 
of section \ref{ss:zm long ana} reveal that the 
power spectrum for the four-dimensional graviton for 
long wavelengths is blue on super-horizon scales, as expected for 
an ekpyrotic scenario. 
\\
The analytical calculations performed in 
section \ref{ss:zm long ana} rely on the assumption that
$y_b \ll y_s$ and $t_{\rm in} \rightarrow -\infty$.
Figure~\ref{f:zeromode dep} shows the behavior of 
the number of generated zero-mode
gravitons of momentum $k=0.01$ in dependence on the inter-brane
distance and the initial integration time. The brane velocity
at the bounce is $v_b = 0.1$ which implies that at the
bounce the moving brane is at
$y_b(0) = \sqrt{v_b} \simeq 0.316$ ($L=1$).
In case of a close encounter of the two branes as for
$y_s=0.35$, the production of massless gravitons is strongly enhanced
compared to the analytical result. But as soon as $y_s \ge 1$,
(i.e. $y_s \ge L$) the numerical result is very well described
by the analytical expression Eq.~(\ref{e:Bogoliubov B00}) 
derived under the assumption $y_s \gg y_b$. 
For $y_s \ge 10$ the agreement between both is very
good.
From panels (b) and (c) one infers that the numerical result
becomes indeed independent of the initial integration time
when increasing $N_{\rm in}$. Note that in the limit
$N_{\rm in} \gg 1$ the numerical result is slightly larger than the
analytical prediction but the difference between both is negligibly
small.
This confirms the correctness and accuracy of the analytical
expressions derived in Section \ref{ss:zm long ana} for the evolution 
of the zero-mode graviton.
\\
%
\subsection{Kaluza-Klein-modes: long wavelengths $k \ll 1$}
\begin{figure}
\begin{center}
\includegraphics[height=6cm]{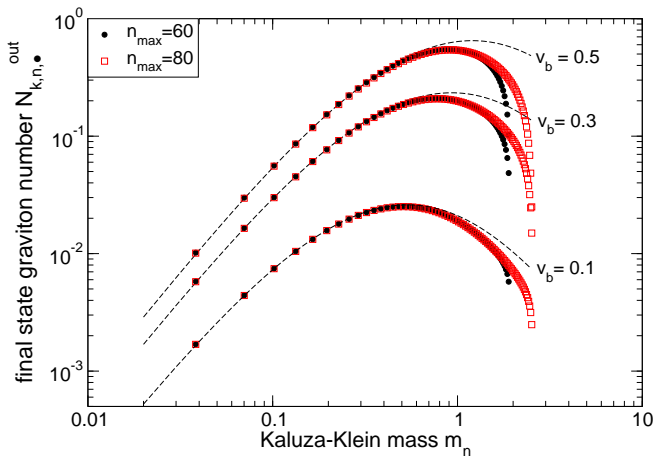}
\caption{Final state KK-graviton spectra for $k=0.001$, $y_s=100$, 
  different maximal brane velocities $v_b$ and $N_{\rm in} =
  1$, $t_{\rm out} = 400$. The numerical results are compared 
  with the analytical prediction Eq.~(\ref{Njapprox}) (dashed line).
\label{f:kkfig1}}
\end{center}
\end{figure}

\begin{figure}
\begin{center}
\includegraphics[height=6cm]{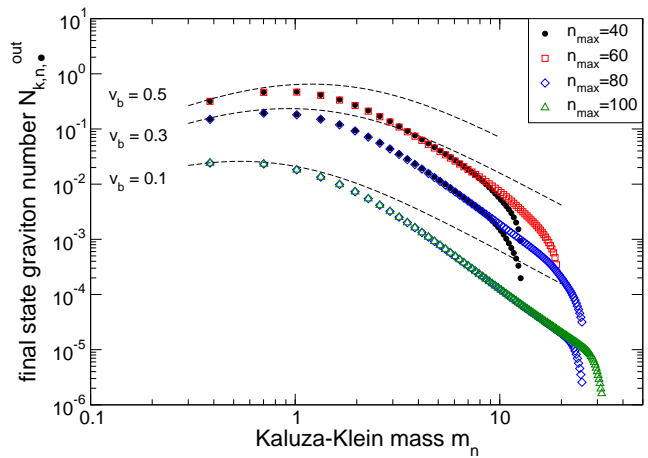}
\caption{Final state KK-graviton spectra for $k=0.01$, $y_s=10$, 
  different maximal brane velocities $v_b$ and $N_{\rm in} =
  2$, $t_{\rm out} = 400$. The numerical results are compared 
  with the analytical prediction Eq.~(\ref{Njapprox}) (dashed line).
\label{f:kkfig3}}
\end{center}
\end{figure}

\begin{figure}
\begin{center}
\includegraphics[height=6cm]{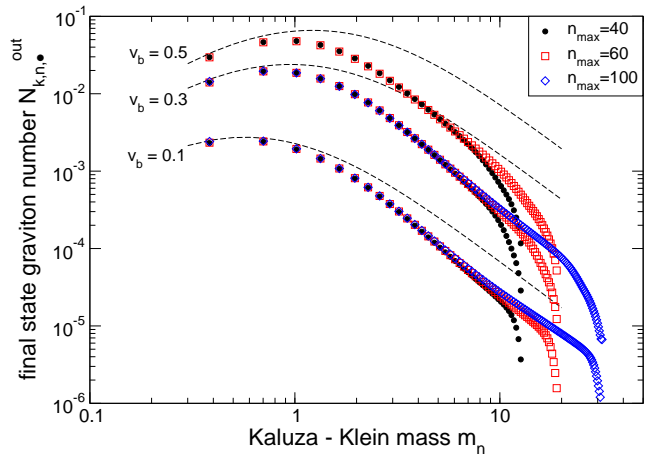}
\caption{Final state KK-graviton spectra for $k=0.1$, $y_s=10$, 
  different maximal brane velocities $v_b$ and $N_{\rm in} =
  2$, $t_{\rm out} = 400$. The numerical results are compared 
  with the analytical prediction Eq.~(\ref{Njapprox}) (dashed line).
\label{f:kkfig4}}
\end{center}
\end{figure}

\begin{figure}
\begin{center}
\includegraphics[height=6cm]{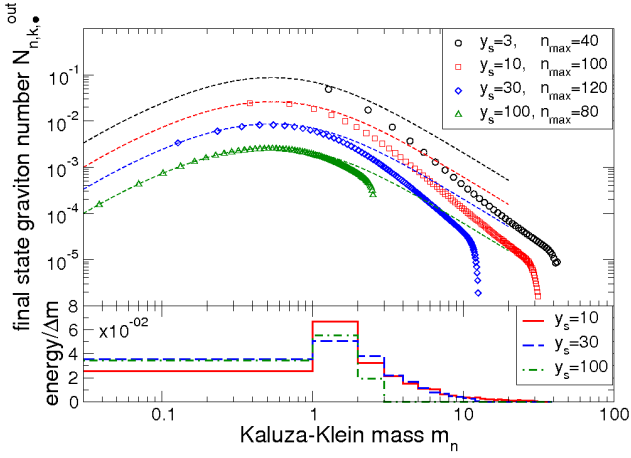}
\caption{Upper panel: Final state KK-particle spectra for
$k=0.01$, $v_b=0.1$ and different $y_s=3,10,30$ and $100$.
The analytical prediction Eq.~(\ref{Njapprox}) is shown as well (dashed line).  
Lower panel: Energy $\omega_{n,k}^{\rm out} {\cal N}_{n,k,\bullet}^{\rm out}$
of the produced final state gravitons binned in
mass intervals $\Delta m = 1$ for $y_s=10,30,100$.
\label{f:kkfig5}}
\end{center}
\end{figure}

\begin{figure}
\begin{center}
\includegraphics[height=6cm]{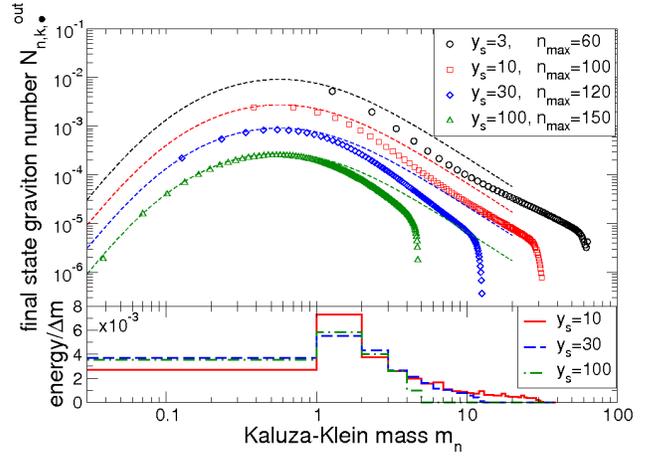}
\caption{Upper panel: Final state KK-particle spectra for
$k=0.1$, $v_b=0.1$ and different $y_s=3,10,30$ and $100$.
The analytical prediction Eq.~(\ref{Njapprox}) is shown as well (dashed line).  
Lower panel: Energy $\omega_{n,k}^{\rm out} {\cal N}_{n,k,\bullet}^{\rm out}$
of the produced final state gravitons binned in
mass intervals $\Delta m = 1$ for $y_s=10,30,100$.
\label{f:kkfig6}}
\end{center}
\end{figure}

\begin{figure}
\begin{center}
\includegraphics[height=6cm]{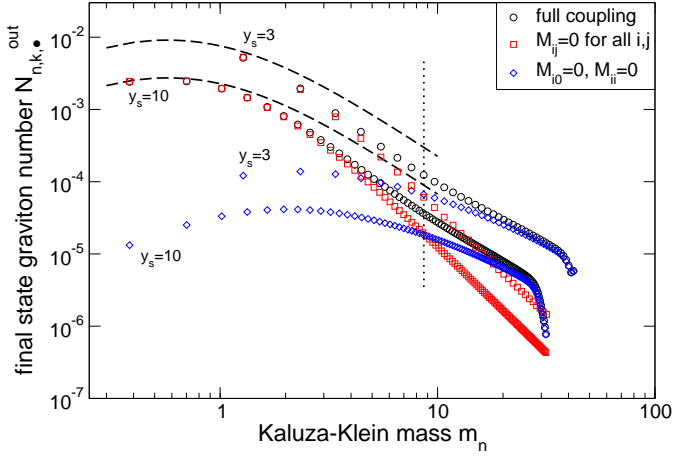}
\caption{KK-particle spectra for three-momentum $k=0.1$,
         maximum brane velocity $v_b=0.1$ and $y_s=3$
         and $10$ with different couplings taken into account.
	 The dashed lines indicates again the analytical expression
	 Eq.~(\ref{Njapprox}). 
\label{f:kkfig7}}
\end{center}
\end{figure}

\begin{figure}
\begin{center}
\includegraphics[height=6cm]{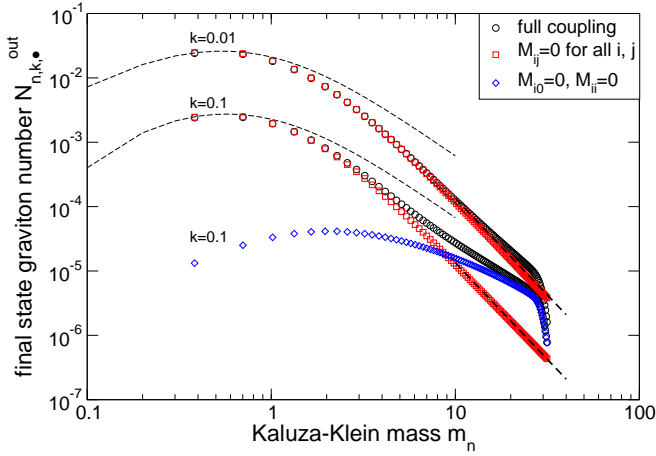}
\caption{KK-particle spectra for $y_s=10$, $v_b=0.1$,
         $n_{\rm max}=100$ and three-momentum $k=0.01$ and $0.1$
         with different couplings taken into account.
	 The thin dashed lines indicates Eq.~\eqref{Njapprox} and the 
	 thick dashed line Eq.~\eqref{e:Nipow}.
\label{f:kkfig10}}
\end{center}
\end{figure}

\begin{figure}
\begin{center}
\includegraphics[height=6cm]{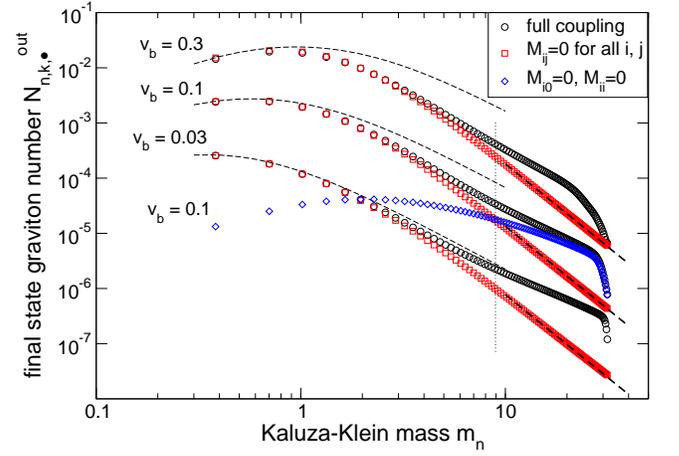}
\caption{KK-particle spectra for three-momentum $k=0.1$,$y_s=10$
         and maximum brane velocities $v_b=0.03,0.1$ and $0.3$
	 with $n_{\rm max}=100$. As in Fig.~ \ref{f:kkfig10} 
	 different couplings have been taken into account and
	 thin dashed lines indicates Eq.~\eqref{Njapprox} and the 
	 thick dashed line Eq.~\eqref{e:Nipow}.
\label{f:kkfig11}}
\end{center}
\end{figure}

\begin{figure}
\begin{center}
\includegraphics[height=6cm]{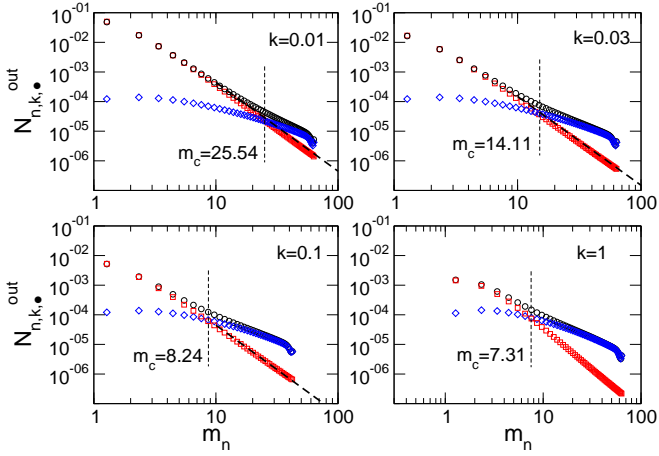}
\caption{KK-particle spectra for three-momentum $k=0.01$, $0.03$,
         $0.1$ and $1$ for $y_s=3$ and maximum brane velocity
         $v_b=0.1$ with different couplings taken into
         account where the notation is like in Fig.~\ref{f:kkfig11}.
	 From the crossing of the $M_{ii}=M_{ij}=0$- and
         $M_{ii}=M_{i0}=0$ results we determine the $k$-dependence
         of $m_c(k,v_b)$. The thick dashed line indicates Eq.~\eqref{e:Nipow}.
\label{f:kkfig12}}
\end{center}
\end{figure}

\begin{figure}
\begin{center}
\includegraphics[height=6cm]{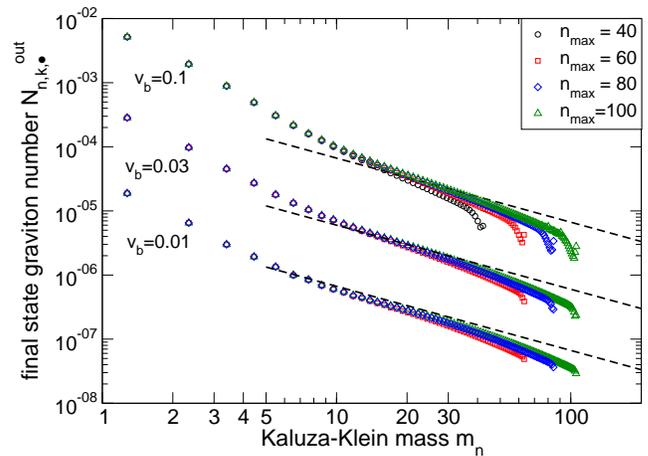}
\caption{KK-particle spectra for $k=0.1$, $y_s=3$ and maximal
brane velocities $v_b = 0.01, 0.03, 0.1$ up to
KK masses $m_n \simeq 100$ compared with an $1/m_n$ decline.
The dashed lines indicate the approximate expression 
(\ref{e:large mass spectrum fit}) which describes 
the asymptotic behavior of the final KK-particle spectra 
reasonably well, in particular for $v_b < 0.1$.
\label{f:kkfig14}}
\end{center}
\end{figure}

%
Because the creation of KK gravitons ceases right after the
bounce [cf Fig.~\ref{genfig1}] one can stop the numerical simulation
and read out the number of produced KK gravitons
${\cal N}_{n,k,\bullet}^{\rm out}$ at times for which 
the zero mode is still super-horizon.
\\
Even though Eq.~(\ref{e:zero equation}) cannot be solved analytically, the
KK masses can be approximated by $m_n \simeq n\pi/y_s$. This expression
is the better the larger the mass.
Consequently, for the massive modes the position of the second brane $y_s$
determines how many KK modes belong to a particular mass 
range $\Delta m$.
\\
In Figure \ref{f:kkfig1} we show the KK-graviton spectra
${\cal N}_{n,k,\bullet}^{\rm out}$ for three-momentum $k=0.001$ and
second brane position $y_s=100$ for maximal brane velocities
$v_b = 0.1,0.3$ and $0.5$. For any velocity $v_b$
two spectra obtained with $n_{\rm max}=60$ and $80$ KK modes taken
into account in the simulation are compared to each other.
This reveals that the numerical results are stable 
up to a KK mass $m_n \simeq 1$. 
\begin{figure}
\begin{center}
\includegraphics[height=6cm]{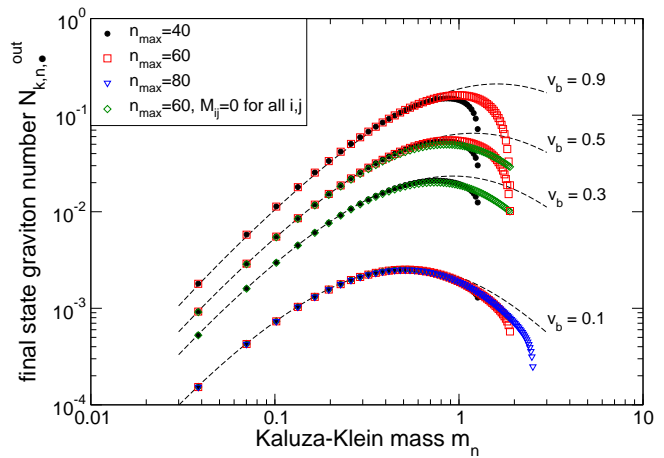}
\caption{Final state KK-graviton spectra for $k=0.01$, 
  $y_s=100$, different $v_b$ and $N_{\rm in} =
  1$, $t_{\rm out} = 400$. The numerical results are compared 
  with the analytical prediction Eq.~(\ref{Njapprox}) (dashed line). 
  For $v_b=0.3,0.5$ the spectra obtained
  without KK-intermode and self-couplings
  ($M_{ij} \equiv 0 \;\forall \,i,j$) are shown as well.
\label{f:kkfig2}}
\end{center}
\end{figure}
One infers that first, ${\cal N}_{n,k,\bullet}^{\rm out}$ grows with
increasing mass until a maximum is reached. The position of the
maximum shifts slightly towards larger masses with increasing
brane velocity $v_b$. Afterwards, ${\cal N}_{n,k,\bullet}^{\rm out}$
declines with growing mass. Until the maximum is reached, the
numerical results for the KK-particle spectrum are very stable.
This already indicates that the KK-intermode couplings mediated by $M_{ij}$
are not very strong in this mass range.
In Figure \ref{f:kkfig2} we show the final KK-particle spectrum
for the same parameters as in Fig. \ref{f:kkfig1} but for
three-momentum $k=0.01$ and the additional velocity
$v_b=0.9$ \footnote{Such a high brane
velocity is of course not consistent with a Neumann boundary
condition Eq.~(\ref{boundary conditions}) at the position of the
moving brane.}.
We observe the same qualitative behavior as in Fig.~\ref{f:kkfig1}.
In addition we show numerical results obtained for $v_b=0.3$ and $0.5$
without the KK-intermode and self couplings, 
i.e. we have set $M_{ij}\equiv 0$ $\forall \,i,j$ by hand.
One infers that for KK masses, depending slightly on the velocity $v_b$
but at least up to $m_n \simeq 1$, the numerical results for
the spectra do not change when the KK-intermode coupling is switched off.
Consequently, the evolution of {\it light}, i.e. $m_n \lsim 1$, KK gravitons
is virtually not affected by the KK-intermode coupling. 
\\
In addition we find that also the time-dependence of the KK masses is
not important for the production of light KK gravitons which
is explicitly demonstrated below.
Thus, production of light KK gravitons is driven by 
the zero-mode evolution only.
This allows us to find an analytical expression, Eq.~(\ref{Njapprox}),
for the number of produced light KK gravitons in terms of exponential 
integrals.
The calculations which are based on several approximations 
are performed in Section \ref{sec.anaKK}.
\\
In Figs.~\ref{f:kkfig1} and \ref{f:kkfig2} the analytical prediction
(\ref{Njapprox}) for the spectrum of final state gravitons 
has already been included (dashed lines).
Within its range of validity it is in excellent agreement with the 
numerical results obtained by including the full KK-intermode
coupling. 
It perfectly describes the dependence of ${\cal N}_{n,k,\bullet}^{\rm out}$
on the three-momentum $k$ and the maximal velocity $v_b$.
For small velocities $v_b \lsim 0.1$ it is also
able to reproduce the position of the maximum.
This reveals that the KK-intermode coupling is negligible for 
light KK gravitons and that their production is entirely driven 
by their coupling to the four-dimensional graviton.
\\
The analytical prediction is very precious for testing the
goodness of the parameters used in the simulations, in particular
the initial time $t_{\rm in}$ (respectively $N_{\rm in}$). 
Since it has been derived for real asymptotic
initial conditions, $t_{\rm in} \rightarrow -\infty$, its  
perfect agreement with the numerical results demonstrates that 
the values for $N_{\rm in}$ used in the numerical simulations 
are large enough.
No spurious initial effects contaminate the numerical results.
\\
Note, that the numerical values for 
${\cal N}_{n,k,\bullet}^{\rm out}$ in the
examples shown are all smaller than one.
However, for smaller values of $k$ than the ones which we consider here
for purely numerical reasons, the number of generated KK-mode
particles is enhanced since ${\cal N}_{n,k,\bullet}^{\rm out}\propto 1/k$
as can be inferred from Eq.~(\ref{Njapprox}) in the limit $k\ll m_n$.
\\
\\
If we go to smaller values of $y_s$, fewer KK modes belong to
a particular mass range. Hence, with the same or similar number of
KK modes as taken into account in the simulations so far, we can
study the behavior of the final particle spectrum for larger masses.
These simulations shall reveal the asymptotical
behavior of ${\cal N}_{n,k,\bullet}^{\rm out}$ for $m_n\rightarrow \infty$
and therefore the behavior of the total graviton number and energy
density. 
Due to the kink in the brane motion we cannot expect that the
energy density of produced KK-mode gravitons is finite when summing
over arbitrarily high frequency modes.
Eventually, we will have to introduce a cutoff setting the scale at
which the kink-approximation 
[cf.~Eqs.~(\ref{e:solexp}) - (\ref{e:vb})]
is no longer valid. This is the scale
where the effects of the underlying unspecified high-energy physics
which drive the transition from contraction to expansion
become important.
The dependence of the final particle spectrum on the kink will be
studied later on in this section in detail.
\\
\\
In Figures \ref{f:kkfig3} and \ref{f:kkfig4} we show final
KK-graviton spectra for $y_s = 10$ and three-momentum
$k=0.01$ and $k=0.1$. 
The analytical expression Eq. (\ref{Njapprox}) is depicted as well
and the spectra are always shown for at least two values of
$n_{\rm max}$ to indicate up to which KK mass stability of the 
the numerical results is guaranteed. 
Now, only two KK modes are lighter than $m=1$.
For these modes the analytical expression Eq.~(\ref{Njapprox}) is valid
and in excellent agreement with the numerical results,
in particular for small brane velocities $v_b\sim 0.1$.
As before, the larger the velocity $v_b$ the more visible is the effect of the
truncation of the system of differential equations at $n_{\rm max}$.
\\
For $k=0.01$ the spectrum seems to follow a power law decrease
right after the maximum in the spectra. In case of
$v_b=0.1$ the spectrum is numerically stable
up to masses $m_n\simeq 20$. In the region $5 \lsim m_n \lsim 20$ the
spectrum is very well fitted by a power law
${\cal N}_{n,k,\bullet}^{\rm out} \propto m_n^{-2.7}$.
Also for larger velocities the decline of the spectrum
is given by the same power within the mass ranges where the
spectrum is numerically stable.
For $k=0.1$, however, the decreasing spectrum bends over
at a mass around $m_n \simeq 10$ towards a less steep decline.
This is in particular visible in the two cases with
$v_b = 0.1$ and $0.3$ where the first $100$
KK modes have been taken into account in the simulation.
The behavior of the KK-mode particle spectrum can therefore
not be described by a single power law decline for masses
$m_n > 1$.
It shows more complicated features instead, which depend on the
parameters.
We shall demonstrate that this bending over of the decline is
related to the coupling properties of the KK modes and to the
kink in the brane motion.
But before we come to a detailed discussion of these issues, 
let us briefly confront numerical results of different $y_s$
to demonstrate a scaling behavior.
\\
\\
In the upper panel of Figures \ref{f:kkfig5} and \ref{f:kkfig6}
we compare the final KK-spectra for several positions of
the second brane $y_s=3,10,30$ and $100$ obtained for a maximal
brane velocity $v_b=0.1$ for $k=0.01$ and $0.1$, respectively.
One observes that the shapes of the spectra are identical.
The bending over in the decline of the spectrum at masses $m_n\sim 1$
is very well visible for $k=0.1$ and $y_s=3,10$.
For a given KK mode $n$ the number of particles
produced in this mode is the larger the smaller $y_s$.
But the smaller $y_s$, the less KK modes belong to
a given mass interval $\Delta m$.
The energy transferred into the system by the moving brane,
which is determined by the maximum brane velocity $v_b$,
is the same in all cases. Therefore, the total energy of 
the produced final state KK gravitons of a given mass interval 
$\Delta m$ should also be the same, 
independent of how many KK modes are contributing to
it.
This is demonstrated in the lower panels of Figs. \ref{f:kkfig5}
and \ref{f:kkfig6} where the energy 
$\omega_{n,k}^{\rm out} {\cal N}_{n,k,\bullet}^{\rm out}$(in units of $L$)
of the generated KK gravitons binned in mass intervals $\Delta m = 1$ is shown
\footnote{The energy for the case $y_s=3$ is not shown because no KK mode
belongs to the first mass interval.}.
One observes that, as expected, the energy transferred into the
production of KK gravitons of a particular mass
range is the same (within the region where the numerical
results are stable), independent of the number of KK modes
lying in the interval. This is in particular evident for
$y_s=30,100$. The discrepancy for $y_s=10$ is due to the
binning.
As we shall discuss below in detail, the particle spectrum
can be split into two different parts. The first part is
dominated by the coupling of the zero mode to the KK modes
(as shown above), whereas the second part is dominated by the 
KK-intermode couplings and is
virtually independent of the wave number $k$.
As long as the coupling of the zero mode to the KK modes is the
dominant contribution to KK-particle production 
it is ${\cal N}^{\rm out}_{n,k,\bullet} \propto 1/k$ 
[cf. Eq.~(\ref{Njapprox})]. Hence,
${\cal E}^{\rm out}_{n,k,\bullet} = \omega_{n,k}^{\rm out}
{\cal N}^{\rm out}_{n,k,\bullet}\propto 1/k$ if $m_n \gg k$.
This explains why the energy per mass interval
$\Delta m$ is one order
larger for $k=0.01$ (cf Fig.~\ref{f:kkfig5})
than for $k=0.1$ (cf Fig.~\ref{f:kkfig6}) .
\\
\\
Let us now discuss the KK-spectrum for large masses.
The qualitative behavior of the
spectrum ${\cal N}^{\rm out}_{n,k,\bullet}$ and 
the mass at which the decline of the spectrum changes 
are independent of $y_s$.
This is demonstrated in Figure~\ref{f:kkfig7} where 
KK-spectra for $v_b=0.1$, $k=0.1$, $y_s=10$ 
[cf Fig.~\ref{f:kkfig4}] and $y_s=3$ [cf Fig.~\ref{f:kkfig6}]
are shown.
The results obtained by taking the full intermode coupling into account 
are compared to results of simulations
where we have switched off the coupling of the KK modes to 
each other as well as their self-coupling
($M_{ij} \equiv 0\;\forall\,i,j$).
Furthermore we display the results for the KK-spectrum
obtained by taking only the KK-intermode couplings
into account, i.e. $M_{i0}=M_{ii}=0 \;\forall\,i$.
One infers that for the lowest masses the spectra obtained
with all couplings are identical to the ones obtained
without the KK-intermode $(M_{ij}=0,i\neq j)$ and self-couplings
$(M_{ii}=0)$. 
Hence, as already seen before, the primary source for the production
of light KK gravitons is their coupling to the evolution of the
four-dimensional graviton.
In this mass range, the contribution to the particle creation
coming from the KK-intermode couplings is very much suppressed
and negligibly small.
\\
For masses $m_n \simeq 4$ a change in the decline of the spectrum sets
in and the spectrum obtained without the coupling of the KK modes to the
zero mode starts to diverge from the spectrum computed by taking all
the couplings into account.
While the spectrum without the KK-intermode couplings
decreases roughly like a power law ${\cal N}^{\rm out}_{n,k,\bullet}\propto m_n^{-3}$ 
the spectrum corresponding to the full coupling case
changes its slope towards a power law decline with less power.
At this point the KK-intermode couplings gain importance and the
coupling of the KK modes to the zero mode looses influence.
For a particular mass $m_c\simeq 9$ the spectrum obtained
including the KK-intermode couplings only, crosses the spectrum
calculated by taking into account exclusively the 
coupling of the KK modes to the zero mode.
After the crossing, the spectrum obtained by using only the
KK-intermode couplings approaches the
spectrum of the full coupling case. 
Both agree for large masses.
Thus for large masses $m_n > m_c$ the production of KK gravitons
is dominated by the couplings of the KK modes to each
other and is not influenced anymore by the evolution of the 
four-dimensional graviton. 
This crossing defines the transition between the two regimes
mentioned before: for masses $m_n < m_c$ the production of KK gravitons 
takes place due to their coupling to the zero mode $M_{i0}$,
while it is entirely caused by the intermode couplings $M_{ij}$
for masses $m_n > m_c$. 
\begin{figure}
\begin{center}
\includegraphics[height=6cm]{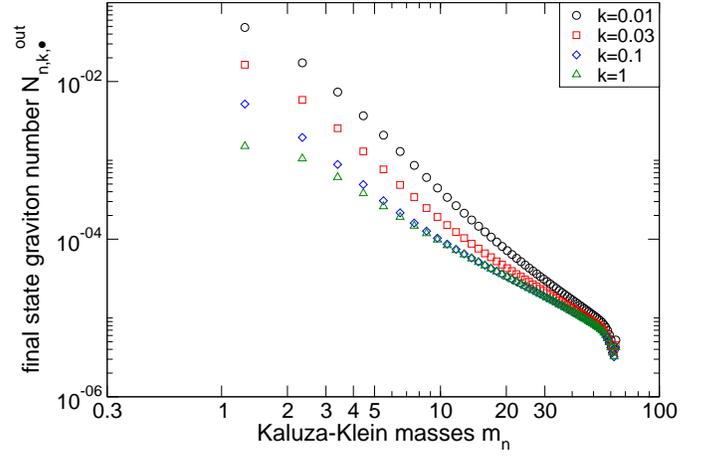}
\caption{Comparison of KK-particle spectra for
$y_s=3$, $v_b=0.1$ and three-momentum
$k=0.01$, $0.03$, $0.1$ and $1$ demonstrating the independence
of the spectrum on $k$ for large masses. $n_{\rm max} = 60$
KK modes have been taken into account in the simulations.
\label{f:kkfig8}}
\end{center}
\end{figure}
\\
Decoupling of the evolution of the KK modes from the dynamics of
the four-dimensional graviton for large masses implies that 
KK-spectra obtained for the same maximal velocity
are independent of the three-momentum $k$. 
This is demonstrated
in Fig.~\ref{f:kkfig8} where we compare spectra obtained
for $v_b=0.1$ and $y_s=3$ but different $k$.
As expected, all spectra converge towards the same behavior for masses
$m_n > m_c$.
\\
\begin{figure}
\begin{center}
\includegraphics[height=6cm]{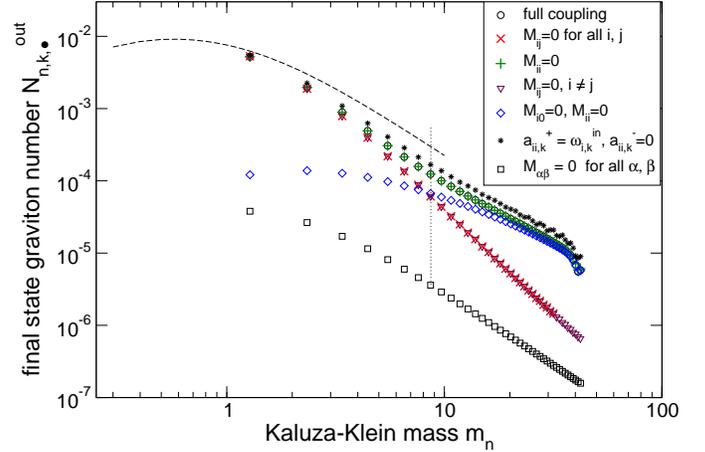}
\caption{KK-particle spectra for three-momentum $k=0.1$,
         maximum brane velocities $v_b=0.1$ and $y_s=3$
	 for $n_{\rm max}=40$ obtained for different 
	 coupling combinations. 
\label{f:kkfig9}}
\end{center}
\end{figure}
\\
Figure.~\ref{f:kkfig9} shows KK-particle spectra for
$k=0.1$,$v_b=0.1$ and $y_s=3$ obtained for different 
couplings.
This plot visualizes how each particular coupling combination 
contributes to the production of KK gravitons.
It shows, as already mentioned before but not shown explicitly,
that the $M_{ii}$ coupling which is the rate of change of the
corresponding KK mass [cf. Eqs.~(\ref{e:m hat}) and
  (\ref{e:T:Mij-exact ii})] 
is not important for the production of KK gravitons.
Switching it off does not affect the final graviton spectrum.
We also show the result obtained with all couplings but 
with $\alpha_{ii}^+(t)=\omega_{i,k}^{\rm in}$ and 
$\alpha_{ii}^-(t)=0$, i.e. the time-dependence of the 
frequency [cf.~Eq.~\eqref{def a matrix}] has been neglected.
One observes that in this case the spectrum for larger masses 
is quantitatively slightly different but has a 
identical qualitative behavior. 
If, on the other hand, all the couplings are switched off 
$M_{\alpha\beta} \equiv 0\;\forall\,\alpha,\beta$ 
and only the time-dependence of the frequency 
$\omega_{i,k}$ is taken into account, the spectrum  
changes drastically. 
Not only the number of produced gravitons is now orders of magnitude
smaller but also the spectral tilt changes. For large masses 
it behaves as ${\cal N}_{n,k,\bullet} \propto m_n^{-2}$.  
Consequently, the time-dependence of the graviton frequency 
itself plays only an inferior role for production of 
KK gravitons.
\\
The bottom line is that the main sources of the production
of KK gravitons is their coupling to the evolution of the
four-dimensional graviton $(M_{i0})$ and their couplings to each other
$(M_{ij},\;i\neq j)$ for small and large masses, respectively.
Both are caused by the time-dependent boundary condition. 
The time-dependence of the oscillator frequency 
$\omega_{j,k}= \sqrt{m^2_j(t)+k^2}$ is virtually
irrelevant. Note that this situation is very different from ordinary
inflation where there are no boundaries and particle production 
is due entirely to the time dependence of the frequency 
\footnote{Note, however, that the time-dependent 
KK mass $m_j(t)$ enters the intermode couplings.}.
\\
\\
The behavior of the KK-spectrum, in particular the mass $m_c$
at which the KK-intermode couplings start to dominate over
the coupling of the KK modes to the zero mode depends only 
on the three-momentum $k=|{\bf k}|$ and the maximal
brane velocity $v_b$.
This is now discussed.
In Figure~\ref{f:kkfig10} we show KK-particle spectra
for $y_s=10$, $v_b=0.1$, $n_{\rm max}=100$
and three-momenta $k=0.01$ and $0.1$.
Again, the spectra obtained by taking all the couplings into account
are compared to the case where only the coupling to the zero mode
is switched on.
One observes that for $k=0.01$ the spectrum is dominated by the
coupling of the KK modes to the zero mode up to larger masses
than it is the case for $k=0.1$. For $k=0.01$ the spectrum
obtained taking into account $M_{i0}$ only is identical to the
spectrum obtained with the full coupling up to $m_n\simeq 10$.
In case of $k=0.1$ instead, the spectrum is purely
zero mode dominated only up to $m_n\simeq 5$.
Hence, the smaller the three-momentum $k$ the larger is the mass
range for which the KK-intermode coupling is suppressed, and the
coupling of the zero mode to the KK modes is the dominant source
for the production of KK gravitons.
As long as the coupling to the zero mode is 
the primary source of particle production, the spectrum 
declines with a power law $\propto m_n^{-3}$.
Therefore, in the limiting case $k\rightarrow 0$ when the
coupling of the zero mode to the KK modes dominates particle
production also for very large masses it is
${\cal N}_{n\gg 1, k\rightarrow 0,\bullet}^{\rm out}
\propto 1/m_n^3.$
\\
\\
Figure~\ref{f:kkfig11} shows KK-graviton spectra 
obtained for the same parameters as in Fig.~\ref{f:kkfig10} 
but for fixed $k=0.1$ and different maximal brane 
velocities $v_b$.
Again, the spectra obtained by taking all the couplings into account are
compared with the spectra to which only the coupling of the KK modes to
the zero mode contributes.
The mass up to which the spectra obtained with
different couplings are identical changes only slightly with
the maximal brane velocity $v_b$.
Therefore, the dependence of $m_c$ on the velocity is rather weak even if
$v_b$ is changed by an order of magnitude, but nevertheless
evident.
\\
This behavior of the spectrum can indeed be understood
qualitatively.
In Section \ref{sec.anaKK} we demonstrate that the coupling 
strength of the KK modes to the zero mode at the bounce $t=0$, where
production of KK gravitons takes place, is proportional to
\begin{equation}
\frac{\sqrt{v_b}}{k}.
\end{equation}
The larger this term the stronger is the coupling of the KK modes to the
zero mode, and thus the larger is the mass up to which this coupling
dominates over the KK-intermode couplings.
Consequently, the mass at which the tilt of the KK-particle
spectrum changes depends strongly on the three-momentum $k$
but only weakly on the maximal brane velocity due to the square
root behavior of the coupling strength.
This explains qualitatively the behavior obtained from the
numerical simulations.
\\
\\
An approximate expression for 
$m_c(k,v_b)$ can be obtained from the numerical simulations.
In Figure~\ref{f:kkfig12} we depict the
KK-particle spectra for three-momentum $k=0.01$, $0.03$,
$0.1$ and $1$ for $y_s=3$ and maximum brane velocity
$v_b=0.1$ with different couplings taken into
account.
The legend is as in Fig.~\ref{f:kkfig11}.
From the crossings of the $M_{ij} = 0,\;i \neq j$ and
$M_{ii}=M_{i0}=0$ results one can determine the $k$-dependence
of $m_c$.
Note that the spectra are not numerically stable for
large masses, but they are stable in the range where
$m_c$ lies [cf., e.g., Fig.~\ref{f:kkfig14}, for $k=0.1$].
Using the data for $k=0.01,0.03$ and $0.1$ one finds $m_c(k,v_b) \propto
1/\sqrt{k}$
.
\begin{figure}
\begin{center}
\includegraphics[height=5.6cm]{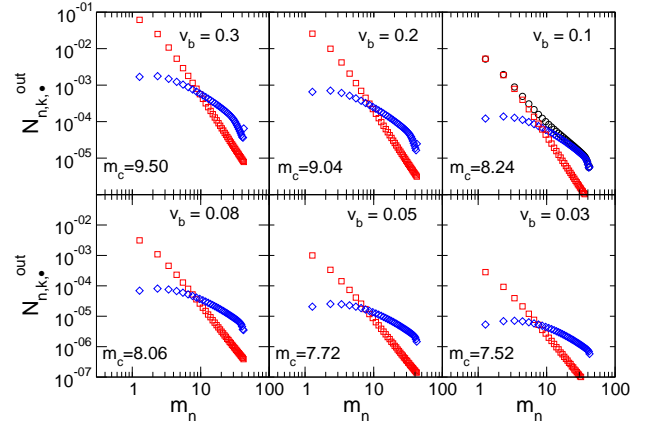}
\caption{KK-graviton spectra for three-momentum $k=0.1$,
         $y_s=3$ and maximum brane velocities
         $v_b=0.3,0.2,0.1,0.08,0.05$ and $0.03$
         with different couplings taken into
         account where the notation is like in Fig.~\ref{f:kkfig11}.
	 From the crossing of the $M_{ii}=M_{ij}=0$- and
         $M_{ii}=M_{i0}=0$ results we determine the $v_b$-dependence
         of $m_c$. 
\label{f:kkfig13}}
\end{center}
\end{figure}
\\
In Fig.~\ref{f:kkfig13} KK-graviton spectra are displayed for  
$k=0.1$, $y_s=3$ and maximal brane velocities
$v_b=0.3,0.2,0.1,0.08,0.05$ and $0.03$
with different couplings taken into account.
It is in principle possible to determine the 
$v_b$-dependence of $m_c$ from the crossings 
of the $M_{ij}=0,\;i\neq j$- and
$M_{ii}=M_{i0}=0$ results as done for the $k$-dependence. 
However, the values for $m_c$ displayed in the Figures indicate 
that the dependence of $m_c$ on $v_b$ is very weak. 
From the given data it is not possible to obtain a good fitting 
formula (as a simple power law) for the $v_b$-dependence of $m_c$. 
(In the range $0.1 \le v_b \le 0.3$ a very good fit is 
$m_c=1.12\pi v_b^{0.13}/\sqrt{k}$.)
The reason is twofold. First of all, given the complicated coupling 
structure, it is {\it a priori} not clear that a simple power law
dependence exists. Recall that also the analytical expression 
for the particle number Eq.~\eqref{Njapprox} has not a 
simple power law velocity dependence. 
Moreover, for the number of modes taken into 
account ($n_{\rm max}=40$) the numerical results are 
not stable enough to resolve the weak dependence 
of $m_c$ on $v_b$ with a high enough accuracy.
(But it is good enough to perfectly resolve the 
$k$-dependence.)  
The reason for the slow convergence of the numerics will become
clear below. 
As we shall see, the corresponding energy density is dominated 
by masses much larger than $m_c$. 
Consequently the weak dependence of $m_c$ on $v_b$ is not very 
important in that respect and therefore does not need to be 
determined more precisely.
However, combining all the data we can give as a fair approximation
\begin{equation}
m_c(k,v_b) \simeq \frac{\pi\,v_b^{\alpha}}{L\,\sqrt{k\,L}}, \quad {\rm
  with} \quad \alpha \simeq 0.1.
\label{e:mc}
\end{equation}
Taking $\alpha = 0.13$ for $0.1 \le v_b \le 0.3$ and 
$\alpha = 0.08$ for $0.03 \le v_b \le 0.1$ fits the given data
reasonably well. 
\\
\\
As we have seen, as long as the zero mode
is the dominant source of KK-particle production, the
final KK-graviton spectrum can be approximated by a power
law decrease $m_n^{-3}$.
We can combine the presented numerical results to obtain a fitting
formula valid in this regime:
\begin{equation}
{\cal N}^{\rm out}_{n \gg 1, k \ll 1,\bullet} =
\frac{\pi}{k\,y_s}\frac{\left(v_b\right)^{2.37}}{(L\,m_n)^3},
\;\;{\rm for}\;\;\frac{1}{L} < m_n < m_c.
\label{e:Nipow}
\end{equation}
This fitting formula is shown in Figs.~\ref{f:kkfig10}
~ \ref{f:kkfig11} and \ref{f:kkfig12} and is in
reasonable good agreement with the numerical results.
Since Eq.~(\ref{e:Nipow}) together with (\ref{e:mc}) is an important
result, we have reintroduced dimensions, 
i.e. the AdS scale $L$ which is set to 
one in the simulations, in both expressions.
\\
\\
Let us now investigate the slope of the KK-graviton spectrum
for masses $m_n\rightarrow \infty$ since it determines the
contribution of the heavy KK modes to the energy density.
In Figure \ref{f:kkfig14} we show KK-graviton spectra obtained for
three-momentum $k=0.1$, second brane position $y_s=3$ and maximal
brane velocities $v_b = 0.01, 0.03$ and $0.1$. 
Up to $n_{\rm max} = 100$ KK modes have been
taken into account in the simulations.
One immediately is confronted with the observation that the 
convergence of the KK-graviton spectra for large
$m_n$ is very slow.
This is since those modes, which are decoupled from the
evolution of the four-dimensional graviton, are 
strongly affected by the kink in the brane motion. 
Recall that the production of light KK gravitons with masses 
$m_n \ll m_c$ is virtually driven entirely by the evolution 
of the massless mode. Those light modes are not so sensitive 
to the discontinuity in the velocity of the brane motion.
To be more precise, their primary source of excitation 
is the evolution of the four-dimensional graviton but not
the kink which, as we shall discuss now, is responsible
for the production of heavy KK gravitons $m_n \gg m_c$.
\\
A discontinuity in the velocity will 
always lead to a divergent total particle number. 
Arbitrary high frequency modes are excited by
the kink since the acceleration diverges there.
Due to the excitation of KK gravitons of arbitrarily high masses,
one cannot expect that the numerical simulations show a
satisfactory convergence behavior which allows to 
determine the slope by fitting the data.
However, it is nevertheless possible to give a quantitative expression
for the behavior of the KK-graviton spectrum for large masses.
The studies of the usual dynamical Casimir effect on a time-dependent 
interval are very useful for this purpose. 
\\
\\
For the usual dynamical Casimir effect it has been shown 
analytically that a discontinuity in the velocity will lead 
to a divergent particle
number \cite{Moore:1970,Castagnino:1984}.
In Appendix \ref{a:dyncas} we discuss in detail the model 
of a massless real scalar field on a time-dependent interval 
$[0,y(t)]$ for the boundary motion $y(t) = y_0 + v\,t$
with $v={\rm const}$, and present numerical results 
for final particle spectra (Fig.~\ref{f:linmo}).
For this motion it was shown in \cite{Castagnino:1984} that 
the particle spectrum behaves as $\propto 
v^2/\omega_n$ where $\omega_n = n\pi/y_0$ is the frequency
of a massless scalar particle.
This divergent behavior is due to the discontinuities 
in the velocity when the motion is switched on and off, and are  
responsible for the slow convergence of the numerical 
results shown in Fig.~\ref{f:linmo} for this scenario.
\\
At the kink in the brane-motion the total change 
of the velocity is $2v_b$, similar to the case for
the linear motion where the discontinuous change 
of the velocity is $2v$. 
Consequently we may conclude that for large
KK masses $m_n \gg m_c$ for which the evolution of the KK modes 
is no longer affected by their coupling to the 
four-dimensional graviton the KK-graviton spectrum behaves as 
\footnote{Note that the discussion in Appendix \ref{a:dyncas}
refers to Dirichlet boundary 
conditions. For Neumann boundary conditions considered 
here, the zero mode and its asymmetric coupling play certainly a
particular role. However, as we have shown, for large masses only
the KK-intermode couplings are important. Consequently, 
there is no reason to expect that the qualitative behavior 
of the spectrum for large masses depends on the particular kind
of boundary condition.} 
\begin{equation}
{\cal N}_{n,k,\bullet}^{\rm out}
\propto  \frac{(v_b)^{2}}{m_n}
\;\;{\rm for}\;\;m_n \gg m_c~.
\label{e:large mass spectrum}
\end{equation}
If we assume that the spectrum declines like $1/m_n$ and
use that the numerical results for masses $m_{n}\simeq 20$
are virtually stable one finds
${\cal N}_{n,k,\bullet}^{\rm out} \propto v_b^{2.08} / m_{n}$
which describes the asymptotics of the numerical results well. 
\\
As for the dynamical Casimir effect for a uniform motion 
discussed in Appendix \ref{a:dyncas} [cf. Fig.~\ref{f:linmo}], 
the slow convergence of the numerical
results towards the $1/m_n$ behavior is well visible
for large masses $m_n \gg m_c$ which do no longer couple 
to the four-dimensional graviton.  
This is a strong indication for the statement that the final
graviton spectrum for large masses behaves indeed
like (\ref{e:large mass spectrum}).
It is therefore possible to give a single simple expression 
for the final KK-particle spectrum for large masses
which comprises all the features of the spectrum even 
quantitatively reasonably well 
[cf.~dashed lines in Fig.~\ref{f:kkfig14}]
\begin{equation}
{\cal N}_{n,k,\bullet}^{\rm out} \simeq 0.2 \frac{v_b^{2}}{
  \omega^{\rm out}_{n,k} \, y_s}
\;\;{\rm for}\;\;m_n \gg m_c~.
\label{e:large mass spectrum fit}
\end{equation}
The $1/y_s$-dependence is compelling. It follows immediately from the
considerations on the energy and the scaling behavior 
discussed above [cf.~Figs.~\ref{f:kkfig5}~and~\ref{f:kkfig6}].
For completeness we now write $1/\omega^{\rm out}_{n,k}$ instead of the KK mass 
$m_n$ only, since what matters is the total energy of a
mode. Throughout this section this has not been important 
since we considered only $k\ll 1$ such that $\omega^{\rm out}_{n,k}$ 
becomes independent of $k$ for large masses
$m_n \gg k$ [cf.~Fig.~\ref{f:kkfig8}].
%
\subsection{Short wavelengths $k \gg 1$}
\begin{figure}
\begin{center}
\includegraphics[height=6cm]{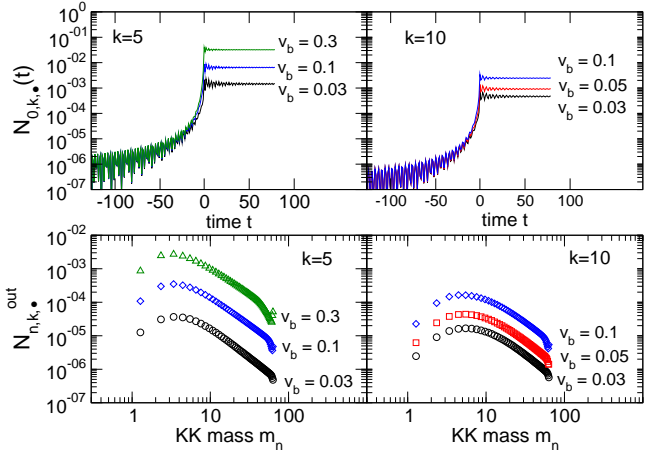}
\caption{Zero-mode particle number ${\cal N}_{0,k,\bullet}(t)$ and 
corresponding final KK-particle spectra 
${\cal N}_{n,k,\bullet}^{\rm out}$ for 
$y_s=3$, $k=5,10$ and different maximal brane velocities $v_b$.
$n_{\rm max}=60$ guarantees numerically stable solutions for the
zero mode. 
\label{f:bkfig3}}
\end{center}
\end{figure}

\begin{figure}
\begin{center}
\includegraphics[height=6cm]{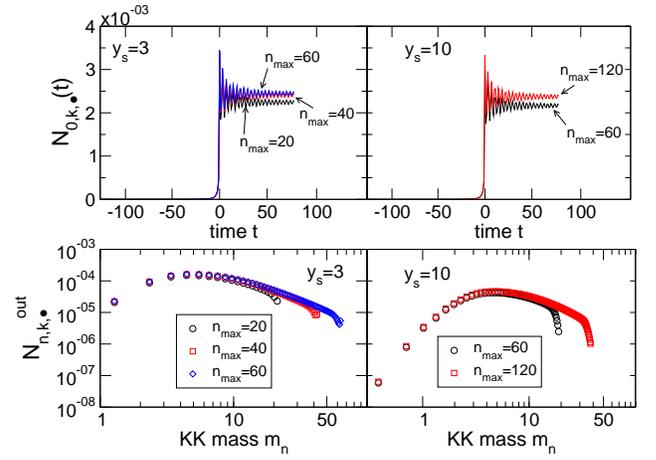}
\caption{Zero-mode particle number ${\cal N}_{0,k\bullet}(t)$ and 
corresponding KK-graviton spectra for $k=10$, $v_b=0.1$ and 2nd brane
positions $y_s=3$ and $10$.
\label{f:bkfig4}}
\end{center}
\end{figure}

\begin{figure}
\begin{center}
\includegraphics[height=6cm]{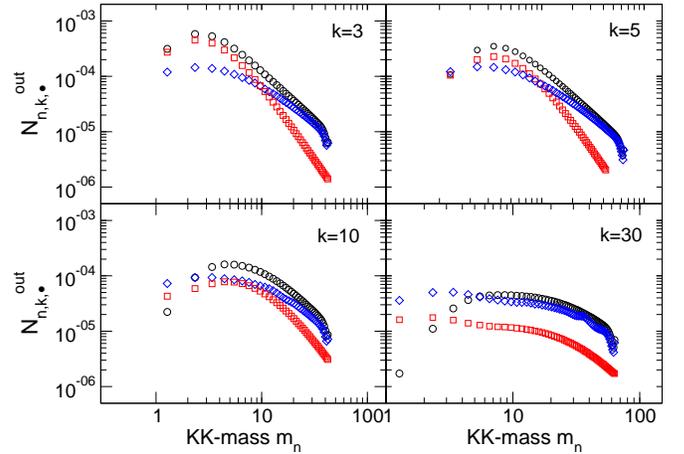}
\caption{Final KK-particle spectra ${\cal N}_{n,k,\bullet}^{\rm out}$ 
for $v_b=0.1$, $y_s=3$ and $k=3,5,10$ and $30$ and different
couplings. Circles correspond to the full coupling case,  
squares indicate the results if $M_{ij}=M_{ii}=0$, 
i.e. no KK-intermode couplings and diamonds
correspond to $M_{i0}=0$, i.e. no coupling of 
KK modes to the zero mode.  
\label{f:bkfig5}}
\end{center}
\end{figure}

\begin{figure}
\begin{center}
\includegraphics[height=6cm]{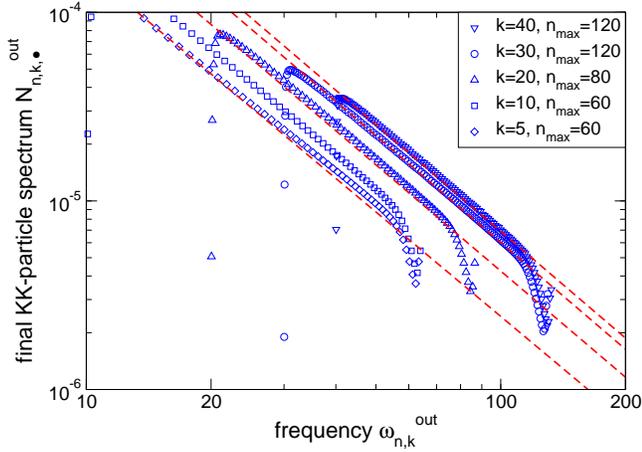}
\caption{Final KK-particle spectra ${\cal N}_{n,k,\bullet}^{\rm out}$ 
for $v_b=0.1$, $y_s=3$ and $k=5,10,20,30$ and $40$. 
The dashed lines indicate Eq.~(\ref{e:asympt pn summary}) for
$k=10,20,30$ and $40$. For $k\ge 20$, the simple analytical 
expression (\ref{e:asympt pn summary}) agrees quite well with the 
numerical results.
\label{f:bkfig6}}
\end{center}
\end{figure}

%
For short wave lengths $k \gg 1$ (short compared 
to the AdS-curvature scale $L$ set to one in the simulations) 
a completely new and very interesting effect appears.  
The behavior of the four-dimensional graviton mode 
changes drastically.
We find that the zero mode now couples to the KK gravitons 
and no longer evolves virtually independently of the 
KK modes, in contrast to the behavior for long wavelengths.
\begin{figure}
\begin{center}
\includegraphics[height=6cm]{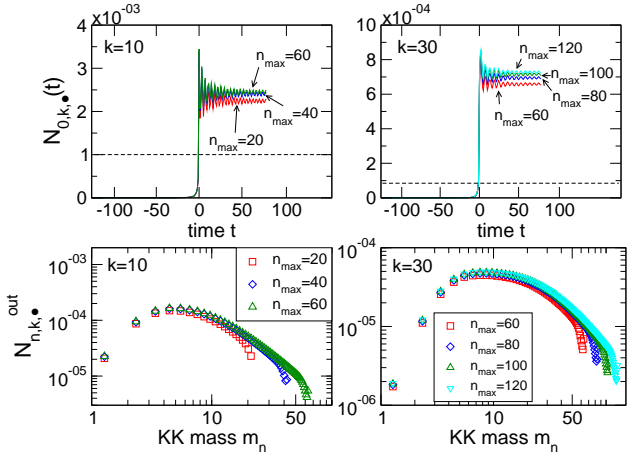}
\caption{Evolution of the zero-mode particle number 
${\cal N}_{0,k,\bullet}(t)$ and final KK-graviton spectra
${\cal N}_{n,k,\bullet}^{\rm out}$ for $y_s=3$, 
maximal brane velocity $v_b=0.1$ and three-momenta 
$k=10$ and $30$.
The dashed line in the upper plots indicate Eq.~(\ref{N0})
(divided by two) demonstrating the value 
of the number of produced zero-mode gravitons 
without coupling to the KK modes. 
\label{f:bkfig1}}
\end{center}
\end{figure}
\\
In Fig.~\ref{f:bkfig1} we show the evolution of the 
zero-mode graviton number ${\cal N}_{0,k,\bullet}(t)$ 
and final KK-graviton spectra 
${\cal N}_{n,k,\bullet}^{\rm out}$ for $y_s=3$, 
maximal brane velocity $v_b=0.1$ and three-momenta 
$k=10$ and $30$.
One observes that the evolution of the four-dimensional 
graviton depends on the number of KK modes $n_{\rm max}$
taken into account, i.e.
the zero mode couples to the KK gravitons. 
For $k=10$ the first $60$ KK modes have to be
included in the simulation in order to obtain a numerically   
stable result for the zero mode.
In the case of $k=30$ one already needs $n_{\rm max}\simeq 100$ 
in order to achieve numerical stability for the zero mode.
\\
Figure \ref{f:bkfig2} displays the time-evolution 
of the number of produced zero-mode gravitons
${\cal N}_{0,k,\bullet}(t)$ for $y_s=3$ and
$v_b=0.1$.
For large $k$ the production of massless gravitons takes place 
 only at the bounce since these short wavelength modes
are sub-horizon right after the bounce. 
Corresponding KK-particle spectra for $k=10,30$ are 
depicted in Figs.~\ref{f:bkfig1} and \ref{f:bkfig3}.
The insert in Fig.~\ref{f:bkfig2} shows the resulting 
final four-dimensional graviton spectrum 
${\cal N}_{0,k,\bullet}^{\rm out}$, which 
is very well fitted by an inverse power law
${\cal N}_{0,k,\bullet}^{\rm out}  =0.02/(k-1.8)$
\footnote{The momenta $k=5,10,20,30$ and
  $40$ have been used  to obtain the fit. Fitting the spectrum for
  $k=20,30$ and $40$ to a power law gives ${\cal
    N}_{0,k,\bullet}^{\rm out} \propto k^{-1.1}$.}.
Consequently, for $k\gg 1$ the zero-mode particle number 
${\cal N}_{0,k,\bullet}^{\rm out}$ declines 
like $1/k$ only, in contrast to the $1/k^2$ behavior 
found for $k \ll 1$.
\\
\\
The dependence of ${\cal N}_{0,k,\bullet}^{\rm out}$ on the 
maximal brane velocity $v_b$ also changes.
In Fig.~\ref{f:bkfig3} we show ${\cal N}_{0,k,\bullet}(t)$ together 
with the corresponding KK-graviton spectra for $y_s=3$, 
$k=5$ and $10$ in each
case for different $v_b$. Using $n_{\rm max}=60$ KK modes
in the simulations guarantees numerical stability for the zero mode.    
\begin{figure}
\begin{center}
\includegraphics[height=6cm]{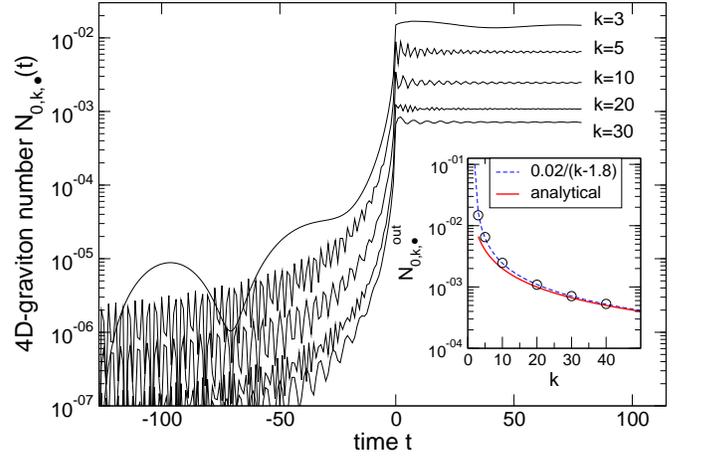}
\caption{4D-graviton number ${\cal N}_{0,k,\bullet}(t)$ for 
$k=3,5,10,20$ and $30$ with $y_s=3$ 
and maximal brane velocity $v_b=0.1$. 
The small plot shows the final graviton spectrum 
${\cal N}_{0,k,\bullet}^{\rm out}$ together with a fit to 
the inverse law $a/(k+b)$ [dashed line] 
and the analytical fitting formula 
Eq.~(\ref{e:N0 small wavelengths})
[solid line].
For $k=10$ and $30$ the corresponding KK-graviton spectra
are shown in Fig.~\ref{f:bkfig1}.
\label{f:bkfig2}}
\end{center}
\end{figure}
The velocity dependence 
of ${\cal N}_{0,k,\bullet}^{\rm out}$ 
is not given by a simple power law as it is the case for $k \ll 1$.
This is not very surprising since now the zero mode couples strongly 
to the KK modes [cf.~Fig.~\ref{f:bkfig1}].
For $k=10$, for example, one finds ${\cal N}_{0,k,\bullet}^{\rm out} \propto
v_b^{1.4}$ if $v_b \lsim 0.1$. 
\\
\\
As in the long wavelengths case, the zero-mode particle
number does not depend on the position of the static brane $y_s$
even though the zero mode now couples to the KK modes.
This is demonstrated in Fig.~\ref{f:bkfig4} where the
evolution of the zero-mode particle number ${\cal N}_{0,k,\bullet}(t)$ and 
the corresponding KK-graviton spectra with $k=10$, $v_b=0.1$ 
for the two values $y_s=3$ and $10$ are shown.
One needs $n_{\rm max}=60$ for $y_s=30$ in order to obtain 
a stable result for the zero mode which is not sufficient
in the case $y_s=10$. Only for $n_{\rm max}\simeq 120$ the zero-mode solution 
approaches the stable result which is identical to the result
obtained for $y_s=3$.  
\\
What is important is not the number of the KK modes the 
four-dimensional graviton couples to, but rather a particular 
mass $m_{\rm zm} \simeq k$.
The zero mode couples to all KK modes of masses below $m_{\rm zm}$
no matter how many KK modes are lighter.  
Recall that the value of $y_s$ just determines how many KK modes
belong to a given mass interval $\Delta m$ since, roughly,
$m_n\simeq n\pi / y_s$.
The KK-spectra for $k \ge 1$ show the same scaling behavior 
as demonstrated for long wavelengths in 
Figs.~\ref{f:kkfig5} and \ref{f:kkfig6}. 
\\
\\
The production of four-dimensional gravitons of short wavelengths 
takes place on the expense of the KK modes. 
In Fig.~\ref{f:bkfig5} we show the numerical results for the final KK-particle
spectra with $v_b=0.1$, $y_s=3$ and $k=3,5,10$ and $30$ obtained 
for different coupling combinations. These spectra should be compared 
with those shown in Fig.~\ref{f:kkfig12} for the long
wavelengths case. 
For $k \gsim 10$ the number of the produced lightest KK gravitons 
is smaller in the full coupling case compared to the situation 
where only the KK-intermode coupling is taken into account. 
In case $k=30$, for instance, the numbers of produced gravitons 
for the first four KK modes are smaller for the full coupling case.
This indicates that the lightest KK modes couple strongly to the
zero mode. Their evolution is damped and graviton 
production in those modes is suppressed.  
The production of zero-mode gravitons on the other hand is enhanced
compared to the long wavelengths case. 
For short wavelengths, the evolution of the KK modes therefore
contributes to the production of zero-mode gravitons. 
This may be interpreted as creation of zero-mode gravitons out 
of KK-mode vacuum fluctuations.
\\
As in the long wavelengths case, the KK-particle spectrum becomes 
independent of $k$ if $m_n \gg k$ and the evolution of the KK modes 
is dominated by the KK-intermode coupling.  
This is visible in Fig.~\ref{f:bkfig5} for $k=3$ and $5$.
Also the bend in the spectrum when the KK-intermode coupling 
starts to dominate is observable.   
For $k=10$ and $30$ this regime with $m_n \gg k$ 
is not reached. 
\\
\\
As we have shown before, in the regime $m_n \gg k$
the KK-particle spectrum behaves as $1/\omega^{\rm out}_{n,k}$ which will 
dominate the energy density of produced KK gravitons.
\\
If $1 \ll m_n \lsim k$, however, the zero mode couples 
to the KK modes and the KK-graviton spectrum does not decay 
like $1/\omega_{n,k}^{\rm out}$. 
This is demonstrated in Fig.~\ref{f:bkfig6} where the 
number of produced final state gravitons 
${\cal N}_{n,k,\bullet}^{\rm out}$ is plotted as function of 
their frequency $\omega_{n,k}^{\rm out}$ for parameters
$v_b=0.1$, $y_s=3$ and $k=5,10,20,30$ and $40$.
\\
While for $k=5$ the KK-intermode coupling dominates for large masses 
[cf.~Fig.~\ref{f:bkfig5}] leading to a bending over in the spectrum
and eventually to an $1/\omega_{n,k}^{\rm out}$-decay, the 
spectra for $k=20,30$ and $40$ show a different behavior. 
All the modes are still coupled to the zero mode leading to 
a power-law decrease $\propto 1/(\omega_{n,k}^{\rm out})^\alpha$ 
with $\alpha\simeq 2$.
The case $k=10$ corresponds to an intermediate regime.
Also shown is the simple analytical expression given in 
Eq.~(\ref{e:asympt pn summary}) which describes the spectra
reasonably well for large $k$ (dashed line).
\\
The KK-particle spectra in the region 
$1 \ll m_n \lsim k$ will also contribute to energy 
density since the cutoff scale is the same
for the integration over $k$ and the 
summation over the KK-tower 
(see Section \ref{s:KK modes energy} below).
%
\subsection{A smooth transition}
\begin{figure}
\begin{center}
\includegraphics[height=6cm]{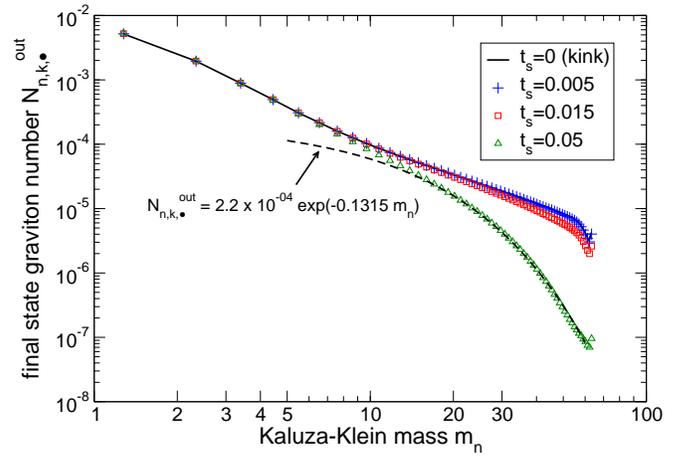}
\caption{KK-particle spectrum for $y_s=3$, $v_b=0.1$ and
  $k=0.1$ for the bouncing as well as smooth motions with
  $t_s=0.005,0.015,$ and $0.05$ to demonstrate the influence of the bounce.
  $n_{\rm max} = 60$ KK modes have been taken into account in the
  simulations and the result for the kink motion is shown as well.
\label{smooth1}}
\end{center}
\end{figure}

%
Let us finally investigate how the KK-graviton spectrum
changes when the kink-motion (\ref{e:yb}) is replaced by
the smooth motion (\ref{e:yb smooth}).
In Fig.~\ref{smooth1} we show the numerical results for 
the final KK-graviton spectrum for $y_s=3$, $v_b=0.1$ and
$k=0.1$ for the smooth motion (\ref{e:yb smooth}) with 
$t_s=0.05, 0.015$ and $0.005$. $n_{\rm max}=60$ modes have been taken
into account in the simulation and the results are compared to the 
spectrum obtained with the kink-motion (\ref{e:yb}).
The parameter $t_s$ defines the scale $L_s\simeq 2t_s$ at which 
the kink is smoothed, i.e. $L_s$ corresponds to the width 
of the transition from contraction to expansion.  
\\
The numerical results reveal that KK gravitons of masses
smaller than $m_s \simeq 1/L_s$ are not affected, but the production of 
KK particles of masses larger than $m_s$ is exponentially 
suppressed.    
This is in particular evident for $t_s=0.05$ where the particle
spectrum for masses $m_n > 10$ has been fitted to a exponential 
decrease.
Going to smaller values of $t_s$, the suppression of 
KK-mode production sets in for larger masses. 
For the example with $t_s=0.005$ the KK-particle
spectrum is identical to the one obtained with the kink-motion
within the depicted mass range.
In this case the exponential suppression of particle production 
sets in only for masses $m_n > 100$. 
\\
Note that the exponential decay of the spectrum for the 
smooth transition from contraction to expansions also 
shows that no additional spurious effects due to the 
discontinuities in the velocity when switching the 
brane dynamics on and off occur. Consequently, 
$t_{\rm in}$ and $t_{\rm out}$ are appropriately chosen.
%
\section{Analytical calculations and estimates}

\subsection{The zero mode: long wavelengths $k \ll 1/L$ }
\label{ss:zm long ana}
%
The numerical simulations show that the evolution of the zero mode
at large wavelengths is not affected by the KK modes.
To find an analytical approximation to the numerical result
for the zero mode, we neglect all the couplings
of the KK modes to the zero mode by setting 
$M_{ij}=0\;\forall \,i,j$ and keeping
$M_{00}$ only.
Then only the evolution equation for $ \epsilon_0^{(\alpha)}
\equiv \delta_0^\alpha\epsilon$ is
important; it decouples and reduces to
\begin{equation}
\ddot\epsilon +[k^2 +{\cal V}(t)]\epsilon =0~,
\label{e:qddot0 with V}
\end{equation}
with ``potential''
\begin{eqnarray}
{\cal V} &=&  \dot{M}_{00} - M_{00}^2 ~.
\end{eqnarray}
The corresponding vacuum initial conditions are
[cf.~Eqs.~(\ref{e:initial conditions for epsilon}),
(\ref{e:initial conditions for epsilon dot});
here we do not consider the unimportant phase]
\begin{eqnarray}
\lim_{t\rightarrow-\infty}\epsilon &=& 1\;,\;\;
\lim_{t\rightarrow-\infty}\dot{\epsilon} = -ik 
\label{in0}.
\end{eqnarray}
\\
A brief calculation using the expression for $M_{00}$ 
(cf.~Appendix \ref{a:MN}) leads to
\begin{eqnarray}
{\cal V} &=& \frac{y_s^2}{y_s^2 - y_b^2}\left[\frac{\ddot{y}_b}{y_b}+
\frac{\dot{y}_b^2}{y_b^2}\frac{3y_b^2 - 2y_s^2}{y_s^2 - y_b^2}\right]\\
&=& -\frac{y_s^2}{y_s^2 - y_b^2}\left[{\cal H}^2
\left(1 - \frac{y_b^2}{y_s^2 - y_b^2}\right) +\dot{{\cal H}}\right].
\end{eqnarray}
If one assumes that the static brane is much further away from the
Cauchy horizon than the physical brane, $y_s \gg y_b$, it is simply
\begin{equation}
{\cal V}=-{\cal H}^2 - \dot{{\cal H}}~,
\end{equation}
and one recovers Eq.~(\ref{e:ddotq0}).
\\
For the particular scale factor (\ref{e:solexp}) one obtains
\begin{eqnarray}
{\cal H} &=& \frac{\dot a}{a} = \frac{\mathrm{sgn}(t)}{|t| + t_b}  \qquad
\mbox{and} \\
\dot{\cal H} &=& \frac{2\delta(t)}{t_b} -\frac{1}{(|t|+t_b)^2} 
\end{eqnarray}
such that
\begin{equation}
\dot{\cal H} + {\cal H}^2 =  \frac{2\delta(t)}{t_b} ~.
\end{equation}
The $\delta$-function in the last equation models the bounce.
Without the bounce, i.e. for an eternally radiation dominated 
dynamics, one has ${\cal V}=0$ and the evolution equation 
for $\epsilon$ would be trivial.  
With the bounce, the potential is just a delta-function 
potential with ``height'' proportional to $-2\sqrt{v_b}/L$
\begin{equation}
{\cal V}=-\frac{2\sqrt{v_b}}{L}\delta(t) ~,
\label{e:delta potential}
\end{equation}
where $v_b$ is given in Eq~\eqref{e:bounce values}.
Equation (\ref{e:qddot0 with V}) with potential 
(\ref{e:delta potential}) can be considered as a Schr{\"o}dinger
equation with $\delta$-function potential.
Its solution is a classical textbook problem. 
\\
Since the approximated potential $\cal V$ vanishes for all $t<0$
one has, with the initial condition \eqref{in0}, 
\begin{equation}
\epsilon(t) =e^ {-ikt}  ~, \quad t<0 ~.
\end{equation}
Assuming continuity of $\epsilon$ through
$t=0$ and integrating the differential equation over  
a small interval $t\in [0^-,0^+]$ around $t=0$ gives 
\begin{eqnarray}
0 &=& \int_{0_-}^{0_+}\left[\ddot{\epsilon} +\left(k^2 -
  \frac{2\sqrt{v_b}}{L}\delta(t)\right)\epsilon\right]\\
&=&
 \dot{\epsilon}(0_+) - \dot{\epsilon}(0_-) - 
\frac{2\sqrt{v_b}}{L}\epsilon(0) ~.
\end{eqnarray}
The jump of the derivative $\dot{\epsilon}$ at
$t=0$ leads to particle creation. Using
$\epsilon(0_+) = \epsilon(0) = \epsilon(0_-)$ and 
$\dot{\epsilon}(0_+) = \dot{\epsilon}(0_-) +
\frac{2\sqrt{v_b}}{L}\epsilon(0)$ as
initial conditions for the solution
for $t>0$, one obtains
\begin{equation}
\epsilon(t) =Ae^ {-ikt} +Be^ {ikt}  ~, \quad t>0
\end{equation}
with
\begin{equation}
A = 1 + i\frac{\sqrt{v_b}}{kL}\;\;,\;\;\;
B =  -i\frac{\sqrt{v_b}}{kL} ~.
\end{equation}
The Bogoliubov coefficient ${\cal B}_{00}$ after the bounce is 
then given by
\begin{equation}
{\cal B}_{00}(t\ge 0)= \frac{e^{-ikt}}{2}\left[\left(1+i\frac{{\cal
      H}}{k}\right)\epsilon(t) - \frac{i}{k}\dot{\epsilon}(t)\right]
\label{e:Bogoliubov B00}
\end{equation}
where we have used that $M_{00} = -{\cal H}$ if $y_s\gg y_b$.
At this point the importance of the coupling matrix $M_{00}$
becomes obvious. Even though the solution $\epsilon$ to the
differential equation (\ref{e:qddot0 with V}) is a plane wave right after
the bounce, $|{\cal B}_{00}(t)|^2$ is not a constant due to
the motion of the brane itself.
Only once the mode is inside the horizon, i.e. ${\cal H}/k \ll 1$,
$|{\cal B}_{00}(t)|^2$ is constant and the number of generated final
state gravitons (for both polarizations) is given by
\begin{eqnarray}
{\cal N}_{0,k}^{\rm out} &=& 2|{\cal B}_{00}(kt\gg 1)|^2
=2 \left( \frac{1}{4}\left[|\epsilon|^2 + \frac{|\dot{\epsilon}|^2}{k^2} \right] -
\frac{1}{2}\right) \nonumber \\
&=&\frac{2v_b}{(kL)^2}
\label{N0}
\end{eqnarray}
where we have used that the Wronskian of $\epsilon, \epsilon^*$ 
is $2ik$.
\\
As illustrated in Fig.~{\ref{f:zeromode numbers} the expression
(\ref{N0}) is indeed in excellent agreement
with the (full) numerical results, not only in its $k$-dependence
but also the amplitude agrees without any fudge factor.
The evolution of the four-dimensional graviton mode and the associated
generation of massless gravitons with momentum $k<1/L$ can therefore
be understood analytically.
\\
Note that the approximation employed here is only valid if
$y_s^2-y_b(0)^2\gg y_b(0)^2$. In the opposite limit, if $\Delta y\equiv
y_s-y_b(0)\ll y_b(0)$ one can also derive an analytical approximation
along the same lines. For $k\le 1/\Delta y$ one obtains instead of
Eq.~(\ref{N0}) 
\begin{eqnarray} \label{N0y}
{\cal N}_{0,k}^{\rm out} &=& \frac{v_b^2}{2(k\Delta y)^2} ~, \\
 &\mbox{ if }& \quad \Delta y\equiv y_s-y_b(0)\ll y_b(0)~, \quad 
k\Delta y \lsim 1   \,. \nonumber 
\end{eqnarray}
\\
In order to calculate the energy density, we have to take
into account that the approximation of an exactly radiation dominated
Universe with an instant transition breaks down on small scales. 
We assume this break down to occur at the string scale 
$L_s$, much smaller than $L$ 
[cf.~Eqs.~(\ref{e:string and Pl scale}),(\ref{e:RS fine tuning})].
$L_s$ is the true width of the transition from collapse to
expansion, which we have set to zero in the treatment. 
Modes with mode numbers $k\gg (2\pi)/L_s$ will not
'feel' the potential and are not generated. 
We therefore choose $k_{\max}= (2\pi)/L_s$
as the cutoff scale. 
Then, with Eq~(\ref{4.15}), one obtains for
the energy density
\begin{equation}
\rho_0 = \frac{1}{2\,\pi^2a^4}\int_0^{2\pi/L_s} dk k^3{\cal N}_{0,k}~.
\label{e:rho0}
\end{equation}
For small wave numbers, $k<1/L$, we can use the above analytical
result for the  zero-mode particle number.
However, as the numerical simulations have revealed,  
as soon as $k \gsim 1/L$, the coupling of the four-dimensional 
graviton to the KK modes becomes important and for large wave 
numbers ${\cal N}^{\rm out}_{0,k}$ decays only like $1/k$. 
Hence the integral (\ref{e:rho0})  
is entirely dominated by the upper cutoff.
The contributions from long wavelengths to the energy 
density are negligible. 
\\
\\
For the power spectrum, on the other hand, we are interested 
in cosmologically large scales, $1/k\simeq $ several Mpc or more,
but not in short wavelengths $kL \gg 1$ dominating the energy density.
Inserting the expression for the number of produced long wavelength
gravitons \eqref{N0} into \eqref{e:Ps}, the gravity wave power 
spectrum at late times becomes
\begin{equation}
{\cal P}_0(k) =
\frac{2\,v_b}{(2\pi)^3}\frac{\kappa_4}{\left(aL\right)^2}
\;\;{\rm for}\;\;kt \gg 1.
\label{e:Ps late sub}
\end{equation}
This is the {\em asymptotic} power spectrum, when $\epsilon$ starts
oscillating, hence inside the Hubble horizon, $kt\gg 1$. 
On super Hubble scales, $kt\ll 1$ when the asymptotic out-state of 
the zero mode is not yet reached, one may  
use~Eq.~\eqref{e:R function with epsilon} with
\begin{equation}
{\cal R}_{0,k}(t) =\frac{|\epsilon(t)|^2 -1}{k} \simeq
\frac{4v_ba^2}{k}~.
\end{equation}
For the $\simeq$ sign we assume $t\gg L$ and $t\gg t_b$ so that one may
neglect terms of order $t/L$ in comparison to $\sqrt{v_b}(t/L)^2$.
We have also approximated $a=(t+t_b)/L \simeq t/L$. 
Inserting this in Eq.~\eqref{e:power spectrum with R} yields
\begin{equation}\label{P0super}
{\cal P}_0(k) =\frac{\kappa_4}{2\,\pi^3}v_b\,k^2 ~, \qquad  kt\ll 1~.
\end{equation}
Both expressions \eqref{e:Ps late sub} and \eqref{P0super} are in very
good agreement with the corresponding numerical results, see
Figs.~\ref{f:zeromode numbers},~\ref{f:zeromode powerspectra} and
\ref{f:zeromode dep}. 
%
\subsection{The zero mode: short wavelengths $k \gg 1/L$ }
%
As we have demonstrated with the numerical analysis, 
as soon as $k\gsim 1/L$, the coupling of the zero mode to 
the KK modes becomes important, and for large wave numbers 
${\cal N}^{\rm out}_{0,k,\bullet}\propto 1/k$.
We obtain a good asymptotic behavior for the 
four-dimensional graviton spectrum if we set 
\begin{equation}
{\cal N}^{\rm out}_{0,k,\bullet} \simeq \frac{v_b}{5(kL)}.
\label{e:N0 small wavelengths}
\end{equation}
This function and Eq.~(\ref{N0}) (divided by two for one polarization)
meet at $kL=5$. Even though the approximation
is not good in the intermediate regime it is very reasonable for large
$k$ [cf. Fig.~\ref{f:bkfig2}]. 
\\
Inserting this approximation into
Eq~(\ref{e:rho0}) for the energy density, one finds that
the integral is dominated entirely by the upper cutoff, i.e.
by the blue, high energy modes:
\begin{equation}
\rho_0 \simeq \frac{16}{30}\frac{\pi}{a^4}\frac{v_b}{LL_s^3}
\simeq \frac{1}{2}\frac{\pi}{a^4}\frac{v_b}{LL_s^3}.
\label{zm energy density}
\end{equation}
The power spectrum associated with the short wavelengths 
$k \gg 1/L$ is not of interest since the gravity wave spectrum is
measured on cosmologically large scales only, $k \ll 1/L$.
%
\subsection{Light Kaluza-Klein modes and long wavelengths $k \ll 1/L$}
\label{sec.anaKK}
The numerics indicates that light ($m_n < 1)$ long wavelength KK 
modes become excited mainly due to their coupling to the zero mode. 
Let us take only this coupling into account and neglect also 
the time-dependence of the frequency, setting
$\omega_{n,k}(t) \equiv \omega_{n,k}^{\rm out}=\omega_{n,k}^{\rm in}$
since it plays an inferior role as shown by the numerics.
\\
The Bogoliubov coefficients are then determined by the equations
\begin{eqnarray}
\dot{\xi}_{n,k} + i\omega_{n,k}^{\rm out} \xi_{n,k} &=& 
\frac{k}{2\omega_{n,k}^{\rm out}} S_n(t;k)\\
\dot{\eta}_{n,k} - i\omega_{n,k}^{\rm out} \eta_{n,k} &=& 
-\frac{k}{2\omega_{n,k}^{\rm out}} S_n(t;k)
\end{eqnarray}
with the ``source''
\begin{equation}
S_n(t;k) =\left(\xi_0 - \eta_0\right) M_{n0}~.
\end{equation}
We have defined $\xi_{n,k} \equiv \xi_{n,k}^{(0)}$, 
$\eta_{n,k} \equiv \eta_{n,k}^{(0)}$,
$\xi_0 \equiv \xi_{0,k}^{(0)}$, and
$\eta_0 \equiv \eta_{0,k}^{(0)}$.
This source is known, since the evolution of the four-dimensional
graviton is know. From the result for $\epsilon$ above and the
definition of $\xi_0$ and $\eta_0$ in terms of $\epsilon$ 
and $\dot{\epsilon}$ one obtains
\begin{eqnarray}
\xi_0 -\eta_0 &=& \frac{2i}{k}\left[-ik+\frac{1}{|t|+t_b}\right]e^{-itk}
~,~  t<0 \\
\xi_0 -\eta_0 &=& 2\left[1+\frac{i}{kt_b} +\frac{1-ikt_b}{k^2t_b(t+t_b)}
\right]e^{-itk} \nonumber \\
&+& 2\left[\frac{i}{kt_b} -\frac{1}{k^2t_b(t+t_b)}
\right]e^{itk}  ~,~  t>0\;~.
\end{eqnarray}
Furthermore, if $y_s \gg y_b$, one has 
[cf.~Eq.~(\ref{e:T:Mij-exact i0})]
\begin{equation}
 M_{n0} = 2\frac{\dot
 y_b}{y_b}\sqrt{\frac{Y_1(m_ny_s)^2}{Y_1(m_ny_b)^2 - Y_1(m_ny_s)^2}}~.
\end{equation}
Assuming $y_sm_n\gg 1$ and $y_bm_n\ll1$ one can expand the Bessel
functions and arrives at
\[ M_{n0} \simeq \sqrt{\pi}\sqrt{\frac{m_n}{y_s}}\dot y_b =
 -\sqrt{\frac{\pi m_nL^2}{y_s}}
   \frac{L\,\mathrm{sgn(t)}}{(|t| +t_b)^2} ~.\]
To determine the number of created final state gravitons 
we only need to calculate $\eta_{n,k}$ 
[cf.~Eq.~(\ref{e:Bout}) with $\Delta_{n,k}^+(|t|\rightarrow \infty)
= 1$ and $\Delta_{n,k}^-(|t|\rightarrow \infty)= 0$],
\begin{equation}
{\cal N}_{n,k,\bullet}^{\rm out} = |{\cal B}_{0n,k}(t_{\rm out})|^2 = 
\frac{1}{4}\frac{\omega_{n,k}^{\rm out}}{k}|\eta_{n,k}|^2
\end{equation}
The vacuum initial conditions require $\lim_{t\rightarrow-\infty}\eta_{n,k} = 0$
so that $\eta_{n,k}$ is given  by the particular solution
\begin{equation}
\eta_{n,k}(t) =\frac{k}{\omega_{n,k}^{\rm out}}\int_{-\infty}^t 
S_j(t';k)e^{-it'\omega_{n,k}^{\rm out}}dt'~,
\end{equation}
and therefore
\begin{equation}
{\cal N}^{\rm out}_{n,k,\bullet}
 =\frac{k}{4\omega_{n,k}^{\rm out}}\left| \int_{-\infty}^{\infty}
S_n(t;k)e^{-it\omega_{n,k}^{\rm out}}dt \right|^2 ~
\end{equation}
where the integration range has been extended from $-\infty$ to
$+\infty$ since the source is very localized around the bounce.
This integral can be solved exactly. A somewhat lengthy but straight
forward calculation gives
\begin{eqnarray}
{\cal N}^{\rm out}_{n,k,\bullet} &=& \frac{\pi m_n^5L^4}{2\omega_{n,k}^{\rm out}ky_s}
\left| 2i\mathrm{Re}\left(
e^{i(\omega_{n,k}^{\rm out}+k)t_b}E_1(i(\omega_{n,k}^{\rm out}+k)t_b)\right) 
\right. \nonumber \\ &&
 + (kt_b)^{-1}e^{i(\omega_{n,k}^{\rm out}-k)t_b} E_1(i(\omega_{n,k}^{\rm out}-k)t_b)
 \nonumber \\ && \label{Njapprox}
 \left. - e^{i(\omega_{n,k}^{\rm out}+k)t_b}E_1(i(\omega_{n,k}^{\rm out}+k)t_b) \right|^2~.
\end{eqnarray}
Here $E_1$ is the exponential integral, $E_1(z)\equiv
\int_z^\infty t^{-1}e^{-t}dt$~. This function is holomorphic in the
complex plane with a cut along the negative real  axis, and the above
expression is therefore well defined.
Note that this expression does not give rise to a simple dependence of
${\cal N}^{\rm out}_{n,k}$ on the velocity $v_b =(L/t_b)^2$.
In the preceding section we have seen that, within its range of
validity, Eq.~(\ref{Njapprox}) is in excellent agreement
with the numerical results (cf., for instance,
Figs.~\ref{f:kkfig1} and \ref{f:kkfig2}).
\\
As already mentioned before, this excellent agreement 
between the numerics and the analytical approximation 
demonstrates that the numerical results are not 
contaminated by any spurious effects. 
%
\subsection{Kaluza-Klein modes: asymptotic behavior and energy density}
\label{s:KK modes energy}
The numerical simulations show that the asymptotic
KK-graviton spectra (i.e. for masses $m_n \gg 1$)
decay like $1/\omega_{n,k}^{\rm out}$  if $m_n \gg k$ 
and like $\left(1/\omega_{n,k}^{\rm out}\right)^\alpha$
with $\alpha \simeq 2$ if $m_n \lsim k$.
The corresponding energy density on the brane is  
given by the summation of   
Eq.~(\ref{late time large mass KK energy density}) 
over all KK modes up to the cutoff. 
Since the mass $m_n$ is simply the momentum into the extra dimension,
it is plausible to choose the same cutoff scale for both,
the $k$-integral and the summation over the KK modes,
namely $2\pi/L_s$. 
The main contribution to the four-dimensional particle
density and energy density comes from $m_n \sim 2\pi/L_s$ and $k\sim
2\pi/L_s$, i.e. the blue end of the spectrum.
\\
The large-frequency behavior of the final KK-spectrum 
can be approximated by 
\begin{equation}
{\cal N}_{n,k,\bullet}^{\rm out} \simeq
\frac{0.2 v_b^2}{y_s}
\left \{
\begin{array}{ll}
\displaystyle
\frac{1}{\omega_{n,k}^{\rm out}} 
& 
\; {\rm if} \;\; 1/L \lsim k \lsim m_n \label{e:asympt pn summary} \\
&\\
\displaystyle
2^{(\alpha-1)/2} \frac{k^{\alpha-1}}{(\omega_{n,k}^{\rm out})^\alpha}
&
\; {\rm if} \;\; m_n \lsim k \lsim 2\pi/L_s
\end{array}
\right .
\end{equation}
with $\alpha \simeq 2$ which is particularly good for large $k$. 
Both expression match at $m_n=k$ and are indicated in 
Figures \ref{f:kkfig14} and \ref{f:bkfig6} as dashed lines. 
Given the complicated coupling structure of the problem and the 
multitude of features visible in the particle spectra, 
these compact expressions describe the numerical results 
reasonable well for all parameters. 
The deviation from the numerical results is 
at most a factor of two.    
This accuracy is sufficient in order to obtain a useful 
expression for the energy density from which  
bounds on the involved energy scales can be derived.    
\\
\\
The energy density on the brane associated with the KK gravitons 
is given by [cf.~Eq.~(\ref{late time large mass KK energy density})]
\begin{equation}
\rho_{{\rm KK}} \simeq \frac{L^2}{\pi a^6y_s} \sum_n
\int dk k^2 \;{\cal N}_{n,k,\bullet}^{\rm out}\,\omega_{n,k}^{\rm out}\,m_n ~.
\label{e:final KK energy density}
\end{equation}
Splitting the momentum integration into two integrations 
from $0$ to $m_n$ and $m_n$ to the cutoff $2\pi/L_s$,
and replacing the sum over the KK masses by an 
integral one obtains 
\footnote{Note that even the transition from the summation 
over the KK-tower to an integration according to (\ref{e:dis to cont})
``eats up'' the $1/y_s$ term in (\ref{e:final KK energy density}),
the final energy density (\ref{e:final KK energy density result}) 
depends on $y_s$ since it explicitly enters the particle number.}
\begin{equation}
\rho_{KK} \simeq C(\alpha) \frac{\pi^5v_b^2}{a^6y_s}\frac{L^2}{L_s^5}.
\label{e:final KK energy density result}
\end{equation}
The power $\alpha$ in Eq.~(\ref{e:asympt pn summary}) enters  
the final result for the energy density only through 
the pre-factor $C(\alpha)$ which is of order unity.
%
\section{Discussion}\label{s:dis}
%
The numerical simulations have revealed many interesting
effects related to the interplay between the 
evolution of the four-dimensional graviton and the KK modes.
All features observed in the numerical results 
have been interpreted entirely on physical grounds 
and many of them are supported by analytical 
calculations and arguments.
Having summarized the results for the power spectrum 
and energy densities in the preceding section,  
we are now in the position to discuss the significance
of these findings for brane cosmology.
%
\subsection{The zero mode}
%
For the zero-mode power-spectrum we have found that
\begin{equation}
{\cal P}_0(k) =  \frac{\kappa_4}{2\,\pi^3} v_b\left\{\begin{array}{ll}k^2
& \mbox{ if }~ kt\ll 1 \\
\frac{1}{2}(La)^{-2} & \mbox{ if }~ kt\gg 1 \end{array} \right.~.
\end{equation}
Therefore, the gravity wave spectrum on large, super Hubble scales is blue
with spectral tilt 
\begin{equation}
n_T =2~,
\end{equation}
a common feature of ekpyrotic and pre-big-bang models.
The amplitude of perturbations on scales at which fluctuations of the
Cosmic Microwave Background (CMB) are observed is of the order of
$(H_0/m_{Pl})^2$, i.e. very suppressed on scales relevant for the
anisotropies of the CMB. 
The fluctuations induced by these Casimir gravitons are much too 
small to leave any observable imprint on the CMB.
\\
\\
For the zero-mode energy density at late times, 
$kt\gg 1$, we have obtained [cf~Eq.~(\ref{zm energy density})]
\begin{equation}
\rho_{h0} \simeq \frac{1}{2}\frac{\pi}{a^4}\frac{v_b}{LL_s^3}.
\end{equation}
In this section we denote the energy density of the zero mode 
by $\rho_{h0}$ in order not to confuse it with the 
present density of the Universe.
Recall that $L_s$ is the scale at which our kinky
approximation (\ref{e:solexp}) of the scale factor breaks 
down, i.e. the width of the bounce. 
If this width is taken to zero, the energy density of
gravitons is very blue and diverges. 
This is not so surprising, since the kink in $a(t)$ 
leads to the generation of gravitons of arbitrary high energies.
However, as the numerical simulations have shown, when we  
smooth the kink at some scale $L_s$, the production of modes 
with energies larger than $\simeq 1/L_s$ is
exponentially suppressed [cf.~Fig.~\ref{smooth1}]. 
This justifies the introduction of $L_s$ 
as a cutoff scale.
\\
\\
In the following we shall determine the density parameter of the 
generated gravitons today and compare it to the
Nucleosynthesis bound.
For this we need the quantities $a_b$ given in Eq~(\ref{e:bounce values}) and
\begin{eqnarray*}
 H_{b} &=&
\left| \frac{\dot a}{a^2}\right|_{t=0} \simeq \frac{v_{b}}{L}~.
\end{eqnarray*}
Here $a_{b}$ is the minimal scale factor and $H_{b}$ is the
maximal Hubble parameter, i.e. the Hubble parameter right after the
bounce.
(Recall that  in the low energy approximation $t=\eta$.) 
During the radiation era, curvature and/or a cosmological 
constant can be neglected so that the density is
\begin{equation}
\rho_\mathrm{rad} =
\frac{3}{\kappa_4}\,H^2 = 
 \frac{3}{\kappa_4L^2}a^{-4}=
\frac{3}{\kappa_4}H_{b}^2\left(\frac{a_{b}}{a}\right)^4 ~.
\end{equation}
In order to determine the density parameter of the generated 
gravitons today, i.e., at $t=t_0$, we use 
\begin{equation}
\Omega_{h0} =
\frac{\rho_{h0}(t_0)}{\rho_{\rm crit}(t_0)} =
\frac{\rho_{h0}(t_0)}{\rho_\mathrm{rad}(t_0)} \, 
\frac{\rho_\mathrm{rad}(t_0)}{\rho_{\rm crit}(t_0)}
= \frac{\rho_{h0}(t_0)}{\rho_\mathrm{rad}(t_0)} \,
\Omega_\mathrm{rad}.
\end{equation}
The second factor $\Omega_\mathrm{rad}$ is the present 
radiation density parameter.
For the factor $\rho_{h0}/\rho_\mathrm{rad}$, 
which is time independent since both $\rho_{h0}$
and $\rho_\mathrm{rad}$ scale like $1/a^4$, we insert the above
results and obtain 
\begin{eqnarray}
\Omega_{h0} &=& \frac{\rho_{h0}}{\rho_\mathrm{rad}}\Omega_\mathrm{rad}
= \frac{1}{2}\frac{\pi}{3} v_b\left(\frac{L_{Pl}}{L_s}\right)^2\frac{L}{L_s}
\Omega_\mathrm{rad} \\
&\simeq& \frac{v_b}{2}\left(\frac{L_{Pl}}{L_s}\right)^2\frac{L}{L_s}
\Omega_\mathrm{rad}
~.
\label{Omho}
\end{eqnarray}
The nucleosynthesis bound~\cite{mm} requests that
\begin{equation}
\Omega_{h0} \lsim 0.1 \, \Omega_\mathrm{rad}~,
\end{equation}
which translates into the relation
\begin{equation}
\frac{v_b}{2}\left({L_{Pl}}/{L_s}\right)^2(L/L_s) \lsim 0.1~
\label{e:nucleo beound 1}
\end{equation}
which, at first sight, relates the different scales involved.
But since we have chosen the cutoff scale $L_s$ to be 
the higher-dimensional fundamental scale (string scale),
Equation (\ref{e:nucleo beound 1}) reduces to 
\begin{equation}
v_b \lsim 0.2~
\label{e:nucleo beound 2}
\end{equation}
by virtue of Equation (\ref{e:RS fine tuning}).
All one has to require to be consistent with the nucleosynthesis 
bound is a small brane velocity which 
justifies the low energy approach.  
In all, we conclude that the model is not severely 
constrained by the zero mode. 
This result itself is remarkable. If there would be no coupling 
of the zero mode to the KK modes for small wavelengths 
the number of produced high energy zero-mode gravitons 
would behave as $\propto k^{-2}$ as it is the case for 
long wavelengths.
The production of high energy zero-mode gravitons from KK gravitons
enhances the total energy density by a factor of about
$L/L_s$. Without this enhancement, the nucleosynthesis bound would not
lead to any meaningful constraint and would not even require $v_b<1$.
%
\subsection{The KK modes}
%
As derived above, the energy density of KK gravitons 
on the brane is dominated by the high energy gravitons 
and can be approximated by [cf.~Eq.~(\ref{e:final KK energy density result})]
\begin{equation}
\rho_{KK} \simeq \frac{\pi^5v_b^2}{a^6y_s}\frac{L^2}{L_s^5}~.
\label{e:rhoKK}
\end{equation}
Let us evaluate the constraint induced from the requirement
that the KK-energy density on the brane be smaller than 
the radiation density $\rho_{\rm KK}(t)<\rho_{\rm rad}(t)$ 
at all times. 
If this is not satisfied, back-reaction cannot be neglected 
and our results are no longer valid. 
Clearly, at early times this condition is
more stringent than at late times since  $\rho_{\rm KK}$ decays
faster then $\rho_{\rm rad}$. 
Inserting the value of the scale factor directly after the 
bounce where the production of KK gravitons takes place, 
$a_b^{-2} =v_b$, one finds, using again the 
RS fine tuning condition (\ref{e:RS fine tuning}), 
\begin{equation}\label{e:constraint0}
\left(\frac{\rho_{\rm KK}}{\rho_{\rm rad}}\right)_{\max} 
\simeq 100\,v_b^3 \left(\frac{L}{y_s}\right)\left(\frac{L}{L_s}\right)^2~.
\end{equation}
If we use the largest value for the brane velocity $v_b$ 
admitted by the nucleosynthesis bound $v_b \simeq 0.2$ and require 
that $\rho_{\rm KK}/\rho_{\rm rad}$ be (much) smaller than one 
for back-reaction effects to be negligible, we obtain the 
very stringent condition
\begin{equation}
\frac{L}{y_s} \ll \left(\frac{L_s}{L}\right)^2~.
\label{e:KK contraint}
\end{equation}
Let us first discuss the largest allowed value for 
$L\simeq 0.1$mm. 
The RS-fine tuning condition (\ref{e:RS fine tuning}) 
then determines 
$
L_s=(LL_{Pl}^2)^{1/3} \simeq 10^{-22}\;{\rm mm}
\simeq 1/(10^6\;{\rm TeV}). 
$
In this case the brane tension is 
$
{\cal T} =6\kappa_4/\kappa_5^2 =6L_{Pl}^2/L_s^6 =6/(LL_s^3) \sim
(10\,{\rm TeV})^4. 
$
Furthermore, we have 
$
(L/L_s)^2 \simeq  10^{42} \;{\rm so}\;{\rm that}\;
y_s > L(L/L_s)^2 \simeq  10^{41}\,{\rm mm} \simeq 3\times 10^{15}\,{\rm Mpc},
$
which is about 12 orders of magnitude larger than the present Hubble
scale.
Also, since  $y_b(t)\ll L$ in the low energy regime, and
$y_s\gg L$ according to the inequality (\ref{e:KK contraint}), the
physical brane and the static brane are very far apart at all times. 
Note that the distance between the physical and the static brane is
$$ 
d = \int_{y_b}^{y_s}\frac{L}{y}dy = L\log(y_s/y_b) \gsim L \gg L_s~. 
$$
This situation is probably not very realistic. 
Some high energy, stringy effects are needed to provoke the 
bounce and one expects these to
be relevant only when the branes are sufficiently close, i.e. at a
distance of order $L_s$. 
But in this case the constraint (\ref{e:KK contraint}) will be 
violated which implies that back-reaction will be relevant.
\\
On the other hand, if one wants that $y_s \simeq L$ 
and back-reaction to be unimportant, 
then Eq.~(\ref{e:constraint0}) implies that the  
bounce velocity has to be exceedingly small, 
$v_b \lsim 10^{-15}$.
\\
\\
A way out of this conclusion is to assume that 
the brane distance at the bounce, $\Delta y=y_s-y_b(0)$, becomes of the order 
of the cutoff $L_s$ or smaller. Then the production
of KK gravitons is suppressed. However, then the
approximation~(\ref{N0y}) has to be used to determine the energy
density of zero-mode gravitons which then becomes
$$\rho_{h0} \simeq \frac{v_b^2}{2}(L_s\Delta y)^{-2} \, . $$
Setting $\Delta y\simeq L_s$, the nucleosynthesis bound, $\rho_{h0} \lsim
0.1 \rho_{\mathrm rad}$, then yields the much more stringent limit on
the brane velocity,
\begin{equation}
v_b^2 < \frac{L_s}{L} ~.
\end{equation}
One might hope to find a way out of these
conclusions by allowing the bounce to happen in the high energy regime.   
But then $v_b \simeq 1$ and the nucleosynthesis bound is violated
since too many zero-mode gravitons are produced. 
Even if one disregards this limit for a moment, saying that the 
calculation presented here only applies in the low energy regime, 
$v_b\ll 1$, the modification coming from the
high energy regime are not expected to alleviate the bounds. 
In the high energy regime one may of course have $y_b(t)\gg L$ and
therefore the physical brane can approach the static brane arbitrarily
closely without the latter having to violate (\ref{e:KK contraint}).
Those results suggest that even in the scenario of a bounce
at low energies, the back reaction from KK gravitons has to 
be taken into account.
But this does not need to exclude the model.
%
\section{Conclusions}\label{s:con}
%
We have studied 
the evolution of tensor perturbations in braneworld 
cosmology using the techniques developed 
for the standard dynamical Casimir effect.
A model consisting of a moving and a fixed $3$-brane embedded
in a five-dimensional static AdS bulk has been 
considered.
Applying the dynamical Casimir effect formulation 
to the study of tensor perturbations in braneworld 
cosmology represents an interesting alternative to 
other approaches existing in the literature so far 
and provides a new perspective on the problem.
The explicit use of coupling matrices 
allows us to obtain detailed information about the
effects of the intermode couplings generated by 
the time-dependent boundary conditions, i.e. the brane
motion. 
\\
Based on the expansion of the tensor perturbations in 
instantaneous eigenfunctions, we have introduced a 
consistent quantum mechanical formulation 
of graviton production by a moving brane.  
Observable quantities like the power spectrum and 
energy density can be directly deduced from 
quantum mechanical expectation values, in particular the 
number of gravitons created from vacuum fluctuations.
The most surprising and at the same time most interesting 
fact which this approach has revealed is that 
the energy density of the massive gravitons decays like $1/a^6$
with the expansion of the Universe.
This is a direct consequence of the localization of gravity: 
five-dimensional aspects of it, like the KK gravitons, become less and less
'visible' on the brane with the expansion of the Universe. 
The $1/a^6$-scaling behavior remains valid also when the fixed brane is
sent off to infinity and one ends up with a single braneworld 
in AdS, like in the original RS II scenario.
Consequently, KK gravitons on a brane moving through 
an AdS bulk cannot play the role of dark matter. 
\\
\\
As an explicit example, we have studied graviton production 
in a generic, ekpyrotic-inspired model of two branes bouncing 
at low energies, assuming that 
the energy density on the moving brane is dominated by a 
radiation component.  
The numerical results have revealed a multitude of interesting
effects.
\\
For long wavelengths $kL\ll 1$ the zero mode 
evolves virtually independently of the KK modes. 
zero-mode gravitons are generated by the self coupling of 
the zero mode to the moving brane.
For the number of produced massless gravitons we have found 
the simple analytical expression $2v_b/(kL)$.  
These long wavelength modes are the once which are 
of interest for the gravitational wave power spectrum. 
As one expects for an ekpyrotic scenario, the 
power spectrum is blue on super-horizon scales
with spectral tilt $n_T=2$.
Hence, the spectrum of these Casimir gravitons has much 
too little power on large scales to affect the fluctuations 
of the cosmic microwave background. 
\\
The situation changes completely for short wavelengths
$kL\gg 1$.
In this wavelength range, the evolution of the zero mode 
couples strongly to the KK modes. 
Production of zero-mode gravitons takes place on the expense of 
KK-graviton production.
The numerical simulation have revealed that the number of 
produced short-wavelength massless gravitons is given by
$2v_b/(5kL)$.
It decays only like $1/k$ instead of the $1/k^2$-behavior 
found for long wavelengths.  
These short wavelength gravitons dominate the energy 
density.
Comparing the energy density with the nucleosynthesis
bound and taking the cutoff scale to be the string scale $L_s$, 
we have shown that the model is not constrained 
by the zero mode. 
As long as $v_b \lsim 0.2$, i.e. a low energy bounce, 
the nucleosynthesis bound is not violated.
\\
\\
More stringent bounds on the model come from the 
KK modes.
Their energy density is dominated by the 
high energy modes which are produced due to the 
kink which models the transition from contraction
to expansion.
Imposing the reasonable requirement that the energy density 
of the KK modes on the brane be (much) smaller than the radiation 
density at all times in order for back reaction effects
to be negligible, has led to two cases. 
On the one hand, allowing the largest values for the AdS curvature 
scale $L\simeq 0.1{\rm mm}$ and the bounce velocity 
$v_b\simeq 0.2$, back reaction can only be neglected 
if the fixed brane is very far away from the physical
brane $y_s \sim 10^{41}{\rm mm}$.
As we have argued, this is not very realistic since some high 
energy, stringy effects provoking the bounce 
are expected to be relevant only when the branes are sufficiently 
close, i.e. $y_s \sim L_s$. 
On the other hand, by only requiring that $y_s \simeq L \gg L_s$, 
the bounce velocity has already to be exceedingly small, 
$v_b \lsim 10^{-15}$, for back reaction to be unimportant.
Therefore, one of the main conclusions to take away from 
this work is that back reaction of 
massive gravitons has to be taken into account 
for a realistic bounce.
\\
\\
Many of the results presented here are based on numerical 
calculations. 
However, since the used approach provides the possibility to 
artificially switch on and off the mode couplings, 
we were able identify the primary sources driving the
time evolution of the perturbations in different
wavelength and KK mass ranges. 
This has allowed us to understand many of the features 
observed in the numerical results on analytical 
grounds.
\\
On the other hand, it is fair to say that most of the 
presented results rely on the low energy approach, i.e.
on the approximation of the junction condition 
(generalized Neumann boundary condition) by 
a Neumann boundary condition. 
Even though we have given arguments for the goodness of this 
approximation, it has eventually to be confirmed by 
calculations which take the exact boundary condition 
into account. 
This is the subject of future work.
%
\section*{Acknowledgment}
%
We thank Cyril Cartier who participated
in the early stages of this work and Kazuya Koyama and David Langlois 
for discussions. We
are grateful for the use of the 
'Myrinet'-cluster of Geneva University on which most of the quite
intensive numerical computations have been performed. This work is supported by the
Swiss National Science Foundation.

\begin{appendix}

\section{Variation of the action}
\label{a:variation}
%
Let us consider the variation of the action 
(\ref{e:action h}) with respect to $h_\bullet$.
It is sufficient to study the
action for a fixed wave number $k$ and polarization 
$\bullet$
\begin{equation}
{\cal S}_{h_\bullet} (k) = \frac{1}{2} \int dt
\int_{y_b(t)}^{y_s} \frac{dy}{y^3} 
\left[|\partial_t h_\bullet|^2 - |\partial_y h_\bullet|^2 
-k^2 |h_\bullet|^2\right] 
\label{e:app action 1}
\end{equation}
and we omit the normalization factor $L^3/\kappa_5$ as well
as the factor two related to ${\mathbb Z}_2$ symmetry.
The variation of (\ref{e:app action 1}) reads
\begin{align}
\delta {\cal S}_{h_\bullet}(k) =
\frac{1}{2} \int_T dt \int_{y_b(t)}^{y_s} & \frac{dy}{y^3} 
\Big[(\partial_t h_\bullet)(\partial_t \delta h_\bullet^*)\\ 
- &(\partial_y h_\bullet) (\partial_y \delta h_\bullet^*) 
-k^2 h_\bullet \delta h_\bullet^* \Big] + {\rm h.c.}~.\nonumber
\end{align}
Here, $T$ denotes a time interval within the variation is
performed and it is assumed in the following that the
variation vanishes at the boundaries of the time interval $T$.
Performing partial integrations and demanding 
that the variation of the action vanishes leads 
to 
\begin{align}
&0= \\
&\int_T dt \int_{y_b(t)}^{y_s} \frac{dy}{y^3}
\Big\{-\partial_t^2 h_\bullet + 
y^3\left[\partial_y \left(\frac{h_\bullet}{y^3} \right)\right] 
-k^2 h_\bullet \Big\} \delta h_\bullet^* \nonumber \\
&+\int_T dt \Big\{ \frac{1}{y^3} 
\left[\left(v \partial_t  + \partial_y \right)h_\bullet\right]
\delta h_\bullet^* |_{y_b(t)} - 
\frac{1}{y^3}(\partial_y h_\bullet) \delta h_\bullet^*|_{y_s}\Big\}
\nonumber
\end{align}
with $v = dy_b(t)/dt$.
The first term in curly brackets is the wave 
operator (\ref{e:T-bulk-eq}).
In order for $h_\bullet$ to satisfy the free wave 
equation (perturbation equation) (\ref{e:T-bulk-eq}) 
the term in curly brackets in the second integral has to vanish. 
Allowing for an evolution of $h_\bullet$ on the branes, 
i.e. in general $\delta h_\bullet|_{\rm brane} \neq 0$, 
enforces the boundary conditions 
\begin{equation}
\left(v \partial_t  + \partial_y \right)h_\bullet|_{y_b(t)} = 0
\;\;{\rm and}\;\;
\partial_y h_\bullet|_{y_s} = 0~, 
\label{e:app bcs}
\end{equation} 
hence, the junction condition (\ref{e:T-JC-simple}).
Consequently, any other boundary conditions than 
(\ref{e:app bcs}) are not compatible with the free 
perturbation equation (\ref{e:T-bulk-eq}) 
under the influence of a moving brane
(provided $\delta h_\bullet \neq 0$ at the branes). 
%
\section{Coupling matrices} \label{a:MN}
%
The use of several identities of Bessel functions leads to
\begin{align}
 M_{00} =&~\hat{y}_b \frac{\yreg^2}{\yreg^2-\ybr^2}~,\\
 M_{0j} =&~ 0~, 
 \label{e:T:Mij-exact 0j}\\
 M_{i0} =&~\frac{4N_i}{\pi m_i} \frac{\hat{y}_b}{y_b}\phi_0 =
\hat{y}_b\, \frac{4}{\pi m_i} \,N_i\, \frac{y_s}{\sqrt{y_s^2-y_b^2}}
~,\label{e:T:Mij-exact i0} \\
 M_{ii} =&~ \TKKR_i~,
 \label{e:T:Mij-exact ii}\\
 M_{ij} =&~ M_{ij}^{\rm A} + M_{ij}^{\rm N}
 \label{e:T:Mij-exact}
\end{align}
with
\begin{align}
&M_{ij}^{\rm A} = (\hat{y}_b+\hat{m}_i) y_b \frac{2\,m^2_i N_i N_j}{m_j^2-
                 m_i^2} \times \label{e:M ij A}\\
                 &\times\left [y_s\,{\cal C}_2(m_jy_s)\,{\cal J}_1(m_iy_s) -
                 y_b\,{\cal C}_2(m_jy_b)\,{\cal J}_1(m_iy_b)\right]\nonumber
\end{align}
where
\begin{equation}
{\cal J}_1 (m_i\,y)=
\left[J_2(m_iy_b)Y_1(m_iy) - Y_2(m_iy_b)J_1(m_iy)\right]
\end{equation}
and
\begin{equation}
M_{ij}^{\rm N} = N_i N_j m_i \hat{m}_i \int_{y_b}^{y_s}
dy y^2{\cal C}_1(m_i y){\cal C}_2(m_j y).
\label{e:M ij N}
\end{equation}
This integral has to be solved numerically.
Note that, because of the boundary conditions, one has the identity
\begin{equation}
\int_{y_b}^{y_s} dy y^2{\cal C}_1(m_i y){\cal C}_2(m_j y) = -
\int_{y_b}^{y_s} dy y^2{\cal C}_1(m_i y){\cal C}_0(m_j y).
\end{equation}
Furthermore, one can simplify
\begin{equation}
{\cal J}_1 (m_i\,y_b) =\frac{2}{\pi m_i y_b}\;,\;\;
{\cal J}_1 (m_i\,y_s) = \frac{2}{\pi m_i y_b}\frac{Y_1(m_iy_s)}{Y_1(m_iy_b)}
\end{equation}
where the limiting value has to be taken for the last term whenever
$Y_1(m_iy_b)=Y_1(m_iy_s)=0$.
%
\section{On power spectrum and energy density calculation} 
%
\subsection{Instantaneous vacuum}
\label{a:inst}
%
In Section~\ref{sec:III} the in - out state approach to particle
creation has been presented. 
The definitions of the in - and out- vacuum states 
Eq.~(\ref{vacuum definitions}) are unique and the 
particle concept is well defined and meaningful.
\\
\\
If we interpret $t_{\rm out}$ as a continuous time variable $t$,
we can write the Bogoliubov transformation Eq.~(\ref{bogoliubov trafo})
as
\begin{equation}
\hat{a}_{\alpha,{\bf k},\bullet}(t) =
\sum_\beta \left[{\cal A}_{\beta\alpha,k}(t)
\hat{a}_{\beta,{\bf k},\bullet}^{\rm in} +
{\cal B}^*_{\beta\alpha,k}(t)
\hat{a}_{\beta,{\bf -k},\bullet}^{\rm in\,\dagger}\right]~,
\label{e:appendix instantaneous Bogoliubov transformation}
\end{equation}
where at any time we have introduced a set of operators
$\{\hat{a}_{\alpha,{\bf k} \bullet}(t),
\hat{a}_{\alpha,{\bf k}, \bullet}^{\dagger}(t)\}$.
Vacuum states defined at any time can be associated with these
operators via 
\begin{equation}
 \hat{a}_{\alpha,{\bf k}, \bullet}(t)|0,t\rangle = 0\;\;
\forall \;\alpha, {\bf k}\, \bullet.
\end{equation}
Similar to Eq.~(\ref{e:graviton number definition}) a "particle number" can be
introduced through 
\begin{eqnarray}
{\cal N}_{\alpha,k}(t) &=& \sum_\bullet \langle0,{\rm in}|  \hat{a}_{\alpha,{\bf k} \bullet}^{\dagger}(t)
\hat{a}_{\alpha,{\bf k}, \bullet}(t)|0,{\rm in}\rangle \nonumber \\ &=&
2 \sum_\beta |{\cal B}_{\beta\alpha,k}(t)|^2~.
\label{e:appendix instantaneous particle number}
\end{eqnarray}
We shall denote $|0,t\rangle$ as the instantaneous vacuum state and the
quantity ${\cal N}_{\alpha,k}(t)$ as instantaneous particle number
\footnote{It could be interpreted as the number of particles 
which would have been created if
the motion of the boundary (the brane) stops at time t.}.
However, even if we call it "particle number" and plot it in section 
\ref{sec:num} for
illustrative reasons, we consider only the particle definitions
for the initial and final state (asymptotic regions)
outlined in section~\ref{sec:III} as physically meaningful. 
%
\subsection{Power spectrum}
\label{a:power}
%
In order to calculate the power spectrum
Eq.~(\ref{def power spectrum}) we need to
evaluate the expectation value
\begin{align}
&\langle \hat{h}_\bb(t,y_b,{\bf k})
\hat{h}^\dagger_\bb(t,y_b,{\bf k'})\rangle_{\rm in} =
\label{e:appendix h expectation value}
\\
&\frac{\kappa_5}{L^3}\sum_{\alpha\alpha'} \phi_\alpha(t,y_b)
\phi_{\alpha'}(t,y_b)
\langle \hat{q}_{\alpha,{\bf k},\bb}(t)
\hat{q}^\dagger_{\alpha',{\bf k'},\bb}(t)
\rangle_{\rm in}
\nonumber
\end{align}
where we have introduced the shortcut
$\langle ... \rangle_{\rm in} =
\langle 0,{\rm in}| ... |0,{\rm in}\rangle$.
Using the expansion (\ref{q expansion}) of
$\hat{q}_{\alpha', {\bf k'}, \bullet}(t)$ in initial state operators
and complex functions $\epsilon_{\alpha,k}^{(\gamma)}(t)$
one finds
\begin{equation}
\langle \hat{q}_{\alpha,{\bf k}, \bullet}(t)
\hat{q}^\dagger_{\alpha', {\bf k'}, \bullet}(t) \rangle_{\rm in}
=\sum_{\beta} \frac{\epsilon_{\alpha,k}^{(\beta)}(t)
\,\epsilon_{\alpha',k}^{(\beta)*}(t)} {2\omega_{\beta,k}^{\rm in}}
\,\delta^{(3)}({\bf k} - {\bf k'}).
\label{e:appendix q expectation value}
\end{equation}
From the initial conditions (\ref{e:initial conditions for epsilon})
it follows that the sum in (\ref{e:appendix h expectation value})  diverges at
$t=t_{\rm in}$.
This divergence is related to the usual normal ordering
problem and can be removed by a subtraction scheme.
However, in order to obtain a well defined power spectrum at all times,
it is not sufficient just to subtract the term
$(1/2)(\delta_{\alpha\alpha'}/\omega_{\alpha,k}^{\rm in})
\delta^{(3)}({\bf k} - {\bf k'})$
which corresponds to $\langle \hat{q}_{\alpha,{\bf k},\bb}(t_{\rm in})
\hat{q}^\dagger_{\alpha',{\bf k'},\bb}(t_{\rm in})
\rangle_{\rm in}$ in the above expression.
In order to identify all terms contained in the power spectrum we use the
instantaneous particle concept which allows us
to treat the Bogoliubov coefficients
(\ref{e:BogA}) and (\ref{e:BogB}) as continuous functions of time.
First we express the complex functions $\epsilon_{\alpha,k}^{(\beta)}$
in (\ref{e:appendix q expectation value}) in terms of
${\cal A}_{\gamma\alpha,k}(t)$ and ${\cal B}_{\gamma\alpha,k}(t)$.
This is of course equivalent to calculating the expectation value
(\ref{e:appendix q expectation value}) using [cf.~Eq.(\ref{e:final q})]
\begin{align}
\hat{q}_{\alpha,k,\bullet}(t) = \frac{1}{\sqrt{2\omega_{\alpha,k}(t)}}
\Big[&\hat{a}_{\alpha,{\bf k},\bullet}(t) \Theta_{\alpha,k}(t) 
\nonumber \\ 
&+\hat{a}^{\dagger}_{\alpha,{\bf -k},\bullet}(t) \Theta^*_{\alpha,k}(t)
\Big]
\end{align}
and the Bogoliubov transformation
Eq.~(\ref{e:appendix instantaneous Bogoliubov transformation}).
The result consists of terms involving the Bogoliubov coefficients
and the factor $(1/2) (\delta_{\alpha\alpha'}/
\omega_{\alpha,k}(t))\delta^{(3)}({\bf k} -{\bf k'})$, leading
potentially to a divergence at all times. This term corresponds to
$\langle 0,t|\hat{q}_{\alpha,{\bf k},\bb}(t)
\hat{q}^\dagger_{\alpha',{\bf k'},\bb}(t)
|0,t\rangle$, and is related to the normal ordering problem
(zero-point energy) with respect to the instantaneous
vacuum state $|0,t\rangle$.
It can be removed by the subtraction scheme
\begin{align}
&\langle \hat{q}_{\alpha,{\bf k}, \bullet}(t)
\hat{q}^\dagger_{\alpha', {\bf k'}, \bullet}(t) \rangle_{\rm in, phys}
\label{e:substraction scheme}
\\
&=\langle \hat{q}_{\alpha,{\bf k}, \bullet}(t)
\hat{q}^\dagger_{\alpha', {\bf k'}, \bullet}(t) \rangle_{\rm in}
-\langle 0,t|\hat{q}_{\alpha,{\bf k}, \bullet}(t)
\hat{q}^\dagger_{\alpha', {\bf k'}, \bullet}(t) |0,t\rangle
\nonumber
\end{align}
where we use the subscript ``${\rm phys}$'' to denote the physically
meaningful expectation value.
\\
Inserting this expectation value into 
(\ref{e:appendix h expectation value}),
and using Eq.~(\ref{e:phi on brane}), we find
\begin{align}
&\langle \hat{h}_{\bullet}(t,y_b,{\bf k})
         \hat{h}_{\bullet}(t,y_b,{\bf k'})\rangle_{\rm in}\\
&=\frac{1}{a^2}\frac{\kappa_5}{L}\sum_{\alpha}
{\cal R}_{\alpha,k}(t) {\cal Y}^2_\alpha(a)
\delta^{(3)}({\bf k} - {\bf k'})
\nonumber
\end{align}
with ${\cal R}_{\alpha,k}(t)$ defined in 
Eq.~(\ref{e:R function with inst particle number}).
The function ${\cal O}^{{\cal N}}_{\alpha,k}$ appearing in
Eq.~(\ref{e:R function with inst particle number})
is explicitely given by
\begin{align}
&{\cal O}^{{\cal N}}_{\alpha,k} =
2\,\Re\;\sum_\beta \Big\{ \Theta_{\alpha,k}^2
{\cal A}_{\beta\alpha,k}{\cal B}_{\beta\alpha,k}^*
 + \Theta_{\alpha,k}\sum_{\alpha'\neq \alpha}
\times \nonumber \\
&\times \sqrt{\frac{\omega_{\alpha,k}}{\omega_{\alpha',k}}}
\frac{{\cal Y}_{\alpha'}(a)}{{\cal Y}_{\alpha}(a)}
\left[
\Theta_{\alpha',k}^*{\cal B}_{\beta\alpha}^*{\cal
  B}_{\beta\alpha'} +
\Theta_{\alpha',k}{\cal A}_{\beta\alpha}{\cal
  B}^*_{\beta\alpha'}
\right]
\Big\}
\label{e:O in terms of A and B}
\end{align}
and ${\cal O}_{\alpha,k}^{\epsilon}$ appearing in
Eq.~(\ref{e:R function with epsilon}) reads
\begin{equation}
{\cal O}_{\alpha,k}^{\epsilon} =
\sum_{\beta,\alpha'\neq\alpha} \frac{{\cal Y}_{\alpha'}(a)}
{{\cal Y}_{\alpha}(a)}\,
\frac{\epsilon_{\alpha,k}^{(\beta)}\epsilon_{\alpha',k}^{(\beta)^*}}
{\omega_{\beta,k}^{\rm in}}.
\label{e:O in terms of epsilon}
\end{equation}
%
\subsection{Energy density}
\label{a:energy}
%
In order to calculate the energy density we need to evaluate
the expectation value $\langle \dot{\hat{h}}_{ij}(t,{\bf x},y_b)
\dot{\hat{h}}^{ij}(t,{\bf x},y_b)\rangle_{\rm in}$.
Using (\ref{e:h fourier decomposition}) and the relation
$e_{ij}^{\bb}({\bf -k}) = (e_{ij}^{\bb}({\bf k}))^*$
we obtain
\begin{align}
&\langle \dot{\hat{h}}_{ij}(t,{\bf x},y_b)
\dot{\hat{h}}^{ij}(t,{\bf x},y_b)\rangle_{\rm in}=
\sum_{\bb \bb'} \int
\frac{d^3k}{(2\pi)^{3/2}}\frac{d^3k'}{(2\pi)^{3/2}}
\times
\label{e:h dot ij expectation value}
\\
&\times \langle \dot{\hat{h}}_\bb(t,y_b,{\bf k})
\dot{\hat{h}}^\dagger_{\bb'}(t,y_b,{\bf k'})\rangle_{\rm in}
e^{i({\bf k} - {\bf k'})\cdot {\bf x}}
e_{ij}^{\bb}({\bf k})\;\left(e^{\bb' \;ij}({\bf k'})\right)^*.
\nonumber
\end{align}
By means of the expansion (\ref{e:h dot as function of p})
the expectation value
$\langle \dot{\hat{h}}_\bb(t,y_b,{\bf k})
\dot{\hat{h}}^\dagger_{\bb'}(t,y_b,{\bf k'})\rangle_{\rm in}$
becomes
\begin{align}
&\langle \dot{\hat{h}}_\bb(t,y_b,{\bf k})
\dot{\hat{h}}^\dagger_{\bb'}(t,y_b,{\bf k'})\rangle_{\rm in}
\label{e:h dot expectation value}\\
&=\frac{\kappa_5}{L^3} \sum_{\alpha\alpha'} \langle
\hat{p}_{\alpha,{\bf k},\bb}(t) \hat{p}^\dagger_{\alpha',{\bf
    k'},\bb'}(t)\rangle_{\rm in}
\phi_\alpha(t,y_b)\phi_{\alpha'}(t,y_b).
\nonumber
\end{align}
From the definition of $\hat{p}_{\alpha,{\bf k},\bb}(t)$
in Eq.~(\ref{e:definition of momentum p}) it is clear that
this expectation value will in general contain terms
proportional to the coupling matrix and its square when
expressed in terms of $\epsilon_{\alpha,k}^{(\beta)}$.
However, we are interested in the expectation value
at late times only when the brane moves very slowly 
such that the mode couplings go to zero
and a physical meaningful particle definition can be given.
In this case we can set
\begin{eqnarray}
\langle
\hat{p}_{\alpha,{\bf k},\bb}(t)
\hat{p}^\dagger_{\alpha',{\bf k'},\bb'}(t)\rangle_{\rm in}
=\left\langle \dot{\hat{q}}_{\alpha,{\bf k},\bb}(t)
\dot{\hat{q}}^\dagger_{\alpha',{\bf k'},\bb'}(t)  \right\rangle_{\rm in}.
\end{eqnarray}
Calculating this expectation value by using
Eq.~(\ref{q expansion}) leads to an expression which,
as for the power spectrum calculation before,
has a divergent part related to the zero-point energy of the
instantaneous vacuum state (normal ordering problem).
We remove this part by a subtraction
scheme similar to Eq~(\ref{e:substraction scheme}).
The final result reads
\begin{align}
&\langle \dot{\hat{q}}_{\alpha,{\bf k}, \bullet}(t)
\dot{\hat{q}}^\dagger_{\alpha', {\bf k'}, \bullet'}(t)
\rangle_{\rm in, phys}\\
&=
\frac{1}{2}\left[\sum_{\beta}\frac{\dot{\epsilon}_{\alpha,k}^{(\beta)}(t)
\dot{\epsilon}_{\alpha',k'}^{(\beta)^*}(t)}
{\sqrt{\omega_{\beta,k}^{\rm in}\omega_{\beta,k'}^{\rm in}}}
-
\omega_{\alpha,k}(t) \delta_{\alpha\alpha'}
\right]\delta_{\bb\bb'}\delta^{(3)}({\bf k} - {\bf k'}).
\nonumber
\end{align}
Inserting this result into Eq.~(\ref{e:h dot expectation value}),
splitting the summations in sums over $\alpha = \alpha'$
and $\alpha \neq \alpha'$ and neglecting the oscillating
$\alpha \neq \alpha'$ contributions (averaging over
several oscillations), leads to
\begin{align}
&\langle \dot{\hat{h}}_\bb(t,y_b,{\bf k})
\dot{\hat{h}}^\dagger_{\bb'}(t,y_b,{\bf k'})\rangle_{\rm in}\\
&=
\frac{1}{a^2}\frac{\kappa_5}{L}
\sum_{\alpha}{\cal K}_{\alpha,k}(t){\cal Y}_{\alpha}^2(a)
\delta_{\bb\bb'}\delta^{(3)}({\bf k} - {\bf k'})
\nonumber
\end{align}
where the function ${\cal K}_{\alpha,k}(t)$ is given by
\begin{equation}
{\cal K}_{\alpha,k}(t) = \sum_\beta \frac{|\dot{\epsilon}_{\alpha,k}^{(\beta)}(t)|^2}
{\omega_{\beta,k}^{\rm in}} - \omega_{\alpha,k}(t)
=\omega_{\alpha,k}(t) {\cal N}_{\alpha,k}(t)~,
\label{K function}
\end{equation}
and we have made use of Eq.~(\ref{e:phi on brane}). 
The relation
between $\sum_{\beta} |\dot{\epsilon}_{\alpha,k}^{(\beta)}(t)|^2
/\omega_{\beta,k}^{\rm in}$ and the number of created
particles can easily be established.
Using this expression in Eq.~(\ref{e:h dot ij expectation value})
leads eventually to
\begin{align}
&\langle\dot{\hat{h}}_{ij}(t,{\bf x},y_b)
\dot{\hat{h}}^{ij}(t,{\bf x},y_b)\rangle_{\rm in}\\
&=\frac{1}{a^2}\frac{\kappa_5}{L}\sum_{\alpha}
\int \frac{d^3k}{(2\pi)^{3}}
{\cal K}_{\alpha,k}(t) {\cal Y}^2_\alpha(a)
\nonumber
\end{align}
where we have used that the polarization tensors satisfy
\begin{equation}
\sum_{\bb} e_{ij}^{\bb}({\bf k})\;\left(e^{\bb \;ij}({\bf k})\right)^*
= 2.
\end{equation}
The final expression for the energy density
Eq.~(\ref{energy density})
is then obtained by exploiting that
$\kappa_5/L = \kappa_4$.
%
\section{Numerics} \label{a:numerics}
%
In order to calculate the number of produced gravitons
the system of coupled differential equations
(\ref{deq for xi}) and (\ref{deq for eta}) is solved numerically.
The complex functions $\xi_{\alpha,k}^{(\beta)}$, 
$\eta_{\alpha,k}^{(\beta)}$
are decomposed into their real and imaginary parts:
\begin{equation}
\xi_{\alpha,k}^{(\beta)} = u_{\alpha,k}^{(\beta)} + i
  v_{\alpha,k}^{(\beta)}\;,\;\;
\eta_{\alpha,k}^{(\beta)} = x_{\alpha,k}^{(\beta)} + i
  y_{\alpha,k}^{(\beta)}.
\end{equation}
The system of coupled differential equations can then
be written in the form (cf.~Eq.~(A2) of \cite{Ruser:2005xg})
\begin{equation} 
{\bf \dot{X}}_k^{(\beta)}(t)={\bf W}_k(t)
{\bf X}^{(\beta)}_k(t)
\label{e:num deq system}
\end{equation}
where
\begin{align}
&{\bf X}^{(\beta)}_k = \nonumber \\
&\left(
u_{0,k}^{(\beta)}... u_{n_{\rm max},k}^{(\beta)}
x_{0,k}^{(\beta)}... x_{n_{\rm max},k}^{(\beta)}
v_{0,k}^{(\beta)}... v_{n_{\rm max},k}^{(\beta)}
y_{0,k}^{(\beta)}... y_{n_{\rm max},k}^{(\beta)}
\right)^{\rm T}.
\end{align}
The matrix ${\bf W}_k(t)$ is given by Eq.~(A4) 
of \cite{Ruser:2005xg} but here indices 
start at zero. 
The number of produced gravitons can be calculated directly 
from the solutions to this system using 
Eqs.~(\ref{particle number}) and (\ref{e:Bout}).
Note that for a given truncation parameter $n_{\rm max}$
the above system of size 
$4(n_{\rm max}+1) \times 4(n_{\rm max}+1)$
has to be solved $n_{\rm max}+1$~-~times, each time 
with different initial conditions 
(\ref{e:vacuum initial conditions for xi,eta}).  
\\
The main difficulty in the numerical simulations is that
most of the entries of the matrix ${\bf W}_k(t)$
[Eq.~(A4) of \cite{Ruser:2005xg}] are not known analytically.
This is due to the fact that Eq.~(\ref{e:zero equation})
which determines the time-dependent KK masses $m_i(t)$
does not have an (exact) analytical solution.
Only the $00$-component of the coupling matrix
$M_{\alpha\beta}$ is known analytically.
We therefore have to determine the
time-dependent KK-spectrum $\{m_i(t)\}_{i=1}^{n_{\rm max}}$
by solving Eq.~(\ref{e:zero equation}) numerically.
In addition, also the part $M_{ij}^N$ [Eq.~(\ref{e:M ij N})] 
has to be calculated numerically since the integral over the
particular combination of Bessel functions can not be 
found analytically. 
\\
\\
We numerically evaluate the KK-spectrum and the 
integral $M_{ij}^N$ for discrete time-values $t_i$ 
and use spline routines to assemble ${\bf W}_k(t)$.
The system (\ref{e:num deq system}) can then be solved using
standard routines.  
We chose the distribution of the $t_i$'s in a non-uniform way. 
A more dense mesh close to the bounce and a less dense mesh at early
and late times. The independence of the numerical results on the 
distribution of the $t_i$'s is checked.   
In order to implement the bounce as realistic as 
possible, we do not spline the KK-spectrum very 
close to the bounce but re-calculate it numerically 
at every time $t$ needed in the differential 
equation solver. 
This minimizes possible artificial effects caused by using 
a spline in the direct vicinity of the bounce. 
The same was done for $M_{ij}^N$ but we found that 
splining $M_{ij}^N$ when propagating through the bounce does not 
affect the numerical results.
\\
Routines provided by the GNU Scientific Library (GSL)
\cite{gsl} have been employed. Different routines for root finding 
and integration have been compared. The code has been parallelized (MPI) 
in order to deal with the intensive numerical computations.    
\\
\\
The accuracy of the numerical simulations can be assessed by
checking the validity of the Bogoliubov relations
\begin{align}
&\sum_\beta\left[
  {\cal A}_{\beta\alpha,k}(t){\cal A}_{\beta\gamma,k}^*(t) - 
  {\cal B}_{\beta\alpha,k}^*(t){\cal
  B}_{\beta\gamma,k}(t)\right]=\delta_{\alpha\gamma} 
\label{e:bogrelation one}\\
&\sum_\beta\left[
  {\cal A}_{\beta\alpha,k}(t){\cal B}_{\beta\gamma,k}^*(t) - 
  {\cal B}_{\beta\alpha,k}^*(t){\cal A}_{\beta\gamma,k}(t)\right]=0.
\label{e:bogrelation two}
\end{align}
In the following we demonstrate the accuracy of the numerical
simulations by considering the diagonal part of 
(\ref{e:bogrelation one}). 
The deviation of the quantity 
\begin{equation}
d_{\alpha,k}(t) = 1 - \sum_\beta\left[ |{\cal A}_{\beta\alpha,k}(t)|^2 
- |{\cal B}_{\beta\alpha,k}(t)|^2\right]
\end{equation}
from zero gives a measure for the accuracy of the numerical result.
We consider this quantity at final times $t_{\rm out}$ and 
compare it with the corresponding final particle spectrum.
In Fig.~\ref{f:accuracy} we compare the final KK-graviton 
spectrum ${\cal N}_{n,k,\bullet}^{\rm out}$ with the 
expression $d_{n,k}(t_{\rm out})$ for two different cases. 
This shows that the accuracy of the numerical simulations is very
good. 
Even if the expectation value for the particle number is only 
of order $10^{-7}$ to $10^{-6}$, 
the deviation of $d_{n,k}(t_{\rm out})$ from zero is 
at least one order of magnitude smaller. 
This demonstrates the reliability of our numerical simulations
and that we can trust the numerical results presented in 
this work. 
\begin{figure}
\begin{center}
\includegraphics[height=6cm]{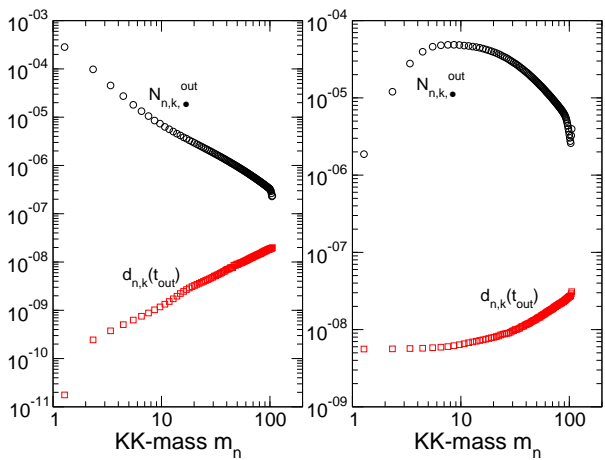}
\caption{Comparison of the final KK-graviton spectrum 
${\cal N}_{n,k,\bullet}^{\rm out}$ with the expression 
$d_{n,k}(t_{\rm out})$ describing to what accuracy the 
diagonal part of the Bogoliubov relation (\ref{e:bogrelation one})
is satisfied. 
Left panel: $y_s=3$, $k=0.1$, $v_b=0.03$ 
and $n_{\rm max}=100$ [cf.~Fig.~\ref{f:kkfig14}].
Right panel: $y_s=3$, $k=30$, $v_b=0.1$ 
and $n_{\rm max}=100$ [cf.~Fig.~\ref{f:bkfig1}].
\label{f:accuracy}}
\end{center}
\end{figure}
\\
%
\section{Dynamical Casimir effect for a uniform motion} \label{a:dyncas}
%
We consider a real massless scalar field on
a time-dependent interval $[0,y(t)]$.
The time evolution of its mode functions are described by a system
of differential equations like (\ref{deq for q}) where the specific
form of $M_{\alpha\beta}$ depends on the particular boundary
condition the field is subject to.
In \cite{Ruser:2006xg,Ruser:2004} a method has been
introduced to study particle creation due to the motion of the
boundary $y(t)$ (i.e. the dynamical Casimir effect) fully
numerically. We refer the reader to these publications
for further details.
\\
If the boundary undergoes a uniform motion $y(t) = 1 + vt$
(in units of some reference length) it was shown in
\cite{Moore:1970,Castagnino:1984} that the total number of created
scalar particles diverges, caused by the discontinuities in the
velocity at the beginning and the end of the motion.
In particular, for Dirichlet boundary conditions (no zero mode),
it was found in \cite{Castagnino:1984} that
$\langle 0, {\rm in}|\hat{N}_n^{\rm out}|0,{\rm in}\rangle\propto v^2/n$
if $n > 6$ and $ v \ll 1$.
Thereby in- and out- vacuum states are defined like in the present
work and the frequency of a mode function is given by
$\omega_n=\pi\,n\;,\;\;n=1,2,...$~.
In Figure~\ref{f:linmo} we show spectra of created scalar particles
obtained numerically with the method of \cite{Ruser:2006xg}
for this particular case. One observes that, as for our bouncing
motion, the convergence is very slow since
the discontinuities in the velocity lead to the excitation 
of arbitrary high frequency modes.
Nevertheless, it is evident from Fig.~\ref{f:linmo}
that the numerically calculated spectra approach the analytical
prediction. 
The linear motion discussed here and the brane-motion (\ref{e:yb}) are
very similar with respect to the discontinuities in the velocity.
In both cases, the total discontinuous change of the velocity is
$2v$ and $2v_b$, respectively. The resulting
divergence of the acceleration is responsible for the 
excitation and therefore creation of particles 
of all frequency modes.
Consequently we expect the same $\propto v^2/\omega_n$ behavior for the
bouncing motion (\ref{e:yb}).
Indeed, comparing the convergence behavior of the final 
graviton spectrum for $v_b=0.01$ shown in Fig.~\ref{f:kkfig14}
with the one of the scalar particle spectrum for $v=0.01$ depicted 
in Fig.~\ref{f:linmo} shows that both are very similar. 
\begin{figure}
\begin{center}
\includegraphics[height=6cm]{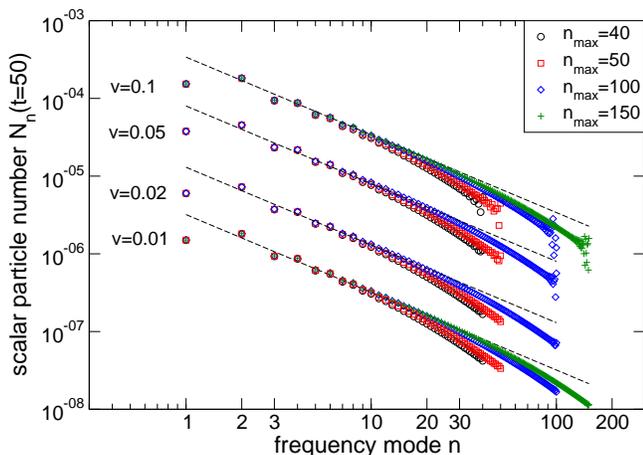}
\caption{Spectra of massless scalar particles produced under the influence
of the uniform motion $y(t) = 1 + vt$ for velocities
$v=0.01,0.02,0.05$ and $0.1$. The numerical results are 
compared to the expression $N_n=0.035 v^2/n$ (dashed lines) 
which agrees with the analytical prediction $N_n\propto v^2/n$.
\label{f:linmo}}
\end{center}
\end{figure}
\end{appendix}

\end{document}